\documentclass[epj]{svjour}
\usepackage{graphicx}

\def\mb#1{\setbox0=\hbox{$#1$}\kern-.025em\copy0\kern-\wd0
\kern-0.05em\copy0\kern-\wd0\kern-.025em\raise.0233em\box0}

\begin{document}
   \title{Dynamics and thermodynamics of a simple model similar to  \\
self-gravitating systems: the HMF model}

 \author{P.H. Chavanis\inst{1}, J. Vatteville\inst{1} and  F. Bouchet\inst{2}}

\institute{ Laboratoire de Physique Th\'eorique, Universit\'e Paul
Sabatier, 118 route de Narbonne 31062 Toulouse, France\\
\email{chavanis@irsamc.ups-tlse.fr} \and  Ecole Normale Sup\'erieure de Lyon, 
46 All\'ee d'Italie, 69364 Lyon, France\\
\email{Freddy.Bouchet@ens-lyon.fr}}

\titlerunning{Dynamics and thermodynamics of the HMF model}

   \date{To be included later }

   \abstract{We discuss the dynamics and thermodynamics of the
   Hamiltonian Mean Field model (HMF) which is a prototypical system
   with long-range interactions. The HMF model can be seen as the one
   Fourier component of a one-dimensional self-gravitating
   system. Interestingly, it exhibits many features of real
   self-gravitating systems (violent relaxation, persistence of
   metaequilibrium states, slow collisional dynamics, phase
   transitions,...) while avoiding complicated problems posed by the
   singularity of the gravitational potential at short distances and
   by the absence of a large-scale confinement. We stress the deep
   analogy between the HMF model and self-gravitating systems by
   developing a complete parallel between these two systems. This
   allows us to apply many technics introduced in plasma physics and
   astrophysics to a new problem and to see how the results depend on
   the dimension of space and on the form of the potential of
   interaction. This comparative study brings new light in the
   statistical mechanics of self-gravitating systems. We also mention
   simple astrophysical applications of the HMF model in relation with
   the formation of bars in spiral galaxies.
       }

   \maketitle
%

\section{Introduction}
\label{sec_introduction}

The statistical mechanics of systems with long-range interactions is
currently a topic of active research in physics because it differs in many
respects from that of more familiar systems with short-range forces that are
extensive (Dauxois et al. 2002a). Among long-range interactions,
gravity is probably the most important and most fundamental example
(Padmanabhan 1990, Chavanis 2002a).  However, the statistical mechanics
of self-gravitating systems initiated by Antonov (1962) and
Lynden-Bell (1968) is complicated due to the divergence of the
gravitational force at short distances and to the absence of shielding
(or confinement) at large distances. These difficulties are specific
to the gravitational force and not to the long-range nature of the
interaction. Therefore, it may be of conceptual interest to consider
simpler systems with long-range interactions to distinguish what is
specific to the gravitational force and what is common to systems with
long-range interactions.

A toy model of systems with long-range interactions is the
so-called HMF (Hamiltonian Mean Field) model. It consists of $N$
particles moving on a circle and interacting via a cosine binary
potential. This can be seen as a one-dimensional plasma where the
potential of interaction is truncated to one mode. This model is
of great conceptual interest because it exhibits many features
present in more realistic systems with long-range interactions
such as gravitational systems. In addition, it is sufficiently
simple to allow for accurate numerical simulations and analytical
results.

To our knowledge, what is now called the HMF model was first
introduced by Konishi \& Kaneko (1992). They found that a cluster is
formed in some cases and that the system remains uniform in other
cases. Inagaki \& Konishi (1993) realized that the Konishi-Kaneko
system is nothing but the one Fourier component of a one-dimensional
self-gravitating system and explained the formation of clusters as an
instability similar to the Jeans instability in self-gravitating
systems described by the Vlasov equation. Inagaki (1993) studied the
thermodynamical stability of the Konishi-Kaneko system and identified
the existence of a second order phase transition at a critical
temperature $T_c$.  Above $T_c$ the only statistical equilibrium state
is uniform, whereas below $T_c$ this uniform state looses its
thermodynamical stability and clustered states appear. In order to
justify his results dynamically, Inagaki (1996) developed a
``collisional'' kinetic theory of the Konishi-Kaneko system based on
results of plasma physics and proposed to model the dynamics of the
system by the Lenard-Balescu equation for a one dimensional plasma
truncated to one mode. However, as we shall see, his conclusions
demand further discussion.

The same model was considered at about the same time by Pichon and
Lynden-Bell (Pichon 1994) who gave an astrophysical
application of this model in relation with the formation of {\it bars}
in galactic disks. In their approach, the stars follow rigid elliptical
orbits with eccentricity $e$. If $\phi_i$ represents the inclination
of ellipse $i$ and $\Omega_{i}$ its angular velocity, the torque
exerted by an orbit to the other can be written
$\alpha^{-1}d\Omega_{1}/dt=\partial\psi_{12}/\partial\phi_{1}$ where
$\alpha^{-1}$ is the adiabatic moment of inertia of the inner Lindblad
orbit and $\psi_{12}=G A^{2}\cos 2(\phi_{1}-\phi_{2})$ is the
effective alignement potential. At high temperatures, the orbits are
almost uniformly distributed in space and the system is in a disk
phase (see Fig. \ref{ellipsgaz}). However, below some critical
temperature $T_c$, the ellipses tend to align to each other and form a
bar (see Fig. \ref{ellipsbarre}). Those bars are reported
observationally in real galaxies. Pichon and Lynden-Bell studied the
linear stability of these bars with respect to the Vlasov equation and
proposed that the clustered phase could result from a process of
violent relaxation, a concept introduced by Lynden-Bell
(1967) to explain the regularity of collisionless stellar systems such
as elliptical galaxies.

The Konishi-Kaneko system, now called the HMF model, also appeared in
statistical mechanics (Antoni \& Ruffo 1995). In that context, the
motivation was to devise a simple model with long-range interactions
keeping the richness of more realistic systems while being amenable to
a full analytical and numerical treatement. Excitingly, this simple
model displays a lot of interesting features (violent relaxation,
persistence of metaequilibrium states, slow collisional relaxation,
phase transitions,...) also present in other systems with long-range
interactions such as stellar systems and 2D vortices (Chavanis
2002a). The properties of the HMF model have been studied in great
detail in a lot of recent papers (see Dauxois et al. 2002b for a
review). Despite its oversimplification, the HMF model can be seen as
a pedagogical model to take a step into the physics of systems with
long-range interactions. It is said sometimes to represent the
``harmonic oscillator'' of systems with long-range interactions. This
probably explains its popularity.

In the present paper, we shall emphasize the connection between the HMF
model and the results established in astrophysics and plasma physics.
In particular, we will adapt the methods developed for 3D
self-gravitating systems to the case of a one-dimensional system of
particles with cosine interactions. The motivation of
this extension is two-fold. The first is to show that the  results
obtained in astrophysics and plasma physics can have applications in
other domains of physics, including the HMF model (this has not been
sufficiently appreciated by workers in that field since the early work
of Inagaki). The second is to stress the analogies and the
differences which appear in long-range systems as we change the
dimension of space and the potential of interaction. Among the
analogies between 3D self-gravitating systems and the HMF model, we note:
the concept of violent relaxation and the slow collisional
dynamics. Among the differences, we note: the equivalence of
statistical ensembles for the HMF model (contrary to 3D
gravitational systems), the existence of second order phase
transitions (instead of first order or zeroth order phase transitions
for 3D gravitational systems) and the vanishing of the collision
operator at the order $1/N$ in the BBGKY hierarchy contrary to the
Coulombian or Newtonian case.

The paper is organized as follows. In Sec. \ref{sec_stat}, we consider
the statistical equilibrium states of the HMF model in both
microcanonical and canonical ensembles. We synthesize previously known
results and we derive explicit criteria of thermodynamical stability
for the uniform phase as well as for the clustered phase. We also
describe corrections to the mean-field approximation close to the
critical point. In Sec. \ref{sec_gas}, we consider a one-dimensional
gaseous system with cosine interactions (the analogue of a ``gaseous
star'') described by the Euler equations with a barotropic equation of
state. We discuss in particular the equivalent of the Jeans
instability.  In Sec. \ref{sec_vlasov}, we consider the collisionless
evolution of the HMF model (the analogue of a ``stellar system'')
described by the Vlasov equation and discuss the concept of violent
relaxation and metaequilibrium states. We interprete these
quasi-equilibrium states as particular stationary solutions of the
Vlasov equation on the coarse-grained scale resulting from phase
mixing and incomplete violent relaxation. We regard Tsallis functional
$S_{q}[f]=-{1\over q-1}\int (f^{q}-f)d\theta dv$ and Boltzmann
functional $S_{B}[f]=-\int f \ln fd\theta dv$ as particular
$H$-functions in the sense of Tremaine {\it et al.} (1986) associated
with polytropic and isothermal distributions. We study the dynamical
stability of stationary solutions of the Vlasov equation and compare
with the dynamical stability of stationary solutions of the barotropic
Euler equations. This is the same type of comparison as between
``gaseous systems" and ``stellar systems" in astrophysics. In that
respect, we discuss the equivalent of the Antonov first law (Binney \&
Tremaine 1987) for the HMF model. We derive a criterion of nonlinear
dynamical stability for steady states of the Vlasov equation of the
form $f=f(\epsilon)$ with $f'(\epsilon)<0$ where $\epsilon$ is the
individual energy, and show that it can be written as a condition on
the velocity of sound in the corresponding barotropic gas. This
criterion is equivalent to the criterion obtained by Yamaguchi {\it et
al.} (2004) but it is expressed differently. We also analyze the
linear dynamical stability of steady states of the Vlasov equation and
study the dispersion relation for isothermal and polytropic
distributions. In Sec. \ref{sec_collisional}, we discuss the
collisional evolution of the HMF model and explain why the kinetic
theory is more complicated than for 3D Newtonian interactions. In
particular, the Landau and the Lenard-Balescu collision terms vanish for
$1D$ systems so that the evolution of the system as a whole is due to
terms of order smaller than $1/N$ in the expansion of the correlation
functions for $N\rightarrow +\infty$. This implies that the relaxation
time is larger than $Nt_{D}$ (where $t_{D}$ is the dynamical time). By
contrast, we can develop a kinetic theory at order $1/N$ to analyze
the relaxation of a ``test particle'' in a bath of ``field particles''
with static distribution $f_{0}(v)$. The evolution of the distribution
function $P(v,t)$ of the velocity of the test particle satisfies a
Fokker-Planck equation. We give explicit expressions for the diffusion
coefficient and the auto-correlation function and we compare their
expressions depending on whether collective effects are taken into
account or not. We also show that the auto-correlation function
decreases exponentially rapidly in time with a rate coinciding with
the damping rate $\gamma$ of a stable perturbed solution of the Vlasov
equation. Finally, in Sec. \ref{sec_brownian}, we discuss the case of
self-attracting Brownian particles described by non-local
Fokker-Planck equations. This stochastic model is the canonical
counterpart of the Hamiltonian $N$-body problem.  It could be called
the BMF (Brownian Mean Field) model. We study the dynamical stability
of steady states of the non-local Smoluchowski equation and solve this
equation numerically to show the formation of a clustered state from
an unstable homogeneous state due to long-range interactions. We also
provide analytical solutions of the dynamics close to the critical
point $T_{c}$ and for $T=0$. All these models have an equivalent in
the astrophysical literature and this paper stresses the analogies and
differences between self-gravitating systems and the HMF model. This
comparative study brings new light in the statistical mechanics of
self-gravitating systems by showing what is specific to gravity and
what is common to systems with long-range interactions.

\section{Statistical equilibrium}
\label{sec_stat}

\subsection{The mean-field approximation}
\label{sec_mf}

We consider a system of $N$ particles moving on a circle and
interacting via a cosine binary potential. This is the so-called
HMF model. As explained in the Introduction, this model can also
describe a system of stars moving on elliptical orbits, each orbit
exerting a torque on the others. Fundamentally, the dynamics of
this system is governed by the Hamilton equations
\begin{eqnarray}
\label{mf1}
m_{i}{d\theta_{i}\over dt}={\partial H\over\partial v_{i}}, \qquad m_{i}{dv_{i}\over dt}=-{\partial H\over\partial \theta_{i}},\nonumber\\
H=\sum_{i=1}^{N}{1\over 2}m_{i}v_{i}^{2}-{k\over 4\pi}\sum_{i\neq j}m_{i}m_{j}\cos(\theta_{i}-\theta_{j}),
\end{eqnarray}
where $\theta_{i}$ is the angle that makes particle/ellipse $i$ with
an axis of reference and $k$ is the coupling constant (similar to the
gravitational constant $G$). In the rest of the paper, we shall refer
to this system as a ``stellar system''; this is to emphasize the
analogies with real 3D stellar systems whose dynamics is also governed
by Hamiltonian equations with long-range interactions.  We have also
generalized the usual HMF model to a population of particles with
different masses $m_{i}$. However, in most of the paper, we shall
assume that all the particles have the same mass $m=1$. The
multi-species HMF model will be discussed specifically in
Sec. \ref{sec_multi}.

The evolution of the $N$-body distribution function is governed by the Liouville equation
\begin{eqnarray}
{\partial P_{N}\over\partial t}+\sum_{i=1}^{N}\biggl (v_{i}{\partial P_{N}\over\partial\theta_{i}}+F_{i}{\partial P_{N}\over\partial v_{i}}\biggr )=0
\label{liouville}
\end{eqnarray}
where $F_{i}=-{k\over 2\pi}\sum_{j=1}^{N}\sin(\theta_{i}-\theta_{j})$
is the force experienced by particle $i$. Any distribution of the form
$P_{N}=\chi(H)\delta(E-H)$ is a stationary solution of the Liouville
equation. For $N\gg 1$ (fixed) and $t\rightarrow +\infty$, this system
is expected to reach a statistical equilibrium state due to the
developement of correlations between particles (this will be refered
to as a ``collisional'' relaxation).  As is customary in statistical
mechanics, we shall {assume} that the equilibrium $N$-body
distribution function is described by the microcanonical distribution
\begin{equation}
\label{mf2} P_N(\theta_{1},v_{1},...,\theta_{N},v_{N})={1\over g(E)}\delta(E-H),
\end{equation}
expressing that all accessible microstates (with the right values of
energy and mass) are equiprobable. Whether this is indeed the case has
not been proved rigorously as it relies on a hypothesis of ergodicity,
so this statement is essentially a {\it postulate}.

For systems with long-range interactions (self-gravitating systems, 2D
vortices, HMF model,...), it can be shown that the mean-field
approximation is {\it exact} in an appropriate thermodynamic
limit. This can be shown for example by considering an equilibrium
BBGKY-like hierarchy (Chavanis 2004b,2005). For the HMF model, the
thermodynamic limit is $N\rightarrow +\infty$ in such a way that the
properly normalized energy $\epsilon={8\pi E/ kM^{2}}$ and temperature
$\eta={\beta kM/4\pi}$ are fixed, where $M=Nm$ is the total
mass. These control parameters are similar to those,
$\epsilon={ER/GM^{2}}$ and $\eta={\beta GMm/ R}$, describing 3D
gravitational systems (see, e.g., Chavanis 2003a). In that limit
$N\rightarrow +\infty$, the two-body distribution function can be
expressed as a product of two one-body distribution functions
\begin{equation}
\label{mf3} P_{2}(\theta_{1},v_{1},\theta_{2},v_{2})=P_{1}(\theta_{1},v_{1})P_{1}(\theta_{2},v_{2})+O(1/N).
\end{equation}
The average density of particles in phase space is given by
$f(\theta,v)=\langle\sum_{i}\delta(\theta-\theta_{i})\delta(v-v_{i})\rangle=NP_{1}(\theta,v)$. The total mass can then be
expressed as
\begin{equation}
\label{mf4} M=\int f d\theta dv.
\end{equation}
On the other hand, the average
energy $E=\langle H\rangle$ is
\begin{eqnarray}
\label{mf5} E=N\int P_{1}(\theta,v){v^{2}\over 2}d\theta dv\qquad\qquad\qquad\qquad\qquad\qquad\nonumber\\
-{k\over 4\pi}N(N-1)\int \cos(\theta-\theta')P_{2}(\theta,v,\theta',v')d\theta dv d\theta' dv'.
\end{eqnarray}
In the mean-field limit, it reduces to
\begin{equation}
\label{mf6} E={1\over 2}\int f v^{2} d\theta dv+{1\over 2} \int
f\Phi d\theta dv ,
\end{equation}
where
\begin{equation}
\label{mf7} \Phi(\theta)=-{k\over
2\pi}\int_{0}^{2\pi}\cos(\theta-\theta')\rho(\theta')d\theta',
\end{equation}
is the potential and $\rho=\int f dv$ is the spatial density. Note
that the average force experienced by a particle located in $\theta$
is $\langle F\rangle=-\Phi'(\theta)$. If $\rho$ is symmetric with respect to the $x$-axis, so that 
$\rho(-\theta)=\rho(\theta)$, the foregoing relation can
be rewritten
\begin{equation}
\label{mf8} \Phi(\theta)=B\cos\theta,
\end{equation}
where
\begin{equation}
\label{mf9} B=-{k\over
2\pi}\int_{0}^{2\pi}\rho(\theta')\cos\theta' d\theta'.
\end{equation}
This parameter $B$ is the equivalent of the magnetization (usually
denoted $M$) in the case of spin systems. Inserting the relation
(\ref{mf8}) in Eq. (\ref{mf6}), we find that the energy can be
rewritten
\begin{equation}
\label{mf10} E={1\over 2}\int f v^{2} d\theta dv-{\pi B^{2}\over
k},
\end{equation}
so that the potential energy is directly expressed in terms of
$B$.

\subsection{The Boltzmann entropy}
\label{sec_boltzmann}

We wish to determine the macroscopic distribution of particles at
statistical equilibrium, assuming that all accessible microstates (with
given $E$ and $M$) are equiprobable. To that purpose, we divide the
$\mu$-space $\lbrace {\theta},{v}\rbrace$ into a very large
number of microcells with size $h$. We do not put any exclusion, so
that a microcell can be occupied by an arbitrary number of particles.
We shall now group these microcells into macrocells each of which
contains many microcells but remains nevertheless small compared to
the phase-space extension of the whole system. We call $\nu$ the
number of microcells in a macrocell. Consider the configuration
$\lbrace n_i \rbrace$ where there are $n_1$ particles in the $1^{\rm
st}$ macrocell, $n_2$ in the $2^{\rm nd}$ macrocell etc. Using the
standard combinatorial procedure introduced by Boltzmann, the number
of microstates corresponding to the macrostate $\lbrace n_{i}\rbrace$,
i.e. its probability, is given by
\begin{equation}
\label{bol1} W(\lbrace n_{i}\rbrace)=N!\prod_{i}{\nu^{n_{i}}\over n_{i}!}.
\end{equation}
This is the Maxwell-Boltzmann statistics. As is customary, we define the
entropy of the state $\lbrace n_i \rbrace$ by
\begin{equation}
S(\lbrace n_i \rbrace)=\ln W(\lbrace n_i \rbrace). \label{bol2}
\end{equation}
It is convenient here to return to a representation in terms of the
distribution function giving the phase-space density in the $i$-th
macrocell $f_i=f({\theta}_i,{v}_i)={n_i/\nu h}$. Using the Stirling
formula and passing to the continuum limit $\nu\rightarrow 0$, we obtain the
usual expression of the Boltzmann entropy
\begin{equation}
\label{bol3} S_{B}[f]=-\int f\ln f d\theta dv,
\end{equation}
up to some unimportant additive constant.  Then, the statistical
equilibrium state, corresponding to the most probable distribution of
particles, is obtained by maximizing the Boltzmann entropy (\ref{bol3}) at
fixed mass $M$ and energy $E$, i.e.
\begin{equation}
\label{bol4} {\rm Max}\ \lbrace S_{B}[f]\quad |\ E[f]=E, M[f]=M\rbrace.
\end{equation}
This maximization problem defines the microcanonical equilibrium
state, which is the correct description for an isolated
Hamiltonian system.

We shall also consider the canonical description which applies to a
system in contact with a thermostat imposing its temperature $T$. We
will give an example of canonical system in Sec. \ref{sec_brownian}
corresponding to Brownian particles in interaction described by
stochastic (not Hamiltonian) equations (BMF model). In the canonical
ensemble, the statistical equilibrium state is obtained by minimizing
the free energy $F_{B}[f]=E[f]-TS_{B}[f]$ at fixed mass $M$ and
temperature $T$, i.e.
\begin{equation}
\label{bol5}  {\rm Min}\ \lbrace F_{B}[f]\quad |\  M[f]=M \rbrace.
\end{equation}
The relation between the Boltzmann entropy $S_{B}[f]$ and the density
of states $g(E)$ in the microcanonical ensemble and between the
Boltzmann free energy $F_{B}[f]$ and the partition function $Z(\beta)$
in the canonical ensemble is discussed in Chavanis (2004b,2005). The
variational problems (\ref{bol4}) and (\ref{bol5}) correspond to a
saddle point approximation in the functional integral representation
of $g(E)$ and $Z(\beta)$.

The variational problem (\ref{bol4}) has been first investigated, in
the HMF context, by Inagaki (1993). We review and precise the main
results of his study and present an alternative derivation of the
condition of thermodynamical stability using methods similar to those
introduced by Padmanabhan (1990) and Chavanis (2002b,2003a) for 3D
self-gravitating systems. This will make the analogy between the two
systems (stellar systems and HMF model) closer. This will also allow
us to study the thermodynamical stability of the clustered phase,
while the analysis of Inagaki (1993) is restricted to the uniform
phase.

\subsection{The first variations: the Maxwell-Boltzmann distribution}
\label{sec_mb}

We need first to determine the critical points of entropy at fixed
mass and energy. We write the variational principle in the form
\begin{equation}
\label{mb1}
\delta S_{B}-\beta\delta E-\alpha\delta M=0,
\end{equation}
where $\beta=1/T$ (inverse temperature) and $\alpha$ (chemical
potential) are Lagrange multipliers enforcing the constraints on
$E$ and $M$. The solution of (\ref{mb1}) is the mean-field Maxwell-Boltzmann
distribution
\begin{equation}
\label{mb2}
f=A' e^{-\beta ({v^{2}\over 2}+\Phi)},
\end{equation}
where $\Phi$ depends on $f$ through Eq. (\ref{mf7}). The foregoing
relation is therefore an integro-differential equation.
Integrating over the velocity, we obtain the mean-field Boltzmann
distribution
\begin{equation}
\label{mb3}
\rho=A e^{-\beta\Phi}.
\end{equation}
The same distributions (\ref{mb2}) and (\ref{mb3}) are obtained in the
canonical ensemble by cancelling the first variations of free energy
at fixed mass, using $\delta F_{B}-\alpha\delta M=0$. Therefore, the
critical points of the variational problems (\ref{bol4}) and
(\ref{bol5}) coincide. In the microcanonical ensemble, we need to
relate the Lagrange multiplier $\beta$ to the energy $E$. This defines
a {\it series of equilibria} $\beta=\beta(E)$. In the canonical
ensemble, the inverse temperature $\beta$ is assumed given and the
corresponding (average) energy is obtained by inversing the graph
$\beta(E)$. Of course, only the stable part of the series of
equilibria is of physical interest, and defines the {\it caloric
curve} (see Sec. \ref{sec_tstab}). We note that the equilibrium
distributions (\ref{mb2}) and (\ref{mb3}) can also be obtained from
an equilibrium BBGKY-like hierarchy in the thermodynamic limit
$N\rightarrow +\infty$ defined in Sec. \ref{sec_mf} (Chavanis
2004b,2005).

Using the relation (\ref{mf8}), the distribution of particles at
statistical equilibrium is given by
\begin{equation}
\label{mb4}
\rho=Ae^{-\beta B\cos\theta}.
\end{equation}
The axis of symmetry is determined by the initial conditions.  If
$B=0$, the density $\rho$ is uniform. This defines the homogeneous
phase. If $B\neq 0$, we have inhomogeneous states with one cluster at
$\theta=0$ (if $B<0$) or at $\theta=\pi$ (if $B>0$).  The constant $A$
is related to the mass by
\begin{equation}
\label{mb5}
M=2\pi A I_{0}(\beta B),
\end{equation}
where $I_{n}$ are the modified Bessel functions
\begin{equation}
\label{mb6}
I_{n}(z)={1\over \pi}\int_{0}^{\pi}e^{z\cos\theta}\cos(n\theta)d\theta.
\end{equation}
For $z\rightarrow 0$,
\begin{equation}
\label{mb7}
I_{n}(z)=({1\over 2}z)^{n}\biggl\lbrack {1\over \Gamma(n+1)}+{z^{2}\over 4\Gamma(n+2)}+...\biggr\rbrack,
\end{equation}
and for $z\rightarrow +\infty$,
\begin{equation}
\label{mb8}
I_{n}(z)={e^{z}\over \sqrt{2\pi z}}\biggl\lbrack 1-{4n^{2}-1\over 8z}+...\biggr\rbrack.
\end{equation}

\begin{figure}
\centering
\includegraphics[width=8cm]{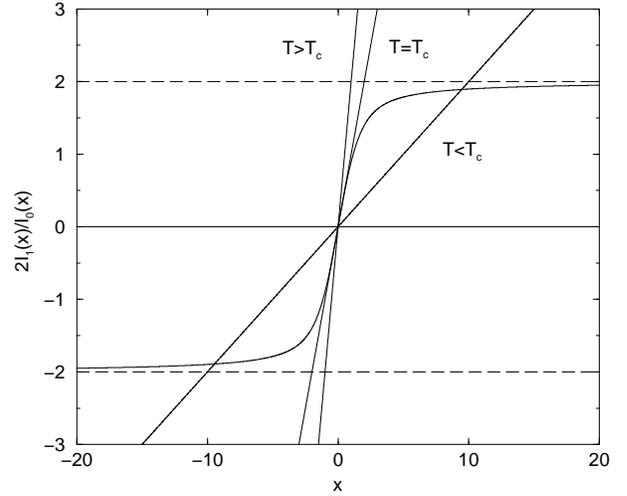}
\caption{Graphical construction showing the appearance of a clustered phase below some critical temperature $T_{c}$. }
\label{construction}
\end{figure}

\begin{figure}
\centering
\includegraphics[width=8cm]{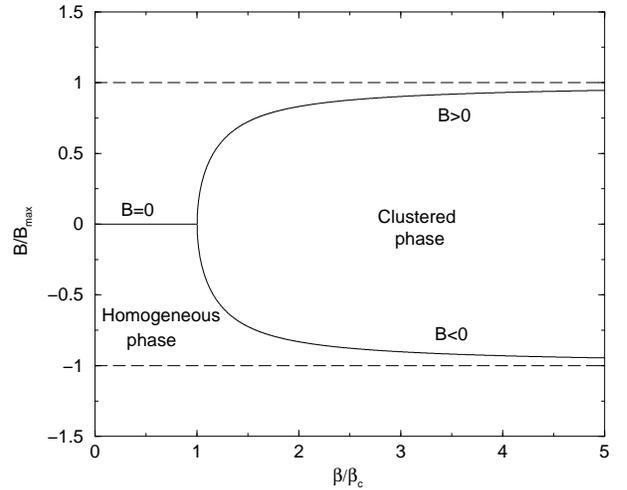}
\caption{Order parameter $B$ (magnetization) as a function of the inverse temperature.}
\label{betaB}
\end{figure}

\begin{figure}
\centering
\includegraphics[width=8cm]{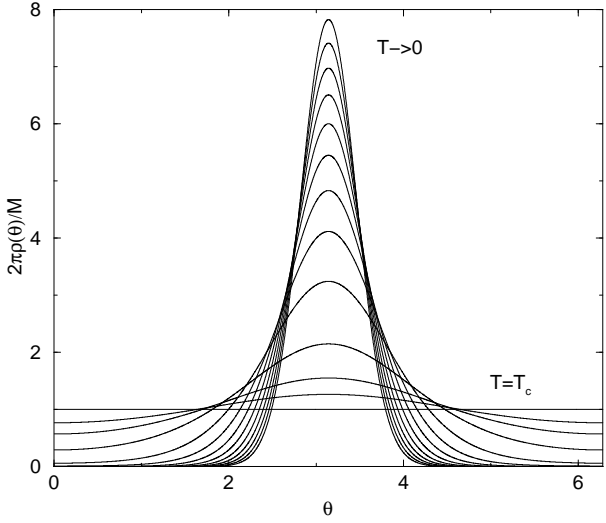}
\caption{Evolution of the density profile as temperature is decreased (from bottom to top).}
\label{rho}
\end{figure}

\begin{figure}
\centering
\includegraphics[width=8cm]{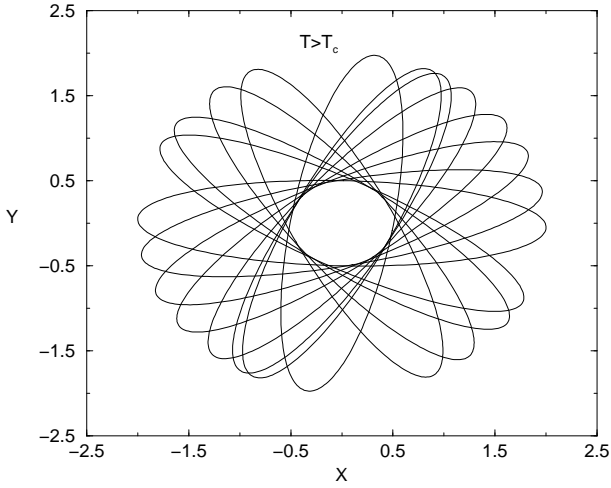}
\caption{Stellar orbits in the ``disk phase'' for  $T>T_{c}$. }
\label{ellipsgaz}
\end{figure}

\begin{figure}
\centering
\includegraphics[width=8cm]{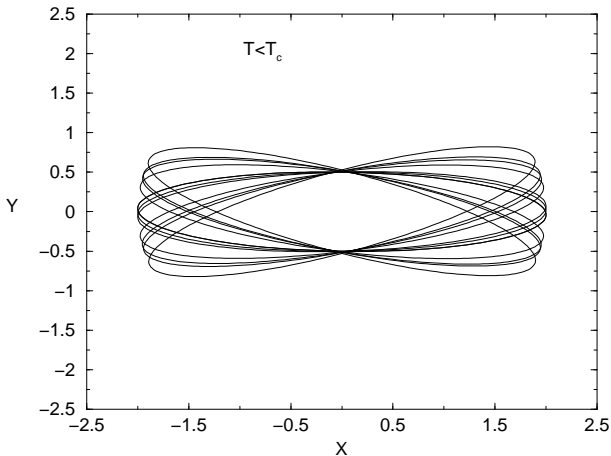}
\caption{Stellar orbits in the ``bar phase'' for $T<T_{c}$.}
\label{ellipsbarre}
\end{figure}

Using Eqs. (\ref{mf9}) and (\ref{mb6}) we find that the order parameter $B$ is determined
as a function of the temperature $\beta$ by the implicit equation
\begin{equation}
\label{mb9}
B={kM\over 2\pi}{I_{1}(\beta B)\over I_{0}(\beta B)}.
\end{equation}
Setting $x=\beta B$, we can rewrite the foregoing relation in the form
\begin{equation}
\label{mb10}
{4\pi T\over kM}x=2{I_{1}(x)\over I_{0}(x)}.
\end{equation}
Then $x$, and consequently $B$, is determined as a function of $T$ by a
simple graphical construction sketched in Fig. \ref{construction}. We
see that $B=0$ is always solution although $B\neq 0$ is possible only
if
\begin{equation}
\label{mb11}
T<{kM\over 4\pi}\equiv T_{c}.
\end{equation}
In terms of the energy (\ref{mf10}) this corresponds to
\begin{equation}
\label{mb12} E<{kM^{2}\over 8\pi}\equiv E_{c}.
\end{equation}
The function $B(T)$ is shown in Fig. \ref{betaB} and its asymptotic
behaviours are given in Sec. \ref{sec_thermop}. Figure \ref{betaB}
displays a second order phase transition. We have a situation similar
to a gravitational collapse below a critical temperature $T_c$ or
below a critical energy $E_c$. For $T>T_{c}$, the system is
homogeneous. For $T<T_{c}$, the system forms one cluster around
$\theta=0$ (for $B<0$) or around $\theta=\pi$ (for $B>0$). At $T=0$,
the equilibrium state is a Dirac peak $\rho=M\delta(\theta-\pi)$ (for
$B=B_{max}$).  Density profiles are plotted in Fig. \ref{rho} for different
values of $x=\beta B(\beta)$. Using the stellar disk interpretation of
Pichon \& Lynden-Bell, we have represented some stellar orbits in
Figs. \ref{ellipsgaz} and \ref{ellipsbarre} by randomly choosing the
orbits' angles with the equilibrium distribution $\rho(\theta)$. The
``disk phase'' for $T>T_{c}$ is represented in Fig. \ref{ellipsgaz}
and the ``bar phase'' for $T<T_{c}$ is represented in
Fig. \ref{ellipsbarre}.

\subsection{The thermodynamical parameters}
\label{sec_thermop}

According to Eq. (\ref{mb10}), the relation between the temperature and the
order parameter can be written in dimensionless form as
\begin{equation}
\label{tp1}
\eta\equiv \beta/\beta_{c}={x\over 2}{I_{0}(x)\over I_{1}(x)}.
\end{equation}
For $x\rightarrow 0$,
\begin{equation}
\label{tp2}
\eta=1+{x^{2}\over 8}+...
\end{equation}
and for $x\rightarrow +\infty$,
\begin{equation}
\label{tp3}
\eta={x\over 2}\biggl (1+{1\over 2x}+...\biggr ).
\end{equation}
Returning to original variables, we deduce that
\begin{equation}
\label{tp4}
{B\over B_{max}}=\pm \sqrt{2\biggl (1-{T\over T_{c}}\biggr )}, \qquad (0<{T_{c}-T\over T_{c}}\ll 1),
\end{equation}
\begin{equation}
\label{tp5}
{B\over B_{max}}=\pm\biggl (1-{T\over 4 T_{c}}\biggr ), \qquad (T\ll T_{c}),
\end{equation}
where $B_{max}=2T_{c}=kM/2\pi$ is the maximum value of the
magnetization obtained for $T=0$ when all the particles are at
$\theta=\pi$. With this notation, the parameter $x$ can be written 
\begin{equation}
\label{tp6}
x=2\eta B/B_{max}.
\end{equation}

\begin{figure}
\centering
\includegraphics[width=8cm]{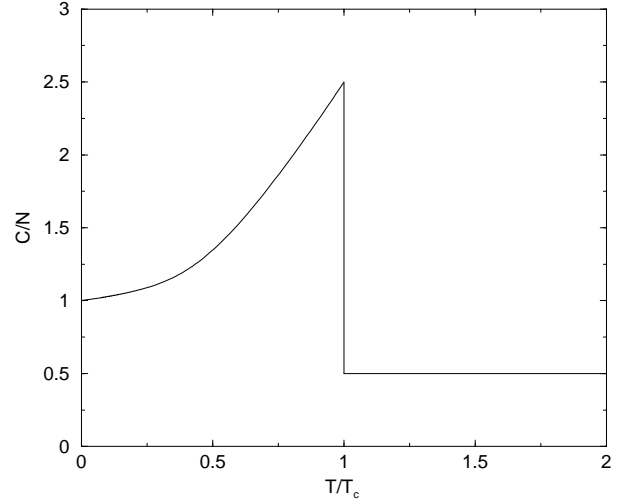}
\caption{Specific heat $C=dE/dT$ as a function of  temperature. It experiences a discontinuity at the critical temperature $T_{c}$.  }
\label{calorique}
\end{figure}

\begin{figure}
\centering
\includegraphics[width=8cm]{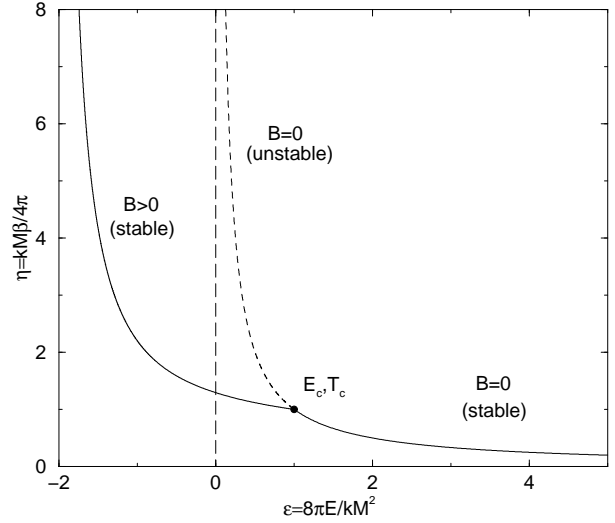}
\caption{Calorique curve (series of equilibria) for the HMF model. The system displays a second order phase transition at a critical point $E_{c},T_{c}$.  }
\label{epseta}
\end{figure}

On the other hand, for the Maxwellian velocity distribution (\ref{mb2}), the
expression of the energy (\ref{mf10}) becomes
\begin{equation}
\label{tp7}
E={1\over 2}MT-{\pi B^{2}\over k}.
\end{equation}
In terms of dimensionless parameters, we get
\begin{equation}
\label{tp8} \epsilon\equiv {E/E_c}={1\over \eta}\biggl
(1-{x^{2}\over 2\eta}\biggr ).
\end{equation}
For the homogeneous phase $B=0$, we simply have
\begin{equation}
\label{tp9}
\epsilon={1\over \eta}.
\end{equation}
For the inhomogeneous phase, we can easily obtain asymptotic
expansions.  For $x\rightarrow 0$,
\begin{equation}
\label{tp10}
\epsilon=1-{5\over 8}x^{2}+...
\end{equation}
and for $x\rightarrow +\infty$,
\begin{equation}
\label{tp11}
\epsilon=-2\biggl (1-{2\over x}+...\biggr ).
\end{equation}
Returning to original variables, we deduce that
\begin{equation}
\label{tp12}
{E\over E_{0}}={1\over 2}{T\over T_{c}}, \qquad (T>T_{c}),
\end{equation}
\begin{equation}
\label{tp13}
{E\over E_{0}}={1\over 2}\biggl (6-5{T\over T_{c}}\biggr ), \qquad (0<{T_{c}-T\over T_{c}}\ll 1),
\end{equation}
\begin{equation}
\label{tp14}
{E\over E_{0}}={T\over  T_{c}}-1, \qquad (T\ll T_{c}),
\end{equation}
where $-E_{0}=-kM^{2}/4\pi$ is the minimum value of energy
obtained for $T=0$ ($\epsilon_{0}=-2$). For $T\rightarrow T_{c}^{-}$,
the specific heat $C=dE/dT$ is given by $C={5\over 2}M$ and for
$T>T_{c}$ by $C={M\over 2}$. Therefore, at the critical point, it experiences a discontinuity (see Fig. \ref{calorique}):
\begin{equation}
\label{tp15}
C(T_{c}^{-})-C(T_{c}^{+})=2M.
\end{equation}
The caloric curve/series of equilibria $\beta(E)$ is shown in
Fig. \ref{epseta}. It displays a second order phase transition at
$(\epsilon,\eta)=(1,1)$. This is different from 3D gravitational
systems which rather display  first order and zeroth order phase
transitions (see, e.g., Chavanis 2002c).

\subsection{The second variations: thermodynamical stability}
\label{sec_tstab}

To analyze the thermodynamical stability of the solutions determined
by the variational problems (\ref{bol4}) and (\ref{bol5}), we use an
approach similar to that followed by Padmanabhan (1990) and Chavanis
(2002b,2003a) in the case of 3D self-gravitating systems. We first maximize
$S_{B}[f]$ at fixed $M[f]$, $E[f]$ {\it and} $\rho(\theta)$. This gives
the Maxwellian
\begin{equation}
\label{tstab1}
f(\theta,v)={1\over \sqrt{2\pi T}}\rho(\theta) e^{-{v^{2}\over 2T}}.
\end{equation}
Then, we can re-express the entropy and the energy as a function of the density in such a way that
\begin{equation}
\label{tstab2}
S_{B}={1\over 2}M \ln T-\int\rho\ln\rho \ d\theta,
\end{equation}
\begin{equation}
\label{tstab3}
E={1\over 2}M T+{1\over 2}\int\rho\Phi d\theta.
\end{equation}
We now take the variations of entropy around an equilibrium solution. To second order
\begin{eqnarray}
\label{tstab4}
\delta S_{B}={M\over 2}{\delta T\over T}-{M\over 4}\biggl ({\delta T\over T}\biggr )^{2}-\int \delta\rho\ln\rho d\theta-{1\over 2}\int {(\delta\rho)^{2}\over \rho}d\theta.\nonumber\\
\end{eqnarray}
Now, the conservation of energy implies
\begin{equation}
\label{tstab5}
0=\delta E={1\over 2}M\delta T+\int \Phi\delta\rho d\theta+{1\over 2}\int \delta\rho\delta\Phi d\theta.
\end{equation}
Eliminating $\delta T$, we find that
\begin{eqnarray}
\label{tstab6}
\delta^{2} S_{B}=-{1\over 2T}\int \delta\rho\delta\Phi d\theta\qquad\qquad\qquad\nonumber\\
-{1\over MT^{2}}\biggl (\int \Phi\delta\rho d\theta\biggr )^{2}-{1\over 2}\int {(\delta\rho)^{2}\over \rho}d\theta.
\end{eqnarray}
We define the quantity $q$ by the relation
\begin{equation}
\label{tstab7}
\delta\rho={dq\over d\theta}.
\end{equation}
Physically, $q=\int_{0}^{\theta}\delta\rho d\theta$ represents the
mass perturbation within the interval $\lbrack 0,\theta\rbrack$. Then, the
conservation of mass is equivalent to $q(0)=q(2\pi)=0$. Inserting this
relation in Eq. (\ref{tstab6}) and using integrations by parts, we can put the
second order variations of entropy in the quadratic form
\begin{equation}
\label{tstab8}
\delta^{2}S_{B}=\int_{0}^{2\pi}\int_{0}^{2\pi}d\theta d\theta' q(\theta)K(\theta,\theta')q(\theta'),
\end{equation}
with
\begin{eqnarray}
\label{tstab9}
K(\theta,\theta')=-{1\over MT^{2}}{d\Phi\over d\theta}(\theta){d\Phi\over d\theta}(\theta')\qquad\qquad\qquad\qquad\nonumber\\
+{k\over 4\pi T}\sin(\theta-\theta'){d\over d\theta'}+{1\over 2}\delta(\theta-\theta'){d\over d\theta'}\biggl \lbrack{1\over \rho(\theta')}{d\over d\theta'}\biggr \rbrack.
\end{eqnarray}
We are thus led to consider the eigenvalue problem
\begin{equation}
\label{tstab10}
\int_{0}^{2\pi}K(\theta,\theta')q_{\lambda}(\theta')d\theta'=\lambda q_{\lambda}(\theta).
\end{equation}
This yields
\begin{eqnarray}
\label{tstab11}
{d\over d\theta}\biggl ({1\over\rho}{dq\over d\theta}\biggr )+{k\over 2\pi T}\int_{0}^{2\pi}q(\theta')\cos(\theta-\theta')d\theta'\nonumber\\
={2V\over MT^{2}}{d\Phi\over d\theta}+2\lambda q,
\end{eqnarray}
where
\begin{equation}
\label{tstab12}
V=\int_{0}^{2\pi}{d\Phi\over d\theta}q(\theta)d\theta.
\end{equation}
The system is stable if all $\lambda<0$ and unstable if at least one
$\lambda>0$. So far, we have worked in the microcanonical ensemble. If
we work in the canonical ensemble, we have to minimize the free energy
$F_{B}=E-TS_{B}$ at fixed mass and temperature. We can easily check
that fixing $T$ in the preceding calculations amounts to taking
$V=0$. Thus, instead of Eq. (\ref{tstab11}), we obtain
\begin{equation}
\label{tstab13}
{d\over d\theta}\biggl ({1\over\rho}{dq\over d\theta}\biggr )+{k\over 2\pi T}\int_{0}^{2\pi}q(\theta')\cos(\theta-\theta')d\theta'=2\lambda q.
\end{equation}

\subsection{The condition of thermodynamical stability}
\label{sec_marg}

\begin{figure}
\centering
\includegraphics[width=8cm]{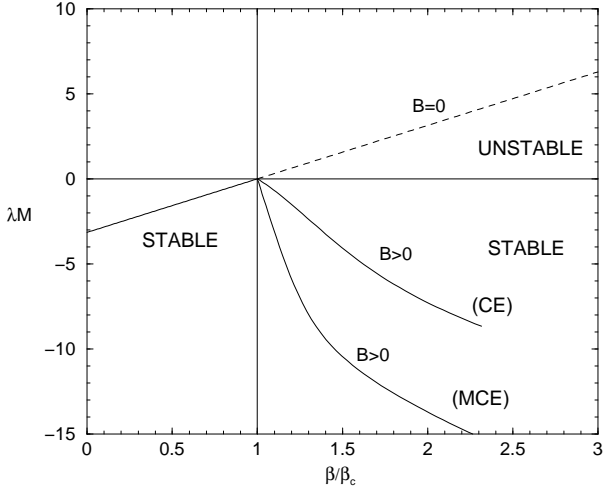}
\caption{Dependence of the largest eigenvalue $\lambda$ with the temperature. A negative value of $\lambda$ corresponds to stability ($\delta^{2}S<0$) and a positive value of $\lambda$ corresponds to instability.}
\label{stabthermo}
\end{figure}

If we consider the stability of the uniform solution $\rho=M/2\pi$ and
$\Phi=0$, the foregoing equations 
simplify into
\begin{equation}
\label{marg1}
{2\pi\over M}{d^{2}q\over d\theta^{2}}+{k\over 2\pi T}\int_{0}^{2\pi}q(\theta')\cos(\theta-\theta')d\theta'=2\lambda q.
\end{equation}
The eigenvalue equation is the same in the two ensembles. Hence, the
stability criteria coincide, implying that the statistical ensembles
(microcanonical and canonical) are equivalent. This is at variance
with the case of 3D stellar systems (Padmanabhan 1990, Chavanis
2002b,2003a).

We can study the solutions of Eq. (\ref{marg1}) by decomposing $q$ in
Fourier series. For the mode $n$, we have $q_{n}=A_{n}\sin
(n\theta)$. For $n\neq 1$, we get $\lambda_{n}=-{\pi n^{2}\over M}<0$
showing that these modes do not induce instability. For $n=1$, we have
$\lambda_{1}={k\over 4 T}-{\pi\over M}$. The uniform solution will be
unstable if $\lambda_{1}>0$ yielding condition
(\ref{mb11}). Therefore, the uniform phase is stable for $T>T_{c}$
while it is unstable for $T<T_{c}$. By using the theory of linear
series of equilibria (Katz 1978, 1980, 2003), applied here to a bifurcation
point, we directly conclude from the inspection of Fig. \ref{epseta}
that the clustered phase will be stable for $T<T_{c}$ when the
homogeneous phase becomes unstable.

More precisely, it is possible to solve Eqs.  (\ref{tstab11}) and
(\ref{tstab13}) analytically for the clustered phase in the limit
$B\rightarrow 0$, which is valid close to the critical point
$(E_{c},T_{c})$. The calculations are detailed in Appendix A. In the
canonical ensemble ($V=0$), it is found that the largest eigenvalue is
\begin{equation}
\label{marg2}
\lambda M=-2\pi\biggl ({\beta\over\beta_{c}}-1\biggr ).
\end{equation}
and in the microcanonical ensemble ($V\neq 0$) that
\begin{equation}
\label{marg3}
\lambda M=-10\pi\biggl ({\beta\over\beta_{c}}-1\biggr ).
\end{equation}
More generally, the exact value of $\lambda$ obtained by solving
Eqs. (\ref{tstab11}) and (\ref{tstab13}) numerically is plotted versus
the inverse temperature in Fig. \ref{stabthermo}.  Since $\lambda<0$,
we check explicitly that the clustered phase is stable.

\subsection{Correction to the mean-field approximation close to the critical point}
\label{sec_corr}

We can obtain the expression of the two-points correlation function
from an equilibrium BBGKY-like hierarchy by closing the second equation of
the hierarchy with the Kirkwood approximation (Chavanis 2004b,2005).  This
is valid to order $1/N$ in the thermodynamic limit defined
previously. For the HMF model, it is then  possible to obtain an explicit
expression of the correlation function in the homogeneous
phase. Writing the two-body distribution function as
\begin{eqnarray}
\label{corr1} N^{2}P_{2}(\theta_{1}-\theta_{2})=\rho^{2}\lbrack 1+h(\theta_{1}-\theta_{2})\rbrack,
\end{eqnarray}
it is found that
\begin{equation}
\label{corr2} h(\theta_{1}-\theta_{2})={2\over N}{\beta/\beta_{c}\over 1-\beta/\beta_{c}}\cos(\theta_{1}-\theta_{2}).
\end{equation}
We note that the correlation function diverges close to the critical
point $\beta\rightarrow \beta_{c}$ where the clustered phase appears
and the homogeneous phase becomes unstable. This implies
that the mean-field approximation ceases to be valid close to the
critical point. We expect a similar result for 3D self-gravitating
systems although the situation is more difficult to analyze as 
(real) self-gravitating systems are always inhomogeneous. 

If we take into account the contribution of non-trivial pair
correlations (\ref{corr2}) in the potential energy, we find
furthermore that Eq. (\ref{tp9}) is replaced by
\begin{equation}
\label{corr3} 
\epsilon={1\over\eta}-{1\over N}{2\eta\over 1-\eta}.
\end{equation}
Therefore, finite $N$ effects modify the shape of the caloric curve in
the vicinity of the critical point. The mean-field approximation is
valid if $N(1-\eta)\gg 1$.  When the mean field approximation is
valid, its order one correction for the specific heat is
\begin{equation}
\label{corr3spac} 
C={N\over 2}\biggl\lbrack 1+{1\over\pi N}{1\over (T/T_{c}-1)^{2}}\biggr\rbrack \qquad (T>T_{c}).
\end{equation}

Finally, it is found that the spatial correlations of the force are given by (Chavanis 2004b,2005)
\begin{equation}
\label{corr4} 
\langle F(0)F(\theta)\rangle={\rho k^{2}\over 4\pi}{1\over 1-\beta/\beta_{c}}\cos\theta.
\end{equation}
In particular, the variance of the force is
\begin{equation}
\label{corr5} 
\langle F^{2}\rangle={\rho k^{2}\over 4\pi}{1\over 1-\beta/\beta_{c}}.
\end{equation}
Note that without the correlations, we would have simply obtained
$\langle F^{2}\rangle={\rho k^{2}\over 4\pi}$ which corresponds to
the high temperature limit ($T\rightarrow +\infty$) of
Eq. (\ref{corr5}). 

For the HMF model, the variance (\ref{corr5}) of the force is finite
while the variance of the Newtonian force for 3D self-gravitating
systems is infinite (Chandrasekhar \& von Neumann 1942). For the HMF
model, the distribution of the force is normal (Gaussian) while the
distribution of the gravitational force in $D=3$ is a particular L\'evy law
called the Holtzmark distribution. On the other hand, for 2D point
vortices, the variance of the velocity is a marginal Gaussian
distribution intermediate between normal and L\'evy laws (Chavanis \&
Sire 2000). Therefore, these three systems with long-range
interactions (self-gravitating systems, 2D vortices and HMF model)
have their own specificities despite their overall analogies.

\section{Gaseous systems}
\label{sec_gas}

As indicated in the Introduction, the HMF model is similar to stellar
systems in astrophysics. In Secs. \ref{sec_vlasov} and
\ref{sec_collisional}, we shall discuss the kinetic theory of the HMF
model and obtain the equivalent of the Vlasov and Landau equations
that are used to describe the dynamics of elliptical galaxies and
globular clusters respectively. However, in order to facilitate the
discussion and the comparison, it is useful to discuss first the
dynamics of a one-dimensional barotropic fluid system with cosine
interactions described by the Euler equations. In astrophysics, these
equations describe the dynamics of barotropic stars.  Stellar systems
and barotropic stars are often treated in parallel due to their
analogies (Binney \& Tremaine 1987). In particular, it is possible to
infer sufficient conditions of dynamical stability for spherical
stellar systems from the dynamical stability of a barotropic star with
the same density distribution. This constitutes the Antonov first
law. Therefore, it is also of interest to develop this parallel in the
case of the HMF model.  To have a similar vocabulary, the systems
considered in this paper will also be called ``stellar systems'' and
``gaseous stars'' although they are only one-dimensional and
correspond to a cosine interaction.

\subsection{Euler equations and energy functional}
\label{sec_ej}

We consider a gaseous system described by the Euler
equations
\begin{eqnarray}
{\partial\rho\over\partial t}+{\partial\over\partial\theta}(\rho u)=0,
\label{ej1}
\end{eqnarray}
\begin{eqnarray}
{\partial u\over\partial t}+u{\partial u\over\partial \theta}=-{1\over\rho}{\partial p\over\partial \theta}-{\partial\Phi\over\partial\theta},
\label{ej2}
\end{eqnarray}
where the potential $\Phi$ is given by (\ref{mf7}). To close the
equations, we consider an arbitrary barotropic equation of state
$p=p(\rho)$. We emphasise that these equations cannot be derived from
the HMF model (\ref{mf1}) which rather leads to kinetic equations like
the Vlasov equation. However, we shall see that there is a close
connection between the stationary states of the Vlasov and the Euler
equations and that the limits of dynamical stability are the same in
the two systems. Thus, the study of the Euler equation (which is
simpler than the Vlasov equation) brings many information about the
stability of stationary states of the HMF model with respect to the
Vlasov equation even if the Euler system does not describe dynamically the
HMF model.

It is straightforward to verify that the energy functional
\begin{eqnarray}
{\cal W}=\int\rho\int_{0}^{\rho}{p(\rho')\over \rho'^{2}}d\rho'd\theta+{1\over 2}\int\rho\Phi d\theta +\int \rho {u^{2}\over 2}d\theta,
\label{ej3}
\end{eqnarray}
is conserved by the Euler equations ($\dot {\cal W}=0$). The first
term is the internal energy, the second the potential energy and the
third the kinetic energy associated with the mean motion. The mass is
also conserved. Therefore, a minimum of ${\cal W}$ at fixed mass
determines a stationary solution of the Euler equations which is
formally nonlinearly dynamically stable in the sense of Holm et
al. (1985).  We are led therefore to consider the minimization problem
\begin{eqnarray}
{\rm Min}\ \lbrace {\cal W}[\rho,u]\quad |\ M[\rho]=M\rbrace . \label{ej4}
\end{eqnarray}

\subsection{First variations: the condition of hydrostatic equilibrium}
\label{sec_hydro}

Cancelling the first order variations of Eq. (\ref{ej3}), we obtain
$u=0$ and the condition of hydrostatic equilibrium
\begin{eqnarray}
{dp\over d\theta}=-\rho {d\Phi\over d\theta}.
\label{hydro1}
\end{eqnarray}
Therefore, extrema of ${\cal W}$ correspond to stationary solutions of
the Euler equations (\ref{ej1})-(\ref{ej2}). On the other hand, combining the condition of hydrostatic equilibrium (\ref{hydro1}) and the equation of state $p=p(\rho)$, we get
\begin{eqnarray}
\int^{\rho}{p'(\rho')\over\rho'}d\rho'=-\Phi,
\label{hydro2}
\end{eqnarray}
so that $\rho$ is a function of $\Phi$ that we note $\rho=\rho(\Phi)$. Using Eq. (\ref{mf8}), we find that
\begin{eqnarray}
\rho=\rho(B\cos\theta), \label{hydro3}
\end{eqnarray}
where $B$ is determined by the implicit equation
\begin{eqnarray}
B=-{k\over 2\pi}\int_{0}^{2\pi} \rho(B\cos\theta')\cos\theta'\  d\theta'.
\label{hydro4}
\end{eqnarray}
Again, $B=0$ is a solution of this equation characterizing a
homogeneous phase $\rho=M/2\pi$. To determine the point of bifurcation to the
inhomogeneous phase, we expand Eq. (\ref{hydro4}) around $B=0$. Then, we
find that cluster solutions appear when
\begin{eqnarray}
1+{k\over 2}{d\rho\over d\Phi}(0)\le 0. \label{hydro5}
\end{eqnarray}
Using the condition of hydrostatic balance (\ref{hydro1}), this can be
rewritten
\begin{eqnarray}
c_{s}^{2}\le (c_{s}^{2})_{crit}={kM\over 4\pi}, \label{hydro6}
\end{eqnarray}
where $c_{s}=(dp/d\rho)^{1/2}$ is the velocity of sound in the
homogeneous phase where $\rho=M/2\pi$.

\subsection{Second variations: the condition of nonlinear dynamical stability}
\label{sec_ejn}

The second variation of ${\cal W}$ due to variation of the velocity is trivially positive. The second variation of
${\cal W}$ due to variation of $\rho$ is 
\begin{eqnarray}
\delta^{2}{\cal W}={1\over 2}\int\delta\rho\delta \Phi d\theta+\int
{p'(\rho)\over 2\rho}(\delta\rho)^{2}d\theta, \label{ejn1}
\end{eqnarray}
which must be positive for stability. Using the same procedure as
in Sec. \ref{sec_tstab}, we find that the eigenvalue equation
determining the stability of the solution is now
\begin{eqnarray}
{d\over d\theta}\biggl ({p'(\rho)\over\rho}{dq\over d\theta}\biggr )+{k\over 2\pi }\int_{0}^{2\pi}q(\theta')\cos(\theta-\theta')d\theta'=2\lambda q,\nonumber\\
\label{ejn2}
\end{eqnarray}
and that the condition of stability is $\lambda< 0$ (this yields a
maximum of $-{\cal W}$). For the uniform solution $\rho=M/2\pi$, we
can repeat exactly the same steps as in Sec. \ref{sec_marg} since
$p'(\rho)$ is a constant $c_{s}^{2}$ which plays the role of $T$ in
the thermodynamical analysis. Therefore, we find that the uniform
phase is formally nonlinearly dynamically stable with respect to the
Euler equations when
\begin{eqnarray}
c_{s}^{2}\ge  {kM\over 4\pi},
\label{ejn3}
\end{eqnarray}
and dynamically unstable otherwise. According to
Eq. (\ref{hydro6}), the onset of dynamical instability coincides with
the point where the clustered phase appears.

The stability of the clustered phase can be investigated by solving
the eigenvalue equation (\ref{ejn2}) for a specified equation of state
$p(\rho)$. This equation is the counterpart of equation (222) of
Chavanis (2003a) for 3D self-gravitating gaseous spheres.

\subsection{The condition of linear stability: Jeans-like criterion}
\label{sec_ejl}

We now study the linear dynamical stability of a stationary solution of the
Euler equation.  We consider a small perturbation
around a stationary solution of Eqs. (\ref{ej1})-(\ref{ej2})  and write
$\rho=\rho+\delta\rho$, $u=\delta u$ etc... The linearized
equations for the perturbation are
\begin{eqnarray}
{\partial\delta\rho\over\partial t}+{\partial\over\partial\theta}(\rho \delta u )=0,
\label{ejl1}
\end{eqnarray}
\begin{eqnarray}
\rho {\partial \delta u\over\partial t}=-{\partial \over\partial \theta} (p'(\rho)\delta\rho )-\rho {\partial\delta \Phi\over\partial\theta}-\delta\rho {\partial\Phi\over\partial\theta},
\label{ejl2}
\end{eqnarray}
\begin{equation}
\label{ejl3}
\delta\Phi(\theta)=-{k\over 2\pi}\int_{0}^{2\pi}\cos(\theta-\theta')\delta \rho(\theta')d\theta'.
\end{equation}
Writing the time dependence in the form $\delta\rho\sim e^{\lambda t}$,..., we get
\begin{eqnarray}
\lambda\delta\rho+{d\over d\theta}(\rho\delta u )=0,
\label{ejl4}
\end{eqnarray}
\begin{eqnarray}
\lambda \rho \delta u=-{d \over d \theta} (p'(\rho)\delta\rho )-\rho {d\delta \Phi\over d\theta}-\delta\rho {d\Phi\over d\theta}.
\label{ejl5}
\end{eqnarray}
Introducing the notation (\ref{tstab7}), the continuity equation can be
integrated into
\begin{eqnarray}
\lambda q+\rho\delta u=0,
\label{ejl6}
\end{eqnarray}
where we have imposed $\delta u(0)=\delta u(2\pi)=0$. Substituting this relation in Eq. (\ref{ejl5}) and using the condition of hydrostatic equilibrium (\ref{hydro1}), we finally obtain
\begin{eqnarray}
{d\over d\theta}\biggl ({p'(\rho)\over\rho}{dq\over d\theta}\biggr )+{k\over 2\pi }\int_{0}^{2\pi}q(\theta')\cos(\theta-\theta')d\theta'={\lambda^{2}\over \rho} q.\nonumber\\
\label{ejl7}
\end{eqnarray}
This equation is the counterpart of the Eddington equation of
pulsations for a barotropic star (see also Eq. (224) of Chavanis
2003a).  We note that Eqs. (\ref{ejn2}) and (\ref{ejl7}) coincide for
the neutral point $\lambda=0$. Therefore, the conditions of
linear stability and formal nonlinear dynamical stability coincide. The same
conclusion holds for 3D barotropic stars (Chavanis 2003a).

Considering the uniform phase $\rho=M/2\pi$ and following a method
similar to that developed in Sec. \ref{sec_marg}, we find that the most
destabilizing mode $(n=1)$ is
\begin{eqnarray}
\delta\rho=a_{1}\cos\theta \ e^{\lambda t}, \quad \delta u=-{2\pi\lambda\over M}a_{1}\sin\theta\ e^{\lambda t},
\label{ejl8}
\end{eqnarray}
where the growth rate is given by
\begin{eqnarray}
\lambda^{2}= {kM\over 4\pi}-c_{s}^{2}.
\label{ejl9}
\end{eqnarray}
When $c_s^2\le kM/4\pi$, then $\lambda=\pm\sqrt{\lambda^{2}}$ and the
perturbation grows exponentially rapidly (unstable case). When $c_s^2\ge
kM/4\pi$, then $\lambda=\pm i\sqrt{-\lambda^{2}}$ and the perturbation
oscillates with a pulsation $\omega=\sqrt{-\lambda^{2}}$ without
attenuation (stable case). Therefore, the uniform phase is linearly
(and also formally nonlinearly) dynamically stable with respect to the Euler equations when
\begin{eqnarray}
c_{s}^{2}\ge  {kM\over 4\pi},
\label{ejl10}
\end{eqnarray}
and linearly dynamically unstable otherwise.

\subsection{Particular examples}
\label{sec_peg}

\subsubsection{Isothermal gas}
\label{sec_isog}

For an isothermal gas, we have
\begin{eqnarray}
p=\rho T, \qquad c_{s}^{2}=T,
\label{peg1}
\end{eqnarray}
and
\begin{eqnarray}
{\cal W}=T\int\rho\ln\rho d\theta+{1\over 2}\int\rho\Phi d\theta
+\int \rho {u^{2}\over 2}d\theta, \label{peg2}
\end{eqnarray}
where the temperature $T$ is uniform.  We note that the energy functional
(\ref{peg2}) of an isothermal gas coincides with the Boltzmann free
energy $F_{B}[\rho]=E[\rho]-TS_{B}[\rho]$ of a $N$-body system in the
canonical ensemble, see Eqs. (\ref{tstab2}) and (\ref{tstab3}) of Sec.
\ref{sec_tstab}. This remark also holds for 3D self-gravitating systems
(Chavanis 2002b). The pulsation equation
(\ref{ejl7}) becomes
\begin{eqnarray}
{d\over d\theta}\biggl ({1\over\rho}{dq\over d\theta}\biggr )+{k\over 2\pi T}\int_{0}^{2\pi}q(\theta')\cos(\theta-\theta')d\theta'={\lambda^{2}\over T\rho} q,\nonumber\\
\label{ejl7new}
\end{eqnarray}
which can be connected to Eq. (\ref{tstab13}). According to the
criteria (\ref{ejn3})-(\ref{ejl10}), the uniform phase is formally
nonlinearly dynamically stable for
\begin{eqnarray}
T\ge T_{c}\equiv {kM\over 4\pi},
\label{peg3}
\end{eqnarray}
and linearly dynamically unstable otherwise. This criterion (or more
generally the criterion (\ref{ejl10})) can be regarded as the
counterpart of the Jeans instability criterion in astrophysics (Binney
\& Tremaine 1987). We emphasize, however, an important difference. In
the case of 3D self-gravitating systems, the Jeans criterion selects a
critical wavelength $\lambda_J$ (increasing with the temperature)
above which the system is unstable against gravitational
collapse. In the present context, where the interaction is truncated
to one Fourier mode $n=1$, the criterion (\ref{peg3}) selects a
critical temperature below which the system is unstable. The
generalization of the Jeans instability criterion for an arbitrary
binary potential of interaction in $D$ dimensions is discussed in
Appendix \ref{sec_disJ} and in Chavanis (2004b). This generalization
clearly shows the connection between the HMF model and 3D
self-gravitating systems.

According to Eq. (\ref{ejl9}), the relation between 
$\lambda$ and the temperature $T$ is
\begin{eqnarray}
\lambda^{2}=T_{c}-T, \qquad T_{c}={kM\over 4\pi}. \label{peg4}
\end{eqnarray}
For $T<T_{c}$, the growth rate is $\lambda=(T_{c}-T)^{1/2}$ and for $T>T_{c}$, the pulsation is $\omega=(T-T_{c})^{1/2}$.  Following
the preceding remark, we stress that, in the present context, $\lambda$ and
$\omega$ only depend on the temperature $T$, while for a
3D gravitational gas, they depend on the wavelength of the perturbation
(Binney \& Tremaine 1987). Here, the unstable mode is fixed to $n=1$.

\begin{figure}
\centering
\includegraphics[width=8cm]{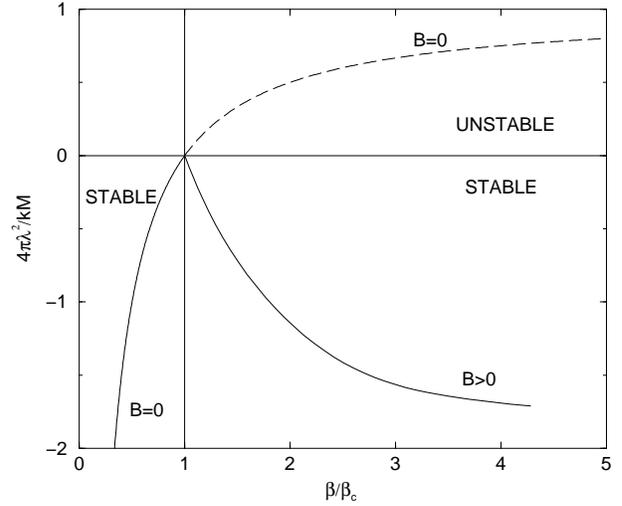}
\caption{Growth rate and pulsation period of an isothermal gas as a function of the temperature.}
\label{lambdaEuler}
\end{figure}

Considering now the clustered phase and using a perturbative approach
similar to that of Appendix \ref{sec_eigen} for $T\rightarrow
T_{c}^{-}$ (not detailed), we find that the pulsation is given by
$\omega=\sqrt{2(T_{c}-T)}$. For smaller temperatures, the eigenvalue
equation can be solved numerically and the results are shown in
Fig. \ref{lambdaEuler}.

\subsubsection{Polytropic gas}
\label{sec_polyg}

For a polytropic gas, we have
\begin{eqnarray}
p=K\rho^{\gamma}, \qquad \gamma=1+{1\over n},
\label{peg5}
\end{eqnarray}
where $K$ is the polytropic constant and $n$ is the polytropic
index. For $n\rightarrow +\infty$, we recover the isothermal case with
$\gamma=1$ and $K=T$. For that reason $K$ is sometimes called a
polytropic temperature.  The energy functional (\ref{ej3}) can be written 
\begin{eqnarray}
{\cal W}={K\over \gamma-1}\int (\rho^{\gamma}-\rho) d\theta+{1\over 2}\int\rho\Phi d\theta
+\int \rho {u^{2}\over 2}d\theta. \label{peg7}
\end{eqnarray}
We have added a constant term (proportional to the total mass) in the
polytropic energy functional (\ref{peg7}) so as to recover the
isothermal energy functional (\ref{peg2}) for $n\rightarrow
+\infty$. Under this form, we note that the energy functional of a
polytropic gas has the same form as the Tsallis free energy
$F_{\gamma}[\rho]=E[\rho]-KS_{\gamma}[\rho]$ where $\gamma$ plays the
role of the $q$-parameter and $K$ the role of a generalized
temperature. The same remark holds for 3D self-gravitating
systems. However, this resemblance is essentially fortuitous and the
mark of a {\it thermodynamical analogy} (Chavanis 2003a; Chavanis \&
Sire 2004b).

If we define the local temperature by $p(\theta)=\rho(\theta)
T(\theta)$, we obtain $T(\theta)=K\rho(\theta)^{1/n}$ and
$c_{s}^{2}(\theta)=\gamma T(\theta)$. We note that, for a polytropic
distribution, the kinetic temperature $T(\theta)$ usually depends on the
position while the polytropic temperature $K$ is uniform as in an
isothermal gas. However, in the uniform phase $T=K\rho^{1/n}$ is a constant that can be called the temperature of the polytropic gas. The velocity of sound in the homogeneous phase is
\begin{eqnarray}
c_{s}^{2}=\gamma T=K\gamma \rho^{\gamma-1}=K{1+n\over n}\biggl ({M\over 2\pi}\biggr )^{1/n}.
\label{peg6}
\end{eqnarray}
The condition of dynamical stability (\ref{ejn3})-(\ref{ejl10}) can be written
\begin{eqnarray}
K\ge K_{n}\equiv {kM\over 4\pi}{n\over 1+n}\biggl ({2\pi\over M}\biggr )^{1/n},
\label{peg7a}
\end{eqnarray}
or, equivalently,
\begin{eqnarray}
T\ge T_{n}\equiv {T_{c}\over\gamma}={kM\over 4\pi\gamma}.
\label{peg7b}
\end{eqnarray}
For $\gamma>1$ (i.e., $n>0$), the critical temperature $T_{n}$ is
smaller than the corresponding one for an isothermal gas
$T_{c}=T_{\infty}$, i.e. the instability is delayed. For $\gamma<1$
(i.e., $n<0$), the instability is advanced. Similar results are
obtained for 3D gravitational systems (Chavanis \& Sire 2004b). According to
Eq. (\ref{ejl9}), the relation between $\lambda$ and the kinetic
temperature $T$ is
\begin{eqnarray}
\lambda^{2}=\gamma(T_{n}-T). \label{peg4grd}
\end{eqnarray}

\subsection{The local Euler equation}
\label{sec_loceuler}

\begin{figure}
\centering
\includegraphics[width=8cm]{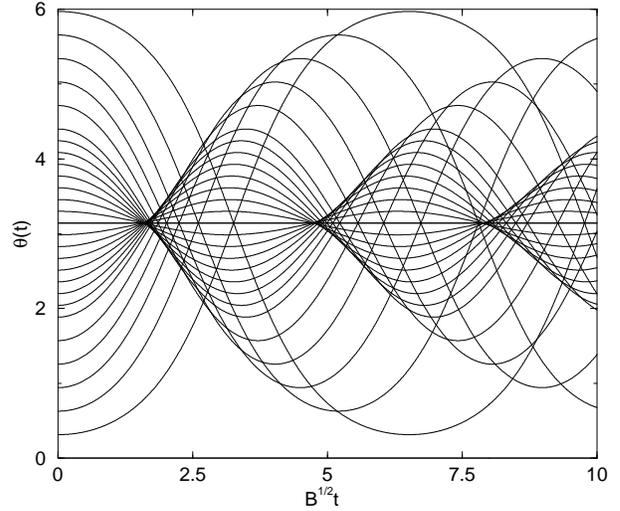}
\caption{Characteristics of the forced Burgers equation (\ref{loceuler2}) showing the appearance of shocks and chevrons.  }
\label{chevrons}
\end{figure}

We can consider a simplified problem where the potential in
Eq. (\ref{ej2}) is fixed to its equilibrium value
$\Phi=B\cos\theta$. In that case, we get the local Euler equation
\begin{eqnarray}
{\partial u\over\partial t}+u{\partial u\over\partial \theta}=-{1\over\rho}{\partial p\over\partial \theta}+B\sin\theta.
\label{loceuler1}
\end{eqnarray}
The stationary solutions are given by (\ref{hydro3}). The linear
stability of a stationary solution amounts to solving the
Sturm-Liouville problem
\begin{eqnarray}
{d\over d\theta}\biggl ({p'(\rho)\over\rho}{dq\over d\theta}\biggr )={\lambda^{2}\over\rho}q.
\label{loceuler1b}
\end{eqnarray}
This problem will be considered in Sec. \ref{sec_lfp} for the
isothermal equation of state.

In the pressureless case, we get the forced Burgers equation 
\begin{eqnarray}
{\partial u\over\partial t}+u{\partial u\over\partial \theta}=B\sin\theta.
\label{loceuler2}
\end{eqnarray}
It can be solved by the method of characteristics, writing the equation of motion of a particle as
\begin{eqnarray}
{d^{2}\theta\over dt^{2}}=B\sin\theta.
\label{loceuler3}
\end{eqnarray}
The trajectories $\theta(t)$ can then be expressed in terms of
elliptic functions. The dynamics of the forced Burgers equation is
interesting as it develops ``shocks'' and ``chevrons'' (caustics)
singularities (see Fig. \ref{chevrons}). The Burgers equation also
appears in cosmology to describe the formation of large-scale
structures in the universe (Vergassola et al. 1994). A detailed
description of this dynamics is given by Barr\'e et al. (2002) in the
context of the repulsive HMF model. In that case, the forced Burgers
equation (\ref{loceuler2}) with $\sin(2\theta)$ instead of
$\sin\theta$ models  the short time dynamics of the
Hamiltonian $N$-body system.  For the attractive HMF model
(\ref{mf1}), the short time dynamics can be modelled by the {\it
non-local} Euler equation (\ref{ej2}) with zero pressure $p=0$. This
is solution of the Vlasov equation (see Sec. \ref{sec_vlasov}) with
$f(\theta,v,t)=\rho(\theta,t)\delta(v-u(\theta,t))$. This single-speed
solution is valid until the first shock.  However, the connection with
the {\it local} Euler equation (\ref{loceuler2}) is not clear in that
case because the homogeneous phase is unstable which precludes the
possibility of deriving (\ref{loceuler2}) from (\ref{ej2}) as is done in
Barr\'e et al. (2002) in the repulsive case.  Note also that in the
attractive HMF model (ferromagnetic) we just have one cluser while the
repulsive HMF model (anti-ferromagnetic) shows a bicluster.

\section{Violent relaxation, metaequilibrium states and dynamical stability of collisionless stellar systems}
\label{sec_vlasov}

We now come back to the HMF model defined by the Hamilton equations
(\ref{mf1}) and develop a kinetic theory by analogy with stellar
systems. In particular, we emphasize the importance of the Vlasov
equation and the concept of violent relaxation introduced by
Lynden-Bell (1967).

\subsection{Vlasov equation and H-functions}
\label{sec_h}

For systems with long-range interactions, the relaxation time toward
the statistical equilibrium state (\ref{mb2}) is larger than $N t_D$,
where $t_D$ is the dynamical time (Chavanis 2004b). Accordingly, for
$N\rightarrow +\infty$, the relaxation time is extremely long and, for
timescales of physical interest, the evolution of the system is
essentially {\it collisionless}. More precisely, for $t\ll t_{relax}$
and $N\rightarrow +\infty$ (Vlasov limit), the time dependence of the
distribution function is governed by the Vlasov equation
\begin{equation}
\label{h1} {\partial f\over\partial t}+v{\partial
f\over\partial\theta}- {\partial\Phi\over\partial \theta}{\partial
f\over\partial v}=0,
\end{equation}
which has to be solved in conjunction with Eq. (\ref{mf7}). This
system of equations is similar to the Vlasov-Poisson system describing
the dynamics of elliptical galaxies and other collisionless stellar
systems in astrophysics. Starting from an unstable initial condition,
the HMF model (\ref{mf1}) will achieve a {\it metaequilibrium} state
(on a coarse-grained scale) as a result of phase mixing and violent
relaxation (Binney \& Tremaine 1987). This metaequilibrium state is a
particular stationary solution of the Vlasov equation. Since it
results from a complex mixing, it is highly robust and nonlinearly
dynamically stable with respect to collisionless perturbations. The
process of violent relaxation and the convergence of the distribution
function toward a stationary solution of the Vlasov equation has been
illustrated numerically by Yamaguchi et al. (2004) for the HMF
model. This is similar to the violent relaxation of stellar systems in
astrophysics and 2D vortices in hydrodynamics (see Chavanis 2002a).

One question of great importance is whether we can {\it predict} the
metaequilibrium state achieved by the system as a result of violent
relaxation. Lynden-Bell (1967) has tried to make such a prediction by
resorting to a new type of statistical mechanics accounting for the
conservation of all the Casimirs imposed by the Vlasov equation. This
theory was developed for the gravitational interaction, but the
general ideas and formalism apply to any system with long-range
interactions described by the Vlasov equation. In the non-degenerate
limit, he predicts a Boltzmann distribution of the form
$\overline{f}\sim e^{-\beta\epsilon}$ where the individual mass of the
particles does not appear. Lynden-Bell (1967) also understood that
his statistical prediction is limited by the concept of {\it
incomplete relaxation}. The system tries to reach the most mixed state
but, as the fluctuations become weaker and weaker as we approach
equilibrium, it can settle on a stable stationary solution of the
Vlasov equation which is not the most mixed state. In
order to quantify the importance of mixing, Tremaine et
al. (1986) have introduced the concept of H-functions
\begin{equation}
\label{h2} S=-\int C(\overline{f})d\theta dv,
\end{equation}
where $C$ is an arbitrary convex function. The H-functions calculated
with the coarse-grained distribution function $\overline{f}$ increase
as a result of phase mixing in the sense that
$S[\overline{f}(\theta,v,t)]\ge S[\overline{f}(\theta,v,0)]$ for $t>0$
where it is assumed that, initially, the system is not mixed so that
$\overline{f}(\theta,v,0)={f}(\theta,v,0)$. This is similar to the
$H$-theorem in kinetic theory.  However, contrary to the Boltzmann
equation, the Vlasov equation does not single out a unique functional
(the above inequality is true for all $H$-functions) and the time
evolution of the $H$-functions is not necessarily monotonic (nothing
is implied concerning the relative values of $H(t)$ and $H(t')$ for
$t,t'>0$). On the other hand, any stationary solution of the Vlasov
equation of the form $f=f(\epsilon)$ with $f'(\epsilon)<0$ extremizes
a $H$-function at fixed mass and energy. If, in addition, it {\it
maximizes} $S$ at fixed $E$, $M$, then it is nonlinearly dynamically
stable with respect to the Vlasov equation.  In astrophysics, such
distribution functions depending only on the energy describe spherical
stellar systems. This is a particular case of the Jeans theorem
(Binney \& Tremaine 1987). For a 1D system such as the HMF model, this
is the general form of inhomogeneous stationary solutions of the
Vlasov equation. Therefore, we expect that the H-functions will
increase during violent relaxation until one of them (non-universal)
reaches its maximum value at fixed mass and energy when a stationary
solution of the Vlasov equation is reached (this is not necessarily
the case in astrophysics since the system can reach a steady state
that does not depend only on energy). Note that the Boltzmann and the
Tsallis functionals are particular H-functions (not thermodynamical
entropies in that context) associated with particular stationary
solutions of the Vlasov equation: isothermal stellar systems and
stellar polytropes (Chavanis \& Sire 2004b). All these ideas, first
developed for stellar systems, apply to other systems with long-range
interactions such as the HMF model.

\subsection{Nonlinear dynamical stability criterion for the Vlasov equation}
\label{sec_nlv}

The theory of violent relaxation explains how a collisionless system
out of mechanical equilibrium can reach a steady  solution
of the Vlasov equation on a very short timescale due to long-range
interactions and chaotic mixing. Since this metaequilibrium state is
stable with respect to collisionless perturbations, it is of interest
to determine a criterion of formal nonlinear dynamical stability for
the Vlasov equation. For the HMF model, this has been considered by
Yamaguchi et al. (2004) using the Casimir-Energy method. We shall
propose another derivation of the stability criterion which uses a
formal analogy with the thermodynamical analysis developed in
Sec. \ref{sec_stat} and which is also applicable to the clustered
phase. This approach is similar to the one developed by Chavanis
(2003a) for 3D stellar systems.

Let us introduce the functional $S=-\int C(f)d\theta dv$ where $C(f)$
is a convex function. This functional is a particular Casimir so it is
conserved by the Vlasov equation. The energy $E$ and the mass $M$ are
also conserved. Therefore, a maximum of $S$ at fixed mass $M$ and
energy $E$ determines a stationary solution $f(\theta,v)$ of the
Vlasov equation that is nonlinearly dynamically stable.  We are led
therefore to consider the maximization problem
\begin{equation}
\label{nlv1} {\rm Max}\ \lbrace  S[f]\quad |\ E[f]=E, M[f]=M\rbrace.
\end{equation}
We also note that $F[f]=E[f]-TS[f]$ (where $T$ is a positive constant) is 
conserved by the Vlasov equation. Therefore, a minimum of $F$ at fixed
mass $M$ is nonlinearly dynamically stable with respect to the
Vlasov equation ($F$ is called an energy-Casimir
functional). This corresponds to the formal stability criterion of Holm et
al. (1985). This criterion can be written
\begin{equation}
\label{nlv2} {\rm Min}\ \lbrace F[f]\quad |\ M[f]=M\rbrace.
\end{equation}

To study the nonlinear dynamical stability of collisionless stellar
systems, we are thus led to consider the maximization problems
(\ref{nlv1}) and (\ref{nlv2}).  These are similar to the conditions of
thermodynamical stability (\ref{bol4}) and (\ref{bol5}) but they
involve a more general functional $S[f]$ than the Boltzmann
entropy. In addition, they have a completely different interpretation
since they determine the nonlinear dynamical stability of a steady
solution of the Vlasov equation, not the thermodynamical stability of
the statistical equilibrium state. Due to this formal resemblance, we
can develop a {\it thermodynamical analogy} (Chavanis 2003c) and use an effective thermodynamical vocabulary to
investigate the nonlinear dynamical stability of a collisionless
stellar system. In this analogy, $S$ plays the role of an effective
entropy, $T$ plays the role of an effective temperature and $F$ plays
the role of an effective free energy. The criterion (\ref{nlv1}) is
similar to a condition of microcanonical stability and the criterion
(\ref{nlv2}) is similar to a condition of canonical stability.

We also note that the stability criterion (\ref{nlv1}) is consistent
with the phenomenology of violent relaxation. Indeed, the H-functions
increase on the coarse-grained scale while the mass and the energy are
approximately conserved. Therefore, the metaequilibrium state is
expected to maximize a certain H-function (non-universal) at fixed
mass and energy.  In that interpretation, ${f}$ has to be viewed as
the coarse-grained distribution function $\overline{f}$, not the
distribution function itself. The point is that during mixing 
$D\overline{f}/Dt\neq 0$ and the $H$-functions $S[\overline{f}]$
increase. Once it has mixed $D\overline{f}/Dt= 0$ so that $\dot
S[\overline{f}]=0$. Since $\overline{f}(\theta,v,t)$ has been
brought to a maximum $\overline{f}_{0}(\theta,v)$ of a certain
$H$-function and since $S[\overline{f}]$ is conserved (after mixing),
then $\overline{f}_{0}$ is a nonlinearly dynamically stable steady
state of the Vlasov equation.

\subsection{First variations: stationary  solutions of the Vlasov equation}
\label{sec_sv}

Introducing Lagrange multipliers as in Sec. \ref{sec_mb}, the critical
points of the variational problems (\ref{nlv1}) and (\ref{nlv2}) are given by
\begin{equation}
\label{sv1}
C'(f)=-\beta\epsilon-\alpha,
\end{equation}
where $\epsilon={v^{2}\over 2}+\Phi$ is the energy of a
particle. Since $C'$ is a monotonically increasing function of $f$, we
can inverse this relation to obtain
\begin{eqnarray}
f=F(\beta\epsilon+\alpha),
\label{sv2}
\end{eqnarray}
where $F(x)=(C')^{-1}(-x)$. We can check that any DF $f=f(\epsilon)$ is a
stationary solution of the Vlasov equation (\ref{h1}). From the identity
\begin{eqnarray}
f'(\epsilon)=-\beta/C''(f),
\label{sv3}
\end{eqnarray}
resulting from Eq. (\ref{sv1}), we see that $f(\epsilon)$ is a monotonic
function of the energy. Assuming that $f(\epsilon)$ is decreasing, which is the physical situation, imposes $\beta=1/T>0$. 

We note also that for each stellar  system with
$f=f(\epsilon)$, there exists a corresponding barotropic star
with the same equilibrium density distribution. Indeed, defining
the density and the pressure by $\rho=\int_{-\infty}^{+\infty}
fdv=\rho(\Phi)$, $p=\int_{-\infty}^{+\infty} fv^{2}dv=p(\Phi)$,
and eliminating the potential $\Phi$ between these two
expressions, we find that $p=p(\rho)$. Writing explicitly the density and
the pressure in the form
\begin{eqnarray}
\rho=2\int_{\Phi}^{+\infty}F(\beta\epsilon+\alpha){1\over \lbrack 2(\epsilon-\Phi)\rbrack^{1/2}}d\epsilon,
\label{sv4}
\end{eqnarray}
\begin{eqnarray}
p=2\int_{\Phi}^{+\infty}F(\beta\epsilon+\alpha)\lbrack 2(\epsilon-\Phi)\rbrack^{1/2}d\epsilon,
\label{sv5}
\end{eqnarray}
and taking the $\theta$-derivative of Eq. (\ref{sv5}), we obtain the condition of hydrostatic equilibrium (\ref{hydro1}).

Due to the analogy between stellar systems and barotropic stars, it
becomes possible to use the results obtained in Sec. \ref{sec_gas} to
study the stationary solutions of the Vlasov equation (\ref{h1}). In
particular, the transition from homogeneous ($B=0$) to inhomogeneous
($B\neq 0$) solutions is again given by the criterion (\ref{hydro6}).
Now, in the case of stellar systems, it is more relevant to express
this criterion in terms of the distribution function.  Using the
identity (\ref{idro9}), the criterion (\ref{hydro6}) determining the
appearance of the clustered phase is equivalent to
\begin{eqnarray}
1+{k\over 2}\int_{-\infty}^{+\infty}{f'(v)\over v}dv\le 0.
\label{sv7}
\end{eqnarray}
We will soon see how this quantity is related to the dielectric
function of a gravitational plasma.

\subsection{Second variations: the condition of nonlinear dynamical stability}
\label{sec_svv}

We shall investigate the formal nonlinear dynamical stability of
stationary solutions of the Vlasov equation by using the criterion
(\ref{nlv2}). This criterion is less refined than the criterion
(\ref{nlv1}) because all solutions of (\ref{nlv2}) are solution of
(\ref{nlv1}), but the reciprocal is wrong in general. In particular,
for long-range interactions, the optimization problems (\ref{nlv1})
and (\ref{nlv2}) may not coincide. In thermodynamics, this corresponds
to a situation of ensemble inequivalence (see Bouchet \& Barr\'e
2004). Therefore, the criterion (\ref{nlv2}) can only give a {\it
sufficient} condition of nonlinear dynamical stability for stationary
solutions of the Vlasov equation of the form $f=f(\epsilon)$ with
$f'(\epsilon)<0$. This corresponds to the criterion of formal
nonlinear dynamical stability given by Holm et al. (1985). The more
refined criterion (\ref{nlv1}) has been introduced by Ellis et
al. (2002) in 2D hydrodynamics (for the 2D Euler-Poisson system) and
applied  by Chavanis (2003a) in stellar dynamics (for the Vlasov-Poisson
system).

To obtain a managable criterion of dynamical stability, we use the
same procedure as the one developed in Chavanis (2003a). We shall not
repeat the steps that are identical. We first minimize the functional
$F[f]$ at fixed temperature {\it and} density $\rho(\theta)$. This
gives an optimal distribution $f_*(\theta,v)$, determined by
$C'(f_*)=-\beta {v^2\over 2}-\lambda(\theta)$, which depends on the
density $\rho(\theta)$ through the Lagrange multiplier
$\lambda(\theta)$. Then, after some manipulations, we can show that
the functional $F[\rho]=F[f_*]$ can be put in the form
\begin{eqnarray}
F={1\over 2}\int\rho\Phi
d\theta+\int\rho\int_{0}^{\rho}{p(\rho')\over
\rho'^{2}}d\rho'd\theta, \label{svv1}
\end{eqnarray}
where $p(\rho)$ is the equation of state determined by $C(f)$
according to Eqs. (\ref{sv4}) and (\ref{sv5}). We now need to minimize
$F[\rho]$ at fixed mass. To that purpose, we just have to observe that
$F[\rho]$ corresponds to the energy functional (\ref{ej3}) of a
barotropic gas with $u=0$. Therefore, the cancellation of the first
variations of Eq. (\ref{svv1}) returns the condition of hydrostatic
equilibrium (\ref{hydro1}) and the positivity of the second variations
leads to the stability criterion $\lambda <0$ linked to the
eigenvalue equation (\ref{ejn2}). Therefore, the criterion of
formal nonlinear dynamical stability (\ref{nlv2}) for the Vlasov equation
(stellar systems) is equivalent to the criterion of formal nonlinear
dynamical stability (\ref{ej4}) for the Euler equations (gaseous barotropic
stars).

Using the results of Sec. \ref{sec_gas}, we conclude that the
uniform phase is formally nonlinearly dynamically stable with respect to the
Vlasov equation when
\begin{eqnarray}
c_{s}^{2}\ge  {kM\over 4\pi}.
\label{svv2}
\end{eqnarray}
When the inequality is reversed, the uniform phase is a saddle point of $F$
at fixed mass $M$ and we shall see that it is linearly dynamically unstable. 
Using the identity (\ref{idro9}), the nonlinear criterion (\ref{svv2}) can be rewritten as
\begin{eqnarray}
1+{k\over 2}\int_{-\infty}^{+\infty}{f'(v)\over v}dv\ge 0,
\label{svv3}
\end{eqnarray}
which was found by Yamaguchi et al. (2004) using a different
method. An advantage of the present approach is that this approach is also
applicable to an inhomogeneous system. Indeed, the stability of the
clustered phase can be investigated by solving the eigenvalue equation
(\ref{ejn2}) for the equation of state specified by the function
$C(f)$, and investigating the sign of $\lambda$.

\subsection{About the Antonov first law}
\label{sec_antonov}

As discussed previously, the criterion (\ref{nlv2}) providing a
condition of nonlinear dynamical stability for a stellar system
with $f=f(\epsilon)$ and $f'(\epsilon)<0$  is
equivalent to the criterion (\ref{ej4}) determining the nonlinear
dynamical stability of a barotropic star with the same equilibrium
density distribution. On the other hand, we have already indicated
that the criterion (\ref{nlv2}) is less refined than the criterion
(\ref{nlv1}) which is believed to be the strongest criterion of
nonlinear dynamical stability for stationary solutions of the Vlasov
equation of the form $f=f(\epsilon)$ with $f'(\epsilon)<0$.  In
general, the criterion (\ref{nlv2}) just provides a sufficient
condition of nonlinear dynamical stability. Thus, we can ``miss''
stable solutions if we use just (\ref{nlv2}) instead of (\ref{nlv1})
[said differently, the set of solutions of (\ref{nlv2}) is included in
(\ref{nlv1})]. From these remarks, we conclude that ``a stellar system
is stable whenever the corresponding barotropic gas is stable'' but
the converse is wrong in general. This is the so-called Antonov first
law in astrophysics (Binney \& Tremaine 1987). Our approach provides an {\it extension} of the Antonov first law
to the case of nonlinear dynamical stability (while the usual Antonov
first law corresponds to linear dynamical stability). Furthermore, by
developing a {\it thermodynamical analogy}, we have provided an
original interpretation of the nonlinear Antonov first law in terms of
 ``ensembles inequivalence'' for systems with
long-range interactions (Chavanis
2003a) . 

For 3D self-gravitating systems, we know that the ensembles are {\it
not} equivalent so that the criterion (\ref{nlv2}) is more restrictive
than the criterion (\ref{nlv1}). In that case, (\ref{nlv2}) is just a
{\it sufficient} condition of nonlinear dynamical stability
(definitely). For example, using the criterion (\ref{ej4}), it can be
shown that polytropic stars with index $n<3$ are nonlinearly
dynamically stable with respect to the Euler equations while
polytropic stars with index $3<n<5$ are dynamically unstable
(polytropes with index $n>5$ have infinite mass). Therefore, using the
``canonical'' criterion (\ref{nlv2}), we can only deduce that stellar
polytropes with index $n<3$ are nonlinearly dynamically stable with
respect to the Vlasov equation. However, using the ``microcanonical''
criterion (\ref{nlv1}), we can prove that all stellar polytropes with
index $n<5$ are nonlinearly dynamically stable with respect to the
Vlasov equation. Polytropes with index $3<n<5$ lie in a region of
``ensemble inequivalence'' (in the thermodynamical analogy) where the
``specific heat'' is negative (Chavanis 2003a).

For the HMF model considered in this paper, we believe that the
criteria (\ref{nlv1}) and (\ref{nlv2}) determine the same set of
solutions so that the ensembles are equivalent in that case (no
solution is forgotten by (\ref{nlv2})). We have shown in
Sec. \ref{sec_stat} that this is at least the case for isothermal
distributions. If we take for granted that this equivalence extends to
any functional (\ref{h2}), we conclude that, for the HMF model, ``a
stellar system is stable if, and only if, the corresponding barotropic
gas is stable''. This would be the HMF version of the nonlinear
Antonov first law. In that case, the stability limits obtained for the
Euler equation in Sec. \ref{sec_gas} can be directly applied to the
Vlasov equation (they coincide).  Clearly, it would be of great
interest to derive general criteria telling when the ensembles are
equivalent or inequivalent, depending on the form of the functional
(\ref{h2}) and on the form of the potential of interaction $u({\bf r}-{\bf
r}')$.

\subsection{The condition of linear stability}
\label{sec_ik}

We now study the linear dynamical stability of a spatially homogeneous
stationary solution of the Vlasov equation described by $f=f(v)$. This
problem was first investigated by Inagaki \& Konishi (1993) and Pichon
(1994) and more recently by Choi \& Choi (2003). We shall complete
here their study. Writing the perturbation in the form $\delta f\sim
e^{i(n\theta-\omega t)}$ and using standard methods of plasma physics,
we obtain the dispersion relation
\begin{eqnarray}
\epsilon(n,\omega)\equiv 1+{k\over 2}(\delta_{n,1}-\delta_{n,-1})\int {f'(v)\over nv-\omega}dv=0, \nonumber\\
\label{ik1}
\end{eqnarray}
where $\epsilon(n,\omega)$ is the dielectric function and the integral
must be performed by using the Landau contour. For the destabilizing
mode $n=1$ ($n=-1$ gives the same result), Eq. (\ref{ik1}) reduces to
\begin{eqnarray}
1+{k\over 2}\int {{f'(v)}\over v-{\omega}}dv=0.
\label{ik2}
\end{eqnarray}
The condition of marginal
stability is $\omega_{i}=0$ where $\omega_{i}$ is the imaginary part
of $\omega=\omega_{r}+i\omega_{i}$. In that case, the integral in
Eq. (\ref{ik2}) can be written as
\begin{eqnarray}
1+{k\over 2}{\cal P}\int {{f'(v)}\over v-\omega_{r}}dv+i\pi {k\over 2}
f'(\omega_{r})=0,
\label{ik4}
\end{eqnarray}
where ${\cal P}$ denotes the principal value. Identifying real and imaginary parts, if follows that 
\begin{eqnarray}
1+{k\over 2}{\cal P}\int {{f'(v)}\over v-\omega_{r}}dv=0,\nonumber\\
f'(\omega_{r})=0.\qquad\qquad\qquad
\label{ik5}
\end{eqnarray}
The second relation fixes the frequency of the perturbation and the first equation determines the point of marginal stability in the series of equilibria. 
The system is linearly dynamically stable if 
\begin{eqnarray}
1+{k\over 2}{\cal P}\int{f'(v)\over v-\omega_{r}}dv\ge 0,
\label{ik6}
\end{eqnarray}
and linearly dynamically unstable otherwise.  Note that $f(v)$ does
not need to be symmetrical. However, if $f(v)$ extremizes a H-function
at fixed mass and energy, then it has a single maximum at
$v=0$. Therefore, $\omega_{r}=0$ according to Eq. (\ref{ik5}) and the
criterion of linear dynamical stability (\ref{ik6}) coincides with the
criterion of formal nonlinear dynamical stability (\ref{svv3}).

\subsection{Particular examples}
\label{sec_lyapunov}

We shall now present explicit results for particular stationary
solutions of the Vlasov equation. We consider the case of
isothermal stellar systems and stellar polytropes.

\subsubsection{Isothermal stellar systems}
\label{sec_we}

We consider the H-function
\begin{eqnarray}
S=-\int f\ln f d\theta dv, \label{we1}
\end{eqnarray}
which is similar to the Boltzmann entropy (\ref{bol3}) in
thermodynamics.  However, as explained in Sec. \ref{sec_h}, its
physical interpretation is different. Its maximization at fixed mass
and energy determines a formally nonlinearly dynamically stable stationary
solution of the Vlasov equation corresponding to the isothermal
distribution function
\begin{eqnarray}
f=A e^{-\beta\epsilon}.
\label{we2}
\end{eqnarray}
This distribution function has the same form (but a different
interpretation) as the statistical equilibrium state (\ref{mb2}) of
the $N$-body system.

The barotropic gas corresponding to the isothermal distribution
function (\ref{we2}) is the isothermal gas with an equation of state
$p=\rho T$ where $T=1/\beta$. Therefore, the velocity dispersion
$\beta^{-1}$ of an isothermal stellar system is equal to the velocity
of sound $c_{s}^{2}=T$ in the corresponding isothermal gas. The
functional (\ref{svv1}) takes the form
\begin{eqnarray}
F[\rho]={1\over 2}\int\rho\Phi d\theta+T\int \rho\ln\rho d\theta,
\label{we4}
\end{eqnarray}
and the density is related to the potential according
to the formula
\begin{eqnarray}
\rho=A' e^{-\beta\Phi},
\label{we5}
\end{eqnarray}
which can be obtained by extremizing $F[\rho]$ at fixed mass. We can also express the distribution function in terms of the density according to
\begin{eqnarray}
f=\biggl ({\beta\over 2\pi}\biggr )^{1/2}\rho(\theta)e^{-\beta {v^{2}\over 2}}.
\label{we5bis}
\end{eqnarray}
According to what has been said in Sec. \ref{sec_antonov} about the
correspondance between stellar systems and barotropic stars, we
conclude that the uniform phase of an isothermal stellar system
(\ref{we2}) is formally nonlinearly dynamically stable with respect to
the Vlasov equation if $T>T_{c}$ and linearly dynamically unstable
otherwise. In terms of the dimensionless parameters $\eta={kM\over
4\pi T}$ and $\epsilon={8\pi E\over kM^{2}}$, the conditions of
dynamical stability can be written
\begin{eqnarray}
\eta\le 1, \qquad \epsilon\ge 1. \label{we7}
\end{eqnarray}
They coincide with the conditions of thermodynamical stability (see
Sec. \ref{sec_stat}).

For the Maxwellian distribution function (\ref{we5bis}) with uniform
density, the dielectric function can be written
\begin{eqnarray}
\epsilon(1,\omega)=1-\eta W(\sqrt{\beta}\omega),
 \label{we8}
\end{eqnarray}
where 
\begin{eqnarray}
W(z)={1\over\sqrt{2\pi}}\int_{-\infty}^{+\infty}{x\over x-z}e^{-{x^{2}\over 2}}dx,
 \label{we8b}
\end{eqnarray}
is the $W$-function of plasma physics (Ichimaru 1973). This is an
analytic function in the upper plane of the complex $z$ plane which is
continued analytically into the lower half plane.  Explicitly,
\begin{eqnarray}
W(z)=1-z e^{-{z^{2}\over 2}}\int_{0}^{z}dy e^{y^{2}\over 2}+i{\sqrt{\pi\over 2}}z e^{-{z^{2}\over 2}}.
\label{lb3}
\end{eqnarray}
We look for solutions of the dispersion relation $\epsilon(1,\omega)=0$ in the form $\omega=i\lambda$ where $\lambda$ is real. First, we note that
\begin{eqnarray}
\epsilon(1,i\lambda)=1-\eta/G\biggl (\sqrt{\beta\over 2}\lambda\biggr ),
\label{tb11neb3}
\end{eqnarray} 
where we have defined the $G$-function
\begin{equation}
\label{ge1g} G(x)={1\over 1-\sqrt{\pi}x e^{x^{2}}{\rm erfc}(x)}.
\end{equation}
For $x\rightarrow 0$, $G(x)=1+\sqrt{\pi}x+...$. For $x\rightarrow +\infty$, $G(x)=2x^{2}(1+{3\over 2x^{2}}+...)$.  For $x\rightarrow -\infty$,  $G(x)\sim -{1\over 2\sqrt{\pi}x}e^{-x^{2}}$. Therefore, the  relation between $\lambda$ and $T$ is given by
\begin{eqnarray}
\eta=G\biggl (\sqrt{\beta\over 2}\lambda\biggr ).
 \label{we8d}
\end{eqnarray}

The case of neutral stability $\omega=0$ corresponds to $T=T_{c}$ (or
$\eta=1$). The case of instability $(\lambda>0)$ corresponds to
$T<T_{c}$. The perturbation grows exponentially rapidly as $\delta
f\sim e^{\lambda t}$. The growth rate $\lambda$ is given by
Eq. (\ref{we8d}) which can be explicitly written
\begin{equation}
\label{we9}1-{T_{c}\over T}\biggl \lbrace 1-\sqrt{\pi \over 2T}{\lambda}e^{{\lambda^{2}\over 2T}}{\rm erfc}\biggl( {\lambda\over \sqrt{2T}}\biggr )\biggr\rbrace=0.
\end{equation}
For $T\rightarrow T_{c}^{-}$, we have
\begin{eqnarray}
\lambda\sim \sqrt{8\over kM}(T_{c}-T),
\label{we11}
\end{eqnarray}
and for $T\rightarrow 0$, we have
\begin{eqnarray}
\lambda\rightarrow \sqrt{kM\over 4\pi}\biggl (1-3{T\over T_{c}}\biggr )^{1/2}.
\label{we12}
\end{eqnarray}
The first term in Eq. (\ref{we12}) can be deduced directly from
Eq. (\ref{ik2}) by using the distribution function $f(v)=\rho \
\delta(v)$ valid at $T=0$ and integrating by parts. The case of stability $(\lambda<0)$
corresponds to $T>T_{c}$.  The perturbation is damped exponentially
rapidly as $\delta f\sim e^{-\gamma t}$ where $\gamma=-\lambda$. This
is similar to the Landau damping in plasma physics, except that here
there is no pulsation ($\omega_{r}=0$). By contrast, in plasma
physics, the pulsation $\omega_{r}$ is much larger than the damping
rate $\gamma$.  The damping rate $\gamma=-\lambda$ is given by
\begin{equation}
\label{gam} \eta=F\biggl (\sqrt{\beta \over 2}{\gamma}\biggr )
\end{equation}
where we have defined the $F$-function
\begin{equation}
\label{ge1} F(x)={1\over 1+\sqrt{\pi}x e^{x^{2}}{\rm erfc}(-x)},
\end{equation}
such that $F(x)=G(-x)$. For $x\rightarrow 0$, $F(x)=1-\sqrt{\pi}x+...$. For $x\rightarrow -\infty$, $F(x)\sim 2x^{2}(1+{3\over 2x^{2}}+...)$.  For $x\rightarrow +\infty$,  $F(x)\sim {1\over 2\sqrt{\pi}x}e^{-x^{2}}$. Explicitly,
\begin{equation}
\label{ge1exp}1-{T_{c}\over T}\biggl \lbrace 1+\sqrt{\pi \over 2T}{\gamma}e^{{\gamma^{2}\over 2T}}{\rm erfc}\biggl(- {\gamma\over \sqrt{2T}}\biggr )\biggr\rbrace=0.
\end{equation} 
For $T\rightarrow
T_{c}^{+}$, we have
\begin{eqnarray}
\gamma\sim \sqrt{8\over kM}(T-T_{c}),
\label{we11b}
\end{eqnarray}
and for $T\rightarrow +\infty$, we have
\begin{eqnarray}
\gamma\sim \sqrt{2T\ln T}.
\label{we12b}
\end{eqnarray}

\begin{figure}
\centering
\includegraphics[width=8cm]{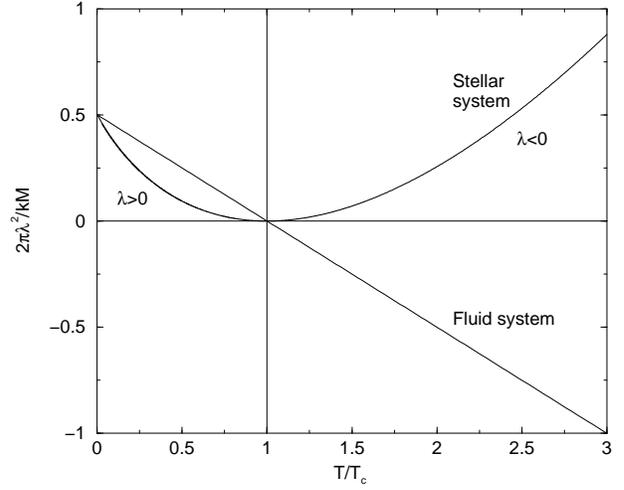}
\caption{Growth rate and decay rate of isothermal stellar systems and isothermal stars as a function of the temperature in the framework of the HMF model. }
\label{TL}
\end{figure}

Obviously, the relation (\ref{we9}) between the growth rate and the
temperature of an isothermal stellar system is different from the
corresponding relation (\ref{peg4}) valid for an isothermal gas (they
coincide only at $T=0$). A similar distinction is noted in the case of
3D self-gravitating systems. In particular, Fig. \ref{TL} can be
compared with Fig. 5.1 of Binney \& Tremaine (1987). Moreover, in the
unstable regime, a gaseous medium supports sound waves with pulsation
$\omega=(T-T_{c})^{1/2}$ that are not attenuated. By contrast, in a
stellar medium at $T<T_{c}$, there exists solutions with no wave
$(\omega_{r}=0)$ for which the perturbation is damped exponentially.
Other solutions with $\omega_{r}\neq 0$ probably exist but they are
more difficult to investigate analytically. This is left for a future
study.

The relations (\ref{we9}) and (\ref{ge1exp}) have been obtained 
previously by Choi \& Choi (2003) using a slightly different approach. Our derivation emphasizes the close link with results in plasma physics. In
addition, our formalism will be used in Sec. \ref{sec_auto} to show 
that the damping rate $\gamma$ of a perturbation is equal to the
exponential decay of the time auto-correlation function of the force.
The generalization of these results for an arbitrary form of long-range potential is given in Chavanis (2004b).

\subsubsection{Stellar polytropes}
\label{sec_spoly}

We consider the H function
\begin{eqnarray}
S_q=-{1\over q-1}\int (f^{q}-f) d\theta dv, \label{we13}
\end{eqnarray}
where $q$ is a real number. This functional has been introduced by
Tsallis (1988) in non-extensive thermodynamics. The aim was to develop
a generalized thermodynamical formalism to describe quasi-equilibrium
structures in complex media that are not described by the Boltzmann
distribution. In this sense, $S_{q}[f]$ is interpreted as a
generalized entropy and its maximization as a condition of
(generalized) thermodynamical stability. In the context of Vlasov
systems, we shall rather interprete $S_{q}[f]$ as a particular
$H$-function (see Sec. \ref{sec_h}) and its maximization as a
condition of nonlinear dynamical stability \footnote{If we were to
apply Tsallis thermodynamics in the context of violent relaxation, we
would need to introduce the fine-grained distribution of phase levels
$\rho(\theta,v,\eta)$ [which is the relevant {\it probability field}
in that context] and replace the Lynden-Bell entropy
$S_{L.B.}[\rho]=-\int
\rho\ln\rho d\theta dvd\eta$ (Lynden-Bell 1967) 
by $S_{q}[\rho]=-{1\over q-1}\int (\rho^{q}-\rho) d\theta dvd\eta$ as
suggested in Brands {\it et al.} (1999).  In that case, $S_{q}[\rho]$
can be regarded as a generalized entropy trying to
take into account non-ergodicity and lack of complete mixing in
collisionless systems with long-range interactions. In that point of view,
the parameter $q$ measures the efficiency of mixing ($q=1$ if the
system mixes well which is implicitly assumed in Lynden-Bell's
statistical theory). In the two-levels approximation $f\in\lbrace 0,
1\rbrace$ and in the dilute limit $\overline{f}\ll 1$, $S_{q}[\rho]$
can be expressed as a functional of the coarse-grained distribution
function $\overline{f}\equiv\int\rho\eta d\eta=\rho$ of the form
$S_q[\overline{f}]=-{1\over q-1}\int (\overline{f}^{q}-\overline{f})
d\theta dv$. In this particular limit, $S_q[\overline{f}]$ can be
interpreted as a thermodynamical entropy generalizing
$S[\overline{f}]=-\int
\overline{f}\ln\overline{f} d\theta dv$ which is a particular case of
the Lynden-Bell entropy for two levels in the dilute limit (see
Chavanis 2004c). In conclusion, Tsallis functional $S_{q}[\rho]$
expressed in terms of $\rho(\theta,v,\eta)$ is an entropy but Tsallis
functional $S_{q}[\overline{f}]$ expressed in terms of
$\overline{f}(\theta,v)$ is either a $H$-function (dynamics) or the
reduced form of entropy $S_{q}[\rho]$ (thermodynamics) for two levels
in the dilute limit. In any case, it is not clear why non-ergodic
effects could be encapsulated in the simple functional $S_{q}[\rho]$
introduced by Tsallis. Tsallis entropy is ``natural'' because it has
mathematical properties very close to those possessed by the Boltzmann
entropy and it is probably relevant to describe a certain type of
mixing and non-ergodic behaviour as in the case of porous media and
weak chaos (it may be seen as an entropy on a fractal
space-time). However, many other types of non-ergodic behaviour can
occur and other functionals $S=-\int C(\rho)d\theta dvd\eta$ could be
relevant. Observation of stellar systems, 2D vortices and
quasi-equilibrium states of the HMF model resulting from incomplete
violent relaxation do not favour Tsallis distributions in a universal
manner. Other distributions can emerge. In fact, we must give up the
hope to {\it predict} the metaequilibrium state in case of incomplete
relaxation. We must rather try to construct stable stationary
solutions of the Vlasov equation in order to {\it reproduce} observed
phenomena. The H-functions (\ref{h2}) can be useful in that
context. An alternative procedure is to keep the Lynden-Bell form of
entropy unchanged and develop a {\it dynamical} theory of violent
relaxation (Chavanis {\it et al.} 1996, Chavanis 1998) in order to
take into account incomplete mixing through a variable diffusion
coefficient related to the strength of the fluctuations. In that case,
non-ergodicity is explained by the decay of the fluctuations of $\Phi$
driving the relaxation, not by a complex structure of
phase-space. Generalized entropies are not necessary in that case.  }.
Its maximization at fixed mass and energy leads to a particular class
of nonlinearly dynamically stable stationary solutions of the Vlasov
equation called stellar polytropes. The fact that the criteria
(\ref{nlv1}) and (\ref{nlv2}) of nonlinear dynamical stability are
similar to criteria of generalized thermodynamical stability is the
mark of a thermodynamical analogy (Chavanis 2003c, Chavanis \& Sire
2004b).

Stellar polytropes are described
by the distribution function
\begin{eqnarray}
f=\biggl\lbrack \mu-{(q-1)\beta\over q}\epsilon\biggr \rbrack ^{1\over q-1},
\label{we14}
\end{eqnarray}
obtained from Eq. (\ref{sv1}).  When the term in brackets is negative,
the distribution function is set equal to $f=0$.  The index $n$ of the
polytrope in one dimension is related to the parameter $q$ by the
relation (Chavanis \& Sire 2004a)
\begin{eqnarray}
n={1\over 2}+{1\over q-1}.
\label{we15}
\end{eqnarray}
For $n\rightarrow +\infty$ (or $q\rightarrow 1$), we recover the
isothermal distribution function (\ref{we2}) and the H-function
(\ref{we1}). Therefore, Tsallis functional connects continuously
isothermal and polytropic distributions. Physical polytropic distribution functions (see Chavanis \& Sire 2004a) have $\beta>0$ and  $q\ge 1$ (i.e. $n\ge 1/2$) or $1/3<q\le 1$ (i.e. $n< -1$).

The barotropic gas corresponding to the polytropic distribution
function (\ref{we14}) is the polytropic gas
\begin{eqnarray}
p=K\rho^{\gamma} ,\qquad \gamma=1+{1\over n},
\label{we16}
\end{eqnarray}
with the polytropic constant
\begin{eqnarray}
K={1\over n+1}\biggl \lbrace \sqrt{2}A{\Gamma(1/2)\Gamma(n+1/2)\over \Gamma(n+1)}\biggr \rbrace^{-1/n}, \ n>{1\over 2}\qquad
\label{we17}
\end{eqnarray}
\begin{eqnarray}
K=-{1\over n+1}\biggl \lbrace \sqrt{2}A{\Gamma(1/2)\Gamma(-n)\over \Gamma(1/2-n)}\biggr \rbrace^{-1/n}, \ n<-1\qquad
\label{we17b}
\end{eqnarray}
where $A=(|q-1|\beta/q)^{1/(q-1)}$.  In the present
context, the polytropic constant $K$ is related to the Lagrange
multiplier $\beta$.  Therefore, $K$ and $T_{0}=\beta^{-1}$ play the
role of effective temperatures (see Chavanis \&  Sire 2004b for a more
detailed discussion). For a polytropic distribution, the functional
(\ref{svv1}) takes the form
\begin{eqnarray}
F[\rho]={1\over 2}\int\rho\Phi d\theta+{K\over \gamma-1}\int (\rho^{\gamma}-\rho)d\theta,
\label{we18}
\end{eqnarray}
and the relation between the density and the potential is
\begin{eqnarray}
\rho=\biggl\lbrack \lambda-{\gamma-1\over K\gamma}\Phi\biggr\rbrack^{1\over\gamma-1}.
\label{we19}
\end{eqnarray}
Comparing Eqs. (\ref{we14}) and (\ref{we19}), we note that a
polytropic distribution with index $q$ in phase space yields a
polytropic distribution with index $\gamma=1+2(q-1)/(q+1)$ in physical
space.  In this sense, polytropic laws are stable laws since they keep
the same structure as we pass from phase space $f=f(\epsilon)$ to
physical space $\rho=\rho(\Phi)$ as noticed in Chavanis (2004d).  This
is probably the only distribution enjoying this property. Similarly,
comparing (\ref{we13})-(\ref{mf6}) and (\ref{we18}) the ``free
energy'' in phase space $F[f]=E[f]-T_{0}S_{q}[f]$ (where
$T_{0}=1/\beta$) becomes $F[\rho]=E[\rho]-KS_{\gamma}[\rho]$ in
physical space. Morphologically, the polytropic temperature $K$ plays
the same role in physical space as the temperature $T_{0}=1/\beta$ in
phase space.

 \begin{figure}
\centering
\includegraphics[width=8cm]{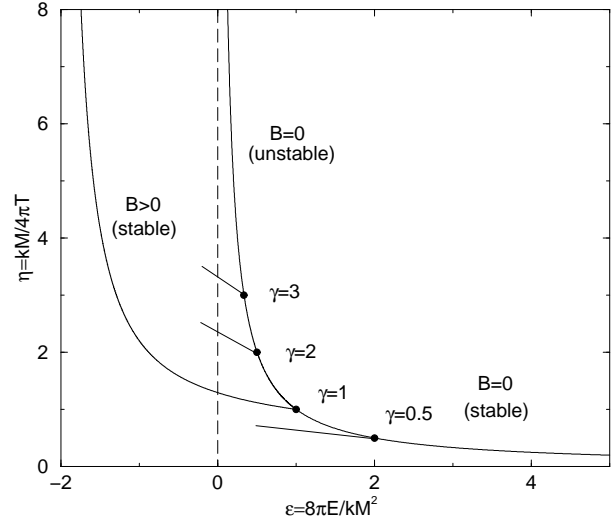}
\caption{Bifurcation diagram of stellar polytropes that 
are stationary solutions of the Vlasov equation. The homogeneous phase
is nonlinearly dynamically stable for $\epsilon\ge
\epsilon_{crit}=1/\gamma$ where $\gamma=1+1/n$ and $n={1\over
2}+{1\over q-1}$. It becomes linearly dynamically unstable for
$\epsilon<\epsilon_{crit}$ where the branch of clustered states
(represented schematically by a line) appears. Isothermal stellar
systems correspond to ($q=1$, $n=\infty$, $\gamma=1$). In ordinate,
$T$ is defined by $K(M/2\pi)^{1/n}$. In the homogeneous phase, it
represents the kinetic temperature of a polytropic stellar
system. Note that $\eta$ is a monotonic function of the Lagrange
multiplier $\beta$ so that the curve can be viewed as a series of
equilibria of polytropic distributions.}
\label{epsetaPOLY}
\end{figure}

We can express the distribution function in terms of the density
according to
\begin{eqnarray}
f={1\over Z}\biggl\lbrack \rho(\theta)^{1/n}-{v^{2}/2\over (n+1)K}\biggr\rbrack^{n-1/2},
\label{we19b}
\end{eqnarray}
with
\begin{eqnarray}
Z=\sqrt{2}{\Gamma(1/2)\Gamma(n+1/2)\over \Gamma(n+1)}   \lbrack K(n+1)\rbrack^{1/2}, \quad n>{1\over 2}\qquad
\label{we19c}
\end{eqnarray}
\begin{eqnarray}
Z=\sqrt{2}{\Gamma(1/2)\Gamma(-n)\over \Gamma(1/2-n)}   \lbrack -K(n+1)\rbrack^{1/2}, \quad n<-1.\qquad
\label{we19cb}
\end{eqnarray}
Introducing the kinetic temperature (velocity dispersion)  $T(\theta)=\langle v^{2}\rangle=p(\theta)/\rho(\theta)=K\rho(\theta)^{1/n}$, this can be rewritten
\begin{eqnarray}
f=B_{n}{\rho(\theta)\over \sqrt{2\pi T(\theta)}}\biggl\lbrack 1-{v^{2}/2\over (n+1)T(\theta)}\biggr\rbrack^{n-1/2},
\label{we19d}
\end{eqnarray}
with
\begin{eqnarray}
B_{n}={\Gamma(n+1)\over \Gamma(n+1/2)({n+1})^{1/2}}, \qquad n>{1\over 2}
\label{we19e}
\end{eqnarray}
\begin{eqnarray}
B_{n}={\Gamma(1/2-n)\over \Gamma(-n)\lbrack-({n+1})\rbrack^{1/2}}, \qquad n<-1.
\label{we19eb}
\end{eqnarray}
Equation (\ref{we19d})  is the counterpart of Eq. (\ref{we5bis}) for isothermal systems. For $n>1/2$, the distribution $f=0$ for $|v|>v_{max}=\sqrt{2(n+1)T}$.

According to what has been said in Sec. \ref{sec_svv} about the
correspondance between stellar systems and barotropic stars, we
conclude that the uniform phase of a polytropic stellar system
(\ref{we14}) with index $n$ is formally nonlinearly dynamically stable
with respect to the Vlasov equation if $K\ge K_{n}$ or $T\ge T_{n}$
and linearly dynamically unstable otherwise. It can be useful to
introduce the dimensionless parameter $\eta={kM/4\pi T}$ where
$T=K\rho^{1/n}$ is the kinetic temperature in the homogeneous phase
where $\rho=M/2\pi$. For $n\rightarrow +\infty$ (isothermal case), we
recover the dimensionless parameter $\eta=kM/4\pi T$ of
Sec. \ref{sec_we}. On the other hand, in the homogeneous phase
($B=0$), the energy (\ref{mf6}) is given by $E={1\over 2}\int pd\theta={1\over
2}MT$. Therefore, the normalized energy $\epsilon={8\pi E/kM^{2}}$ is
expressed in term of $\eta$ according to $\epsilon={1\over\eta}$. In
terms of these dimensionless parameters, the uniform phase is
formally nonlinearly dynamically stable for
\begin{eqnarray}
\eta\le \eta_{crit}=\gamma,\qquad \epsilon\ge \epsilon_{crit}={1\over \gamma}, \label{we24}
\end{eqnarray}
and linearly dynamically unstable otherwise. Note how the critical
energy and temperature are simply expressed in terms of the
polytropic index $\gamma$. For $n\rightarrow +\infty$, we recover the
case of isothermal stellar systems with $\epsilon_{crit}=1$ and
$\eta_{crit}=1$. Note that the line $(\epsilon_{crit},\eta_{crit})$
coincides with the line $B=0$ in Fig. \ref{epseta}. We thus clearly
see how the series of equilibria for polytropic distributions places
itself in the $(\epsilon,\eta)$ plane (we just have to displace the
critical point $(\epsilon_{crit},\eta_{crit})$ along the line $B=0$ as
sketched in Fig. \ref{epsetaPOLY}).

We also emphasize that, using the criterion (\ref{svv2}) we have
obtained the condition of nonlinear dynamical instability (\ref{we24})
for stellar polytropes with almost no calculation.  Of course, the
same result can be obtained from the criterion (\ref{svv3}) by
explicitly performing the integral. The criterion of nonlinear dynamical
stability (\ref{svv2}) that we have found is simpler, albeit
equivalent. Moreover, it has a more physical interpretation since it
is expressed as a condition on the velocity of sound in a gas with the
same density distribution as the original kinetic system.

\begin{figure}
\centering
\includegraphics[width=8cm]{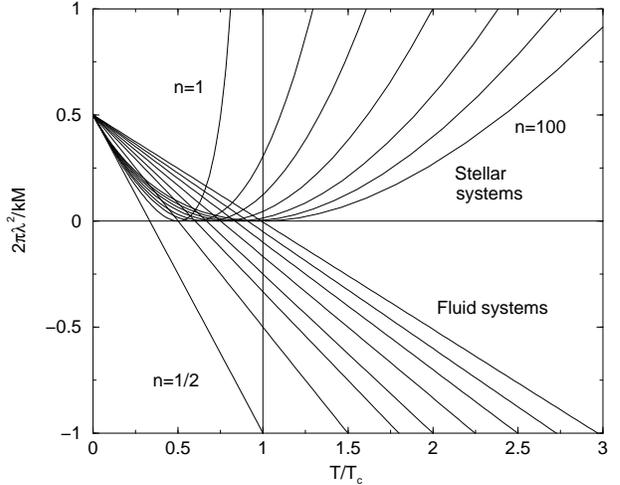}
\caption{Growth rate and decay rate for stellar polytropes and polytropic stars in the framework of the HMF model. The index goes from $n=1$ to $n=100$. We have also shown the case $n=1/2$ (water-bag distribution). The critical temperature is smaller than for an isothermal gas ($T_{n}<T_{c}$). }
\label{gr}
\end{figure}

\begin{figure}
\centering
\includegraphics[width=8cm]{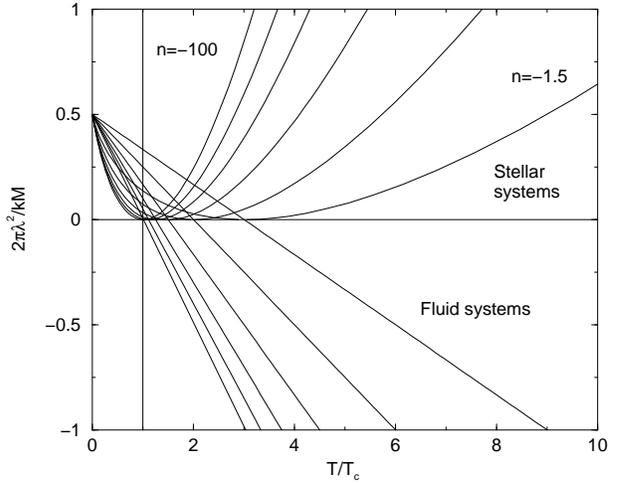}
\caption{Growth rate and decay rate for stellar polytropes and polytropic stars in the framework of the HMF model. The index goes from $n=-1.5$ to $n=-100$. The critical temperature is larger than for an isothermal gas ($T_{n}>T_{c}$). }
\label{disneg}
\end{figure}

For the polytropic distribution
(\ref{we19d}) with uniform density, the dielectric function can be written
\begin{eqnarray}
\epsilon(1,\omega)=1-{T_{n}\over T}W_{n}(\omega/\sqrt{T}),
 \label{mer1}
\end{eqnarray}
where we have introduced the function
\begin{eqnarray}
W_{n}(z)={1\over\sqrt{2\pi}}{B_{n}\over n}(n-1/2)\int {x\lbrack 1-{x^{2}/2\over n+1} \rbrack^{n-3/2}\over x-z}dx \qquad
 \label{mer2}
\end{eqnarray}
with $W_{n}(0)=1$. The range of integration is such
that the term in brackets remains positive. For $n\rightarrow
+\infty$, we recover the $W$-function (\ref{we8b}). For $\omega=0$, we
obtain the critical temperature $T=T_{n}$ as in a polytropic
gas. As in Sec. \ref{sec_we}, we look for solutions of the dispersion relation
$\epsilon(1,\omega)=0$ in the form $\omega=i\lambda$ where $\lambda$
is real. For $T<T_{n}$, the system is unstable and the growth rate
$\lambda>0$ is given by
\begin{eqnarray}
{T_{n}\over T}=G_{n}\biggl ({\lambda\over\sqrt{2T}}\biggr ),
 \label{mer3}
\end{eqnarray} 
where
\begin{eqnarray}
G_{n}(x)=\biggl\lbrace {B_{n}\over n\sqrt{\pi}}\biggl (n-{1\over 2}\biggr )\int {t^{2}\over t^{2}+x^{2}}\biggl \lbrack  1-{t^{2}\over n+1} \biggr \rbrack^{n-{3\over 2}}dt\biggr\rbrace^{-1}.\nonumber
 \label{mer4}
\end{eqnarray}  
For $T>T_{n}$, the system is stable and the damping rate $\gamma=-\lambda>0$ is given by
\begin{eqnarray}
{T_{n}\over T}=F_{n}\biggl ({\gamma\over\sqrt{2T}}\biggr ),
 \label{mer5}
\end{eqnarray} 
where 
\begin{eqnarray}
F_{n}(x)={1\over G_{n}(x)^{-1}+R_{n}(x)}
 \label{mer6}
\end{eqnarray} 
with
\begin{eqnarray}
R_{n}(x)={2\sqrt{\pi}B_{n}\over n}\biggl (n-{1\over 2}\biggr )x\biggl \lbrack  1+{x^{2}\over n+1} \biggr \rbrack^{n-{3\over 2}}.
 \label{mer7}
\end{eqnarray} 
This additional term comes from the residue theorem when the pulsation
$\omega=-i\gamma$ lies in the lower half of the complex plane. For
$n\rightarrow +\infty$, we recover the $G$ and $F$ functions
(\ref{ge1g}) and (\ref{ge1}).  The dependence of the growth rate and
decay rate with the temperature is shown in Figs. \ref{gr} and
\ref{disneg}. For $n=\infty$, we recover the isothermal case of
Fig. \ref{TL}.

\subsubsection{Fermi or water-bag distribution}
\label{sec_wb}

For $n=1/2$, the distribution function (\ref{we14}) is a step
function: $f(\epsilon)=\eta_{0}$ if $-v_{0}(\theta)\le v\le
v_{0}(\theta)$ and $f(\epsilon)=0$ otherwise. This is similar to the
Fermi distribution at $T=0$ describing cold white dwarf stars in
astrophysics (Chandrasekhar 1939).  This is also called the water-bag
distribution in plasma physics (when $v_{0}$ is independent on
$\theta$). The density and the pressure are given by
$\rho=2\eta_{0}v_{0}$ and $p=(2/3)\eta_{0}v_{0}^{3}$. This leads to a
polytropic equation of state $p=K\rho^{3}$ of index $n=1/2$ and
polytropic constant $K=1/(12\eta_{0}^{2})$.  For a homogeneous system,
we have the relation $M=4\pi\eta_{0}v_{0}$. Then, combining the
preceding relations, we find that the velocity of sound is
$c_{s}=v_{0}$. Therefore, the system is formally nonlinearly dynamically stable if
\begin{eqnarray}
v_{0}^{2}\le {kM\over 4\pi}, \label{wb1}
\end{eqnarray}
and linearly dynamically unstable otherwise.  Noting that the kinetic
temperature is $T=v_{0}^{2}/3$, we check that the above result returns
(\ref{peg7b}) with $\gamma=3$. Thus, $\eta_{crit}=3$ and
$\epsilon_{crit}=1/3$. Once again, these results have been obtained with
almost no calculation.  This is an advantage of formula (\ref{svv2})
with respect to formula (\ref{svv3}).

On the other hand, using $f'(v)=\eta_{0}\lbrack
\delta(v+v_{0})-\delta(v-v_{0})\rbrack$,  the
dielectric function (\ref{ik1}) can be written
\begin{eqnarray}
\epsilon(1,\omega)=1-{T_{1/2}\over T}W_{1/2}(\omega/\sqrt{T}),
\label{mer8}
\end{eqnarray}
with
\begin{eqnarray}
W_{1/2}(z)={1\over 1-{1\over 3}z^{2}}.
\label{mer9}
\end{eqnarray}
We look for solutions of the dispersion relation $\epsilon(1,\omega)=0$ in the form $\omega=\Omega$ where $\Omega$ is real. This solution only exists for $T>T_{1/2}$ and corresponds to an oscillatory solution $\delta f\sim e^{i\Omega t}$. The pulsation is given by
\begin{eqnarray}
\Omega=\pm\sqrt{3}(T-T_{1/2})^{1/2}=\pm\sqrt{v_{0}^{2}-{kM\over 4\pi}}.
\label{mer10}
\end{eqnarray}
We now consider the case $\omega=i\lambda$ where $\lambda$ is real.  This solution only exists for $T<T_{1/2}$ and 
\begin{eqnarray}
\lambda=\pm\sqrt{3}(T_{1/2}-T)^{1/2}=\pm\sqrt{{kM\over 4\pi}-v_{0}^{2}}.
\label{mer11}
\end{eqnarray}
The case $\lambda>0$ corresponds to a growing (unstable) mode $\delta
f\sim e^{\lambda t}$ and the case $\lambda=-\gamma<0$ corresponds to a
damped mode $\delta f\sim e^{-\gamma t}$.  We note that for this special case
$n=1/2$, the growth rate and the pulsation period of the stellar
system are the same as for the corresponding barotropic gas, see
Eq. (\ref{ejl9}). The results (\ref{mer10}) and (\ref{mer11}) have been
previously  derived by Choi \& Choi (2003). They are recalled here for sake of
completeness and because we will need them in Sec. \ref{sec_lb}.

\section{Collisional relaxation of stellar systems}
\label{sec_collisional}

The Vlasov equation (\ref{h1}) can be obtained from the BBGKY
hierarchy, issued from the Liouville equation (\ref{liouville}), by
using the mean-field approximation (\ref{mf3}) which is valid in the
limit $N\rightarrow +\infty$ with $\eta$ and $\epsilon$ fixed. We
would like now to take into account the effect of correlations between
particles in order to describe the ``collisional'' relaxation. We
shall develop a kinetic theory which takes into account terms of order
$1/N$ in the correlation function.

\subsection{The evolution of the whole system: the Landau equation}
\label{sec_ev}

There are different methods to obtain a kinetic equation for the
distribution function $f(\theta,v,t)$. One possibility is to start
from the N-body Liouville equation and use projection operator
technics. This method has been followed by Kandrup (1981) for stellar
systems and by Chavanis (2001) for the point vortex gas. We shall
first consider an application of this theory to the HMF model
(Chavanis 2003b). In the large $N$ limit and  neglecting collective
effects, the projection operator formalism leads to a kinetic equation
of the form
\begin{eqnarray}
{\partial f\over\partial t}+v{\partial f\over\partial\theta}+\langle F\rangle {\partial f\over\partial v}={\partial\over\partial v}\int_{0}^{t}d\tau\int d\theta_{1}dv_{1} {\cal F}(1\rightarrow 0,t)\nonumber\\
\times \biggl\lbrace 
 {\cal F}(1\rightarrow 0, t-\tau){\partial\over\partial v}+{\cal F}(0\rightarrow 1, t-\tau){\partial\over\partial v_{1}}\biggr\rbrace\nonumber\\
\times  f(\theta_{1},v_{1},t-\tau)f(\theta,v,t-\tau).\qquad
 \label{ev1}
\end{eqnarray}
Here, $f(\theta,v,t)=NP_{1}(\theta,v,t)$ is the distribution function,
$\langle F\rangle(\theta,t)$ is the (smooth) mean-field force and
${\cal F}(1\rightarrow 0,t)=F(1\rightarrow 0,t)-\langle
F\rangle(\theta,t)$ is the fluctuating force created by particle $1$
(located at $\theta_{1},v_{1}$) on particle $0$ (located at
$\theta,v$) at time $t$. Between $t$ and $t-\tau$, the particles are
assumed to follow the trajectories determined by the slowly evolving
mean-field $\langle F\rangle(\theta,t)$.  Equation (\ref{ev1}) is a
non-Markovian integrodifferential equation. We insist on the fact that
this equation is valid for an inhomogeneous system while the kinetic
equations presented below will only apply to homogeneous
systems. Unfortunately, Eq. (\ref{ev1}) remains too complicated for
practical purposes and we will have to make simplifications. If we
consider a spatialy homogeneous system for which the distribution
function $f=f(v,t)$ depends only on the velocity, and if we implement a Markovian approximation, the foregoing equation reduces to
\begin{eqnarray}
{\partial f\over\partial t}={\partial\over\partial v}\int_{0}^{+\infty}d\tau\int d\theta_{1}dv_{1} F(1\rightarrow 0,t)\nonumber\\
\times F(1\rightarrow 0, t-\tau)\biggl ({\partial\over\partial v}-{\partial\over\partial v_{1}}\biggr )f(v_{1},t)f(v,t),
 \label{ev1bis}
\end{eqnarray}
where $F(1\rightarrow 0, t)=-{k\over 2\pi}\sin(\theta(t)-\theta_{1}(t))$. We thus need to calculate the memory function
\begin{eqnarray}
M=\int_{0}^{+\infty}d\tau\int d\theta_{1}F(1\rightarrow 0,t)F(1\rightarrow 0,t-\tau)\qquad\qquad\nonumber\\
={k^{2}\over 4\pi^{2}}\int_{0}^{+\infty}d\tau \int d\theta_{1} \sin(\theta-\theta_{1})\sin\lbrack \theta(t-\tau)-\theta_{1}(t-\tau)\rbrack,\nonumber\\
\label{ev3}
\end{eqnarray}
where $\theta_{i}(t-\tau)$ is the position at time $t-\tau$ of the $i$-th particle located at $\theta_{i}=\theta_{i}(t)$ at time $t$. Since the system is homogeneous, the mean force acting on a particle vanishes and the average equations of motion are $\theta(t-\tau)=\theta-v\tau$
and $\theta_1(t-\tau)=\theta_1-v_1\tau$. Thus,  we get
\begin{eqnarray}
M={k^{2}\over 4\pi^{2}}\int_{0}^{+\infty}d\tau \int_{0}^{2\pi} d\phi \sin\phi 
\sin(\phi-u\tau), \label{ev4}
\end{eqnarray}
where $\phi=\theta-\theta_{1}$ and $u=v-v_{1}$. The integration yields
\begin{eqnarray}
M={k^{2}\over 4\pi}\int_{0}^{+\infty}d\tau \cos(u\tau)={k^{2}\over 4}\delta(u).\label{ev5}
\end{eqnarray}
Therefore, the kinetic equation (\ref{ev1bis}) becomes
\begin{eqnarray}
{\partial f\over\partial t}={k^{2}\over 4}{\partial\over\partial v}\int dv_{1}\delta(v-v_{1})\biggl (f_{1}{\partial f\over\partial v}-f{\partial f_{1}\over\partial v_{1}}\biggr )=0. \qquad\label{ev6}
\end{eqnarray}
This equation can be considered as the counterpart of the Landau
equation describing the ``collisional'' evolution of stellar systems
such as globular clusters (in that case, the system is not homogeneous
but the collision term is often calculated by making a local
approximation). The Landau collision term can also be obtained from
the BBGKY hierarchy at the order $O(1/N)$ by neglecting the cumulant
of the three-body distribution function of order $1/N^{2}$ (see
Chavanis 2004b).  For the HMF model, and for one dimensional systems
in general, we find that the Landau collision term vanishes. This is
because the diffusion term (first term in the r.h.s.) is equally
balanced by the friction term (second term in the r.h.s.). A similar
cancellation of the collision term at order $1/N$ is found in the case
of 2D point vortices when the profile of angular velocity is
monotonic (Dubin \& O'Neil 1988, Chavanis 2001, Dubin
2003). Therefore, after a phase of violent relaxation, the system can
remain frozen in a stationary solution of the Vlasov equation for a
very long time, larger than $Nt_{D}$. Only non-trivial three-body
correlations can induce further evolution of the system. However,
their effect is difficult to estimate. Note that the collision term of
order $1/N$ may not cancel out in the case of inhomogeneous systems
and for the multi-species HMF model (see Sec. \ref{sec_multi}).

\subsection{The evolution of a test particle in a thermal bath: the Fokker-Planck equation}
\label{sec_tb}

Equations (\ref{ev1}) and (\ref{ev1bis}) can also be used to describe
the evolution of the distribution function $P(v,t)$ of a test particle
evolving in a bath of field particles with static distribution function
$f_1(v_{1})$. In that case, we have to consider that the distribution
function of the bath is {\it given}, i.e. $f_1=f_0(v_1)$. The
evolution of $P(v,t)$ is then governed by the equation
\begin{eqnarray}
{\partial P\over\partial t}={\partial\over\partial v}\int_{0}^{+\infty}d\tau\int d\theta_{1}dv_{1} F(1\rightarrow 0,t)\nonumber\\
\times F(1\rightarrow 0, t-\tau)\biggl ({\partial\over\partial v}-{\partial\over\partial v_{1}}\biggr )f_{0}(v_{1})P(v,t).
 \label{nm1}
\end{eqnarray}
Equation (\ref{nm1}) can be written in the form of a
Fokker-Planck equation
\begin{eqnarray}
{\partial P\over\partial t}={\partial \over\partial v}\biggl (D{\partial P\over\partial v}+P\eta\biggr ). \label{tb2}
\end{eqnarray}
The two terms of this equation correspond to a diffusion and a
friction.  The diffusion coefficient is given by the Kubo formula
\begin{eqnarray}
D=\int_{0}^{+\infty} d\tau\langle F(t)F(t-\tau)\rangle d\tau.
\label{tb5}
\end{eqnarray}
The friction can be understood physically by developing a linear
response theory. It arises from  the response of the field particles to
the perturbation induced by the test particle, as in a polarization
process (see Kandrup 1983).

Introducing the memory function (\ref{ev5}), the Fokker-Planck equation (\ref{nm1}) can be rewritten 
\begin{eqnarray}
{\partial P\over\partial t}={k^{2}\over 4}{\partial\over\partial v}\int dv_{1}\delta(v-v_{1})\biggl ({\partial \over\partial v}-{\partial \over\partial v_{1}}\biggr )f_{0}(v_{1})P(v,t).\nonumber\\
 \label{tb1}
\end{eqnarray}
It can be put in the form (\ref{tb2}) where the coefficients of
diffusion and friction are explicitly given by (Chavanis 2003b): 
\begin{eqnarray}
D={k^{2}\over 4}\int dv_{1}\delta(v-v_{1})f_{0}(v_{1})={k^{2}\over 4}f_{0}(v),\label{tb3}
\end{eqnarray}
\begin{eqnarray}
\eta=-{k^{2}\over 4}\int dv_{1}\delta(v-v_{1}){df_{0}\over dv}(v_{1})=-{k^{2}\over 4}f'_{0}(v). \label{tb4}
\end{eqnarray}
More precisely, comparing Eq. (\ref{tb2})
with the general Fokker-Planck equation
\begin{eqnarray}
{\partial P\over\partial t}={1\over 2}{\partial^{2}\over\partial v^{2}}\biggl (P{\langle (\Delta v)^{2}\rangle\over \Delta t}\biggr )-{\partial\over\partial v}\biggl (P{\langle \Delta v\rangle\over \Delta t}\biggr ). \label{tb6}
\end{eqnarray}
we find that
\begin{eqnarray}
{\langle (\Delta v)^{2}\rangle\over \Delta t}=2D, \qquad {\langle \Delta v\rangle\over \Delta t}={dD\over dv}-\eta.
\label{tb7}
\end{eqnarray}
Using Eqs. (\ref{tb3}) and (\ref{tb4}), we obtain
\begin{eqnarray}
\eta=-{1\over 2}  {\langle \Delta v\rangle\over \Delta t}.
\label{tb8}
\end{eqnarray}
Therefore, $\eta$ represents half the friction force. This is the same
result as in the case of Coulombian or Newtonian interactions (see,
e.g., Chavanis 2004a). Using Eqs. (\ref{tb3}) and (\ref{tb4}), the Fokker-Planck equation (\ref{tb2}) can be rewritten
\begin{eqnarray}
{\partial P\over\partial t}={\partial\over\partial v}\biggl\lbrack D\biggl ({\partial P\over\partial v}-P{d\ln f_{0}\over dv}\biggr )\biggr\rbrack, 
\label{tb9}
\end{eqnarray}
with the initial condition $P(v,t=0)=\delta (v-v_{0})$.  It describes
the evolution of a test particle in a potential $U(v)=-\ln f_0(v)$
created by the field particles. A similar equation is found for 2D
point vortices in position space where the friction is replaced by a drift
(Chavanis 2001).  When $f_0(v)$ is the Maxwellian (\ref{mb2}),
corresponding to a statistical equilibrium state (thermal bath
approximation), Eq. (\ref{tb9}) takes the form of the Kramers equation
\begin{eqnarray}
\label{tb10}{\partial P\over\partial t}={\partial\over\partial v}\biggl\lbrack 
D(v)\biggl ({\partial P\over\partial v}+\beta P v\biggr )\biggr\rbrack,
\end{eqnarray}
as in the theory of Brownian motion (Risken 1989). We note in
particular that the friction coefficient is given by the Einstein
formula $\xi=D\beta$. However, in the present context, the diffusion
coefficient depends on the velocity and, in the ballistic approach
that we have considered, is given by
\begin{eqnarray}
D(v)={\rho k^{2}\over 4}\biggl ({\beta\over 2\pi}\biggr )^{1/2}e^{-\beta {v^{2}\over 2}}.\label{tb11}
\end{eqnarray} 
We note that $P(v,t)$ always converges to the distribution function of the bath $NP(v,t)\rightarrow f_{0}(v)$ for $t\rightarrow +\infty$ while for $3D$ self-gravitating system, this is the case only when $f_{0}$ is the statistical equilibrium distribution.

Finally, again neglecting collective effects, a simple
calculation shows that the temporal correlations of the force are
\begin{eqnarray}
\langle F(0)F(t)\rangle ={k^{2}\over 4\pi}\int_{-\infty}^{+\infty}\cos\lbrack (v-v_{1})t\rbrack f_{0}(v_{1})dv_{1}\nonumber\\
={k^{2}\over 2}\cos(vt)\hat{f}_{0}(t)=2\cos(vt)\hat{D}(t),
\label{tb11ne}
\end{eqnarray}
where $\hat{D}(t)$ is the Fourier transform of $D(v)$. For the Maxwellian distribution, we get 
\begin{eqnarray}
\langle F(0)F(t)\rangle = {\rho k^{2}\over 4\pi}\cos(vt)e^{-{t^{2}\over 2\beta}},
\label{tb11b}
\end{eqnarray}
which seems to indicate a gaussian decay of the correlations.

\subsection{Collective effects: the Lenard-Balescu equation}
\label{sec_lb}

\begin{figure}
\centering
\includegraphics[width=8cm]{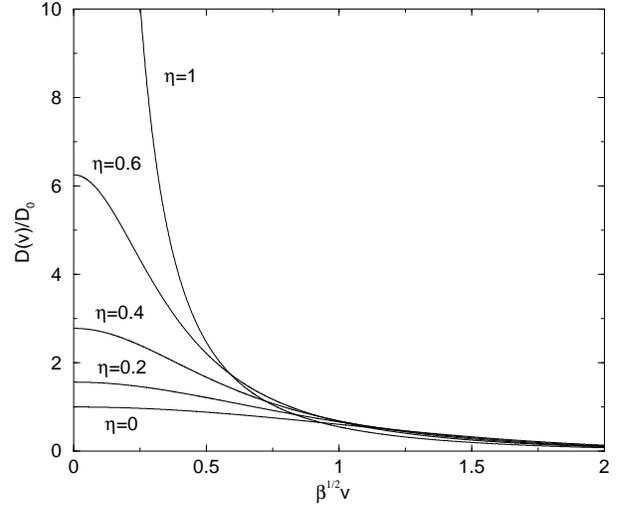}
\caption{Dependence of the diffusion coefficient $D(v)$ on the velocity $v$ for different values of the inverse temperature $\eta=\beta/\beta_{c}$. The normalization constant is $D_{0}={nk^{2}\over 4}({\beta/ 2\pi})^{1/2}$ corresponding to $D(0)$ for $\eta=0$. We note that $D(0)$ increases as $\eta$ increases and that it diverges  at the critical point $\eta=1$.}
\label{difw}
\end{figure}

The kinetic theory developed previously (Chavanis 2003b), while useful
as a first step, is however inaccurate because it is based on a
ballistic approximation and ignores collective effects which are
non-negligible for the HMF model close to the critical temperature
(Bouchet 2004). In the case of 3D stellar systems, collective effects
have only a weak influence on the kinetic theory and they are often
neglected. This implicitly assumes that the size of the system
 is smaller than the Jeans length (recall that the
Jeans length plays the role of the critical temperature in the present
context). In general, collective effects can be taken into account by
developing a kinetic theory as in the case of plasmas (Ichimaru
1973). Noting that the HMF model is the one Fourier mode of a one
dimensional plasma, Inagaki (1996) proposed to describe the
collisional evolution of the system by the corresponding form of the
Lenard-Balescu equation. It can be written
\begin{eqnarray}
{\partial f\over\partial t}={k^{2}\over 4}{\partial\over\partial v}\int dv_{1}{\delta(v-v_{1})\over |\epsilon(1,v)|^{2}}\biggl (f_{1}{\partial f\over\partial v}-f{\partial f_{1}\over\partial v_{1}}\biggr )=0,\qquad \label{lb1}
\end{eqnarray}
where $\epsilon(1,v)$ is the dielectric function (\ref{ik1}). We note
that the collision term again cancels out. However, if we use this
equation to describe the evolution of a test particle in a thermal
bath, as we did in Sec. \ref{sec_tb} by replacing $f_{1}$ by
$f_{0}(v_{1})$, we obtain Eq. (\ref{tb9}) where the expression of the diffusion coefficient is now given by
\begin{eqnarray}
D={k^{2}\over 4}{f_{0}(v)\over |\epsilon(1,v)|^{2}}.
\label{lb1b}
\end{eqnarray}
It differs from the preceding expression (\ref{tb3}) due to the
occurence of the term $|\epsilon(1,v)|^{2}$ in the denominator which
takes into account collective effects.  For the Maxwellian
distribution, the dielectric function is given by Eqs. (\ref{we8}) and
(\ref{lb3}). This leads to the expression of the diffusion coefficient
\begin{eqnarray}
D(v)={{\rho k^{2}\over 4}({\beta\over 2\pi})^{1/2}e^{-\beta {v^{2}\over 2}}\over \lbrack 1-\eta A(\sqrt{\beta}v))\rbrack^{2}+{\pi\over 2}\eta^{2}\beta v^{2}e^{-\beta v^{2}}}
\label{lb3new}
\end{eqnarray}
with $A(x)=1-xe^{-{x^{2}\over 2}}\int_{0}^{x}e^{u^{2}\over 2}du$. We note that $A(x)=1-x^{2}+...$ for $x\rightarrow 0$ and $A(x)\sim -1/x^{2}$ for $x\rightarrow +\infty$. Therefore, the diffusion coefficient behaves as Eq. (\ref{tb11}) for $v\rightarrow +\infty$ and tends to ${\rho k^{2}\over 4}({\beta/ 2\pi})^{1/2}/(1-\eta)^{2}$ for $v\rightarrow 0$ and $\eta<1$. At the critical temperature $\eta=1$ it diverges as $D(v)\sim {\rho k^{2}\over 2\pi}({\beta/ 2\pi})^{1/2}/(\beta v^{2})$ for $v\rightarrow 0$. Its behaviour is represented in Fig. \ref{difw}.

\begin{figure}
\centering
\includegraphics[width=8cm]{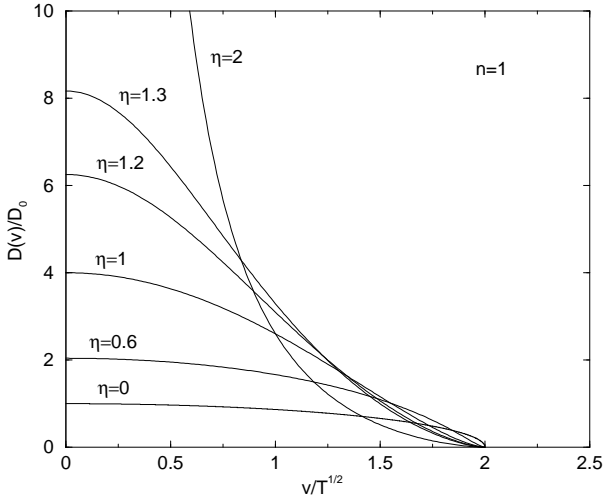}
\caption{Same as Fig. \ref{difw} for a polytropic distribution with index $n=1$. Here $\eta=kM/4\pi T$ where $T$ is the kinetic temperature. In this case, the critical point is $\eta_{1}=2$. The diffusion coefficient vanishes at the maximum velocity $v_{max}=\sqrt{4T}$.}
\label{diffTsa}
\end{figure}

\begin{figure}
\centering
\includegraphics[width=8cm]{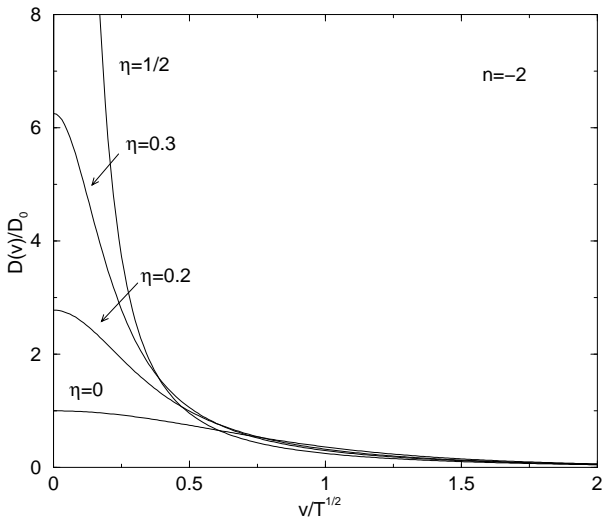}
\caption{Same as Fig. \ref{difw} for a polytropic distribution with index $n=-2$. Here $\eta=kM/4\pi T$ where $T$ is the kinetic temperature. In this case, the critical point is $\eta_{-2}=1/2$. This figure is very similar to  Fig. \ref{difw} except that the diffusion coefficient decreases algebraically. }
\label{diph}
\end{figure}

The expression (\ref{lb3new}) of the diffusion coefficient was
obtained by Bouchet (2004) in a different manner, by analyzing the
stochastic process of equilibrium fluctuations. This approach was then
generalized to an arbitrary distribution function by Bouchet \&
Dauxois (2004), leading to equation (\ref{lb1b}). In fact, formulae
(\ref{lb1b}) and (\ref{lb3new}) correspond to the one dimensional
version of the diffusion coefficient computed in plasma physics
(Ichimaru 1973) when the potential of interaction is truncated to one
Fourier mode. The general expression of the diffusion coefficient and of
the Fokker-Planck equation is given  in Chavanis (2004b) for
an arbitrary potential of interaction in $D$ dimensions.

We emphasize that the previous results are valid for an
arbitrary distribution function $f_{0}(v)$ of the bath provided that
it is stable with respect to the Vlasov equation. This is because, as
we shall see, the relaxation time of the ``field particles'' (bath)
towards statistical equilibrium (Maxwellian) is longer than the
relaxation time of a ``test particle'' towards $f_{0}$, so that the
distribution function $f_{0}$ of the bath can be considered as
``frozen''. The general expression of the diffusion coefficient can be
written
\begin{eqnarray}
\label{ven1} D(v)={{k^{2}\over 4}f_{0}(v)\over \lbrack 1+{k\over 2}{\cal P}\int_{-\infty}^{+\infty} {f_{0}'(w)\over w-v}dw\rbrack^{2}+{k^{2}\pi^{2}\over 4}f_{0}'(v)^{2}},
\end{eqnarray}  
where ${\cal P}$ stands for the principal value.  Its asymptotic
behaviour for $v\rightarrow +\infty$ is always given by
Eq. (\ref{tb3}). As a complementary example to Fig. \ref{difw}, we
have plotted in Figs. \ref{diffTsa} and \ref{diph} the diffusion
coefficient corresponding to a polytropic distribution of index $n=1$
and $n=-2$ (in that case $f_{0}$ is given by Eq. (\ref{we19d}) where
$\rho$ and $T$ are uniform). Another example is provided by the
water-bag model for which an explicit expression of $D(v)$ can be
given. Using Eqs. (\ref{mer8}), (\ref{mer9}) and (\ref{mer10}), it can
be written conveniently as
\begin{eqnarray}
\label{ven1conv} D(v)={k^{2}\over 4}\eta_{0}\biggl\lbrack {v_{0}^{2}-v^{2}\over\Omega^{2}-v^{2}}\biggr\rbrack^{2},
\end{eqnarray} 
for $v \le v_{0}$ and $D=0$ otherwise. The diffusion coefficient
diverges when the velocity of the particle $v$ is in phase with the
frequency $\Omega<v_{0}$ of the wave arising from the slightly
perturbed distribution of field particles (see
Sec. \ref{sec_wb}). This divergence occurs because, for the water bag
distribution, there exists purely oscillatory modes for a wide range
of temperatures. In general, the diffusion coefficient (\ref{lb1b})
diverges only at the critical point for $v=\omega_{r}$ with
$f'(\omega_{r})=0$; this precisely correspond to the criterion of
marginal stability (\ref{ik5}). For example, for the gaussian
distribution, we recover the divergence at $T=T_{c}$ for
$v=\omega_{r}=0$.

\subsection{The auto-correlation function}
\label{sec_auto}

We note that for high temperatures (i.e., $\eta\rightarrow 0$),
Eq. (\ref{lb3new}) reduces to the expression (\ref{tb11}) found in the
ballistic approach developed in Sec. \ref{sec_tb}. This is because, in
that limit, collective effects are weak with respect to the pure
ballistic motion and $\epsilon\simeq 1$ according to
Eq. (\ref{we8}). However, the behaviour of the correlation function is
different. Indeed, a direct analysis (Chavanis 2004b) shows that the
temporal auto-correlation function of the force is given by
\begin{eqnarray}
\langle F(0)F(t)\rangle ={k^{2}\over 4\pi}\int_{-\infty}^{+\infty}{\cos\lbrack (v-v_{1})t\rbrack\over |\epsilon(1,v_{1})|^{2}} f_{0}(v_{1})dv_{1}\nonumber\\
=2\cos(vt)\hat{D}(t).
\label{tb11neb}
\end{eqnarray} 
Substituting this expression in the Kubo formula (\ref{tb5}) and using
$\delta(v-v_{1})={1\over\pi}\int_{0}^{+\infty}\cos\lbrack
(v-v_{1})t\rbrack dt$, we recover Eq. (\ref{lb1b}). In the ballistic
approximation, the correlation function (\ref{tb11b}) is gaussian
while the exact treatment taking into account collective effects shows
that the decay of the fluctuations is in fact exponential with a decay
exponent $\gamma(\beta)=(2/\beta)^{1/2}F^{-1}(\eta)$ where
$F(x)=1/(1+\sqrt{\pi}x e^{x^{2}}{\rm erfc}(-x))$ is the $F$-function
(\ref{ge1}). This result was obtained by Bouchet (2004) by working out
the integro-differential equation satisfied by the auto-correlation
function. We shall present here an alternative derivation (Chavanis
2004b) which will make a clear link with the decay exponent appearing
in the linear stability analysis of the Vlasov equation in
Sec. \ref{sec_we}. Noting that the correlation function is
proportional to the Fourier transform of the diffusion coefficient,
according to Eq. (\ref{tb11neb}), we can obtain the expression of
$\gamma$ by determining the pole of $D(v)$ in
Eq. (\ref{lb1b}). Setting $v=i\lambda$ where $\lambda$ is real, we
find after some calculations that
\begin{eqnarray}
|\epsilon|^{2}(1,i\lambda)=\epsilon(1,i\lambda)\epsilon(1,-i\lambda)
\label{tb11neb2}
\end{eqnarray} 
where we recall that $\epsilon(1,i\lambda)$ is given by Eq. (\ref{tb11neb3}).
Clearly, $|\epsilon|^{2}(1,i\lambda)$ is an even function of
$\lambda$. We need to determine the values of $\lambda$ for which this
function vanishes. Since $\eta<1$, we find that $\lambda=\pm
\gamma$ where $\gamma>0$ is determined by
$\epsilon(1,-i\gamma)=0$. Therefore, $\gamma$ is the damping rate of
the stable perturbed solutions of the Vlasov equation; it is given by
Eq. (\ref{gam}). Next, we consider
$\lambda=\pm\gamma+\epsilon$. Expanding Eq. (\ref{tb11neb2}) for
$\epsilon\ll 1$, we find after elementary
calculations that
\begin{equation}
\label{lq1} D(v)\sim {K(\gamma)\over v^{2}+\gamma^{2}}, \quad K(\gamma)={2T\over\sqrt{\pi}}{1\over |F' ({\gamma\over \sqrt{2T}})|}
\end{equation}
for $v\rightarrow \pm i\gamma$. Therefore, for $t\rightarrow +\infty$,
the correlation function (\ref{tb11neb}) is the Fourier transform of a
Lorentzian so it decays like
\begin{eqnarray}
\label{lq2} \langle F(0)F(t)\rangle\sim {k^{2}M\over 8\pi^{2}}{\sqrt{2T}\over\gamma}{1\over |F'({\gamma\over\sqrt{2T}})|}\cos(vt)e^{-\gamma t},
\end{eqnarray} 
with
\begin{eqnarray}
\label{lq3}\gamma=({2/ \beta})^{1/2} F^{-1}(\eta).
\end{eqnarray}  
The exponential decay of the correlation function corresponds to the
Markovian limit of the stochastic process, thereby justifying the
Markovian approximation in the kinetic theory. This is quite different
from the correlations of the gravitational force which decay as
$t^{-1}$ (Chandrasekhar 1944). This slow decay may throw doubts on the
validity of the ordinary Landau equation, based on a Markovian
approximation, used to describe stellar systems (see Kandrup 1981 for
a detailed discussion). For the HMF model, the decay exponent
$\gamma(\beta)$ depends on the temperature. It diverges like
$\gamma\sim\sqrt{2T\ln T}$ as $T\rightarrow +\infty$. This is
why the correlation function is gaussian in the treatement neglecting
collective effects (see Sec. \ref{sec_tb}). On the other hand,
$\gamma\sim (8/kM)^{1/2}(T-T_{c})$ as $T\rightarrow
T_{c}^{+}$, so that the correlation function decreases very slowly
close to the critical temperature. This may invalidate the Markovian
approximation close to the critical point. The Fokker-Planck equation
(\ref{tb10}) with the diffusion coefficient (\ref{lb3new}) has been
recently investigated by Bouchet \& Dauxois (2004). In this paper,
using the rapid decay for large $v$ of the diffusion coefficient
(\ref{lb3new}), the numerically observed (Yamaguchi 2003, Pluchino et al
2004) anomalous algebraic decay of the velocity autocorrelation
function is explained and algebraic exponents are explicitely
computed. For a large class of bath distribution function $f_0$, this
may also lead to anomalous diffusion of angles $\theta$.

\subsection{The relaxation timescale}
\label{sec_relaxtime}

Let us consider the relaxation of a test particle in a thermal
bath. Due to the rapid decrease of $D(v)$ for large $v$, the spectrum
of the Fokker-Planck equation (\ref{tb10}) has no gap, and it exists
no exponential relaxation time (see Bouchet \& Dauxois 2004 for a
detailed discussion). A time scale will however describe relative
relaxation speeds for different values of the temperature
$T$. We shall obtain an estimate of this time scale. Ignoring
collective effects in a first step, this process is described by
Eqs. (\ref{tb10})-(\ref{tb11}). If the diffusion coefficient were
constant, we would deduce that the dispersion of the particles
increases as $\langle (\Delta v)^{2}\rangle=2Dt$. Introducing the
r.m.s velocity $v_{m}=\langle v^{2}\rangle^{1/2}$, we define the
relaxation timescale $t_{r}$ such that $\langle (\Delta
v)^{2}\rangle=v_{m}^{2}$. This leads to $t_{r}=v_{m}^{2}/2D$. Since
$D$ depends on $v$, the description of the diffusion process is more
complex. However, the formula
\begin{eqnarray}
t_{r}={v_{m}^{2}\over 2D(v_{m})},
\label{relaxtime1}
\end{eqnarray}
provides a useful estimate of the speed of relaxation. For the Maxwellian distribution for which
$v_{m}=1/\sqrt{\beta}$, we get
\begin{eqnarray}
t_{r}={v_{m}^{3}\over 0.121 \rho k^{2}}.
\label{relaxtime2}
\end{eqnarray}  
We can also estimate the relaxation timescale by $t_{r}'=1/\xi$, where
$\xi$ is the friction coefficient. Using the Einstein relation
$\xi=D\beta$, this yields $t_{r}'=2t_{r}$. Finally, setting
$w=v/(\sqrt{2}v_{m})$, we can rewrite Eq. (\ref{tb10}) in the
dimensionless form
\begin{eqnarray}
{\partial P\over\partial t}={1\over t_{R}}{\partial\over\partial w}\biggl\lbrack G(w)\biggl ({\partial P\over\partial w}+2Pw\biggr )\biggr\rbrack,
\label{relaxtime3}
\end{eqnarray} 
where $G(w)=e^{-w^{2}}$ and
\begin{eqnarray}
t_{R}={v_{m}^{3}\over 0.05 \rho k^{2}},
\label{relaxtime4}
\end{eqnarray}
which provides a useful time renormalization factor in the Fokker-Planck
equation. If we use expression (\ref{lb3new}) of the diffusion coefficient, we
need to multiply the relaxation timescale  by
$|\epsilon(1,v_{m})|^{2}=(1+2.25\eta)^{2}+0.578\eta^{2}$. In
particular, close to the critical temperature (i.e. $\eta=1$), the
relaxation timescale is multiplied by $\simeq 11.1$.

If we now introduce a dynamical timescale through the relation
\begin{eqnarray}
t_{D}={2\pi\over v_{m}},
\label{relaxtime5}
\end{eqnarray}
we find that $t_{R}/t_{D}\sim v_{m}^{4}/Nk^{2}\sim 1/Nk^{2}\beta^{2}\sim N/\eta^{2}\sim N$ in the thermodynamic limit of Sec. \ref{sec_mf}. Therefore, the relaxation time of a test particle in a thermal bath increases linearly with the number $N$ of field particles. More precisely, we have
\begin{eqnarray}
t_{R}={0.127\over \eta^{2}}Nt_{D}.
\label{relaxtime6}
\end{eqnarray}
The collision term in Eq. (\ref{relaxtime3}) is of order $1/N$ in the
thermodynamic limit $N\rightarrow +\infty$ with $\eta$ fixed. This
scaling also holds for the collision term in the Landau equation
(\ref{ev6}) noting that $k\sim 1/N$ and $f\sim N$ if we take
$v_{m}\sim 1$. It represents therefore the first correction to the
Vlasov limit in an expansion in $1/N$ of the correlation function.
However, contrary to the relaxation of a single particle in a thermal
bath, the relaxation time of the whole system is {\it not}
$t_{relax}\sim Nt_{D}$ because, as we have seen, the collision term
cancels out at the order $1/N$.  Therefore, the relaxation time of the
whole system is larger than $Nt_{D}$. This is consistent with the
finding of Yamaguchi et al. (2004) who numerically obtain
$t_{relax}\sim N^{1.7}t_{D}$.

It is interesting to compare this result with other systems with
long-range interactions. For stellar systems, the Chandrasekhar
relaxation time scales as $t_{relax}\sim {N\over \ln N}t_{D}$. It
corresponds to the $Nt_{D}$ scaling polluted by logarithmic
corrections. This is the relaxation time of a test star immersed in a
bath of field stars as well as the relaxation time of the whole
cluster itself (in the absence of gravothermal catastrophe). For
the point vortex gas, the collision term in the kinetic equation
cancels out at the order $1/N$ when the profile of angular velocity is
monotonic. Therefore, the relaxation time of the whole system is
larger than $Nt_{D}$. In fact, it is not clear whether the point
vortex gas ever relaxes towards statistical equilibrium (Chavanis
2002a, Dubin 2003). By contrast, the relaxation time of a test vortex
in a thermal bath of field vortices scales as 
$t_{relax}\sim {N\over\ln N}t_{D}$ (Chavanis 2002a).

\section{Self-attracting Brownian particles: the BMF model}
\label{sec_brownian}

The Hamilton equations (\ref{mf1}) describe an isolated system of
particles with long-range interactions. Since energy is conserved, the
fundamental statistical description of the system is based on the
microcanonical ensemble. It can be of interest to consider in parallel
the case where the system is stochastically forced by an external
medium. We thus introduce a system of Brownian particles with
long-range interactions which is the canonical version of the
Hamiltonian system (\ref{mf1}). This could be called the BMF (Brownian
Mean Field) model. Likewise in the case of 3D Newtonian interactions,
a system of self-gravitating Brownian particles has been recently
introduced and studied (see Chavanis et al. 2002 and subsequent
papers).

\subsection{Non-local Kramers and Smoluchowski equations}
\label{sec_nlocal}

We consider a one-dimensional system of self-attracting Brownian
particles with cosine interaction whose dynamics is governed by the
$N$-coupled stochastic equations
\begin{eqnarray}
{d\theta_{i}\over dt}=v_{i}, \qquad\qquad\qquad\qquad\qquad\qquad\qquad\quad\nonumber\\
{dv_{i}\over dt}=-{\partial\over\partial \theta_{i}}U(\theta_{1},...,\theta_{N})-\xi v_{i}+\sqrt{2D}R_{i}(t),
\label{nlocal1}
\end{eqnarray}
where $-\xi v_{i}$ is a friction force and $R_{i}(t)$ is a white noise satisfying
\begin{eqnarray}
\langle R_{i}(t)\rangle=0, \qquad \langle R_{i}(t)R_{j}(t')\rangle=\delta_{ij}\delta(t-t'),
\label{nlocal2}
\end{eqnarray}
where $i=1,...,N$ refer to the particles. The particles interact
through the potential
$U(\theta_{1},...,\theta_{N})=\sum_{i<j}u(\theta_{i}-\theta_{j})$
where $u(\theta_{i}-\theta_{j})=-{k\over
2\pi}\cos(\theta_{i}-\theta_{j})$.  We define the inverse temperature
$\beta=1/T$ through the Einstein relation $\xi=D\beta$. The stochastic
model (\ref{nlocal1})-(\ref{nlocal2}) is analogous to the model of
self-gravitating Brownian particles introduced by Chavanis et
al. (2002). For this system, the relevant ensemble is the canonical
ensemble where the temperature measures the strength of the stochastic
force. The evolution of
the $N$-body distribution function is governed by the $N$-body
Fokker-Planck equation 
\begin{eqnarray}
{\partial P_{N}\over\partial t}+\sum_{i=1}^{N}\biggl (v_{i}{\partial P_{N}\over\partial\theta_{i}}+F_{i}{\partial P_{N}\over\partial v_{i}}\biggr )=\nonumber\\\sum_{i=1}^{N}{\partial\over\partial v_{i}}\biggl (D{\partial P_{N}\over\partial v_{i}}+\xi P_{N}v_{i}\biggr ),
\label{nfp1}
\end{eqnarray}
where $F_{i}=-{\partial U\over\partial\theta_{i}}$.  The stationary
solution corresponds to the canonical distribution
\begin{eqnarray}
P_{N}={1\over Z}e^{-\beta (\sum_{i=1}^{N}{v_{i}^{2}\over 2}+U(\theta_{1},...,\theta_{N}))}.
\label{nfp2}
\end{eqnarray}
We note that the canonical distribution (\ref{nfp2}) is the {\it only}
stationary solution of the $N$-body Fokker-Planck equation while the
microcanonical distribution (\ref{mf2}) is just a {\it particular}
stationary solution of the Liouville equation (see
Sec. \ref{sec_mf}). For the system (\ref{nlocal1})-(\ref{nlocal2}),
the equilibrium canonical distribution does not rely, therefore, on a
postulate. In the thermodynamic limit $N\rightarrow +\infty$ with
$\eta=\beta kM/4\pi$ fixed, one can prove that the $N$ particules
distribution function factorizes and that the mean-field approximation
is exact (Chavanis et al. 2004b, Chavanis 2004b). The evolution of the
one particle distribution function is then governed by the non-local
Kramers equation
\begin{equation}
\label{nlocal3}
{\partial f\over\partial t}+v{\partial f\over\partial\theta}-{\partial\Phi\over\partial \theta}{\partial f\over\partial v}={\partial \over\partial v}\biggl (D{\partial f\over\partial v}+\xi fv\biggr ),
\end{equation}
which has to be solved in conjunction with Eq. (\ref{mf7}). These equations have been considered, independently, by Choi \& Choi (2003). 

To simplify the problem further, we shall consider a strong friction
limit $\xi\rightarrow +\infty$ or, equivalently, a long time limit
$t\gg\xi^{-1}$. In that case, we can neglect the inertia of the
particles and the stochastic equations (\ref{nlocal1}) reduce to
\begin{eqnarray}
{d\theta_{i}\over dt}=-\mu{\partial\over\partial \theta_{i}}U(\theta_{1},...,\theta_{N})+\sqrt{2D_{*}}R_{i}(t),
\label{nlocal4}
\end{eqnarray}
where $\mu=1/\xi$ is the mobility and $D_{*}=D/\xi^{2}=T/\xi$ is the
diffusion coefficient in physical space.  The evolution of the
$N$-body distribution function is governed by the $N$-body
Fokker-Planck equation
\begin{eqnarray}
{\partial P_{N}\over\partial t}=\sum_{i=1}^{N}{\partial\over\partial \theta_{i}}\biggl\lbrack D_{*}{\partial P_{N}\over\partial\theta_{i}}+\mu P_{N}{\partial\over\partial \theta_{i}}U(\theta_{1},...,\theta_{N})\biggr\rbrack.\qquad
\label{nfp1s}
\end{eqnarray}
The stationary solution corresponds to the canonical distribution in configuration space
\begin{eqnarray}
P_{N}={1\over Z}e^{-\beta U(\theta_{1},...,\theta_{N})}.
\label{nfp2s}
\end{eqnarray}
In the thermodynamic limit $N\rightarrow +\infty$ with $\eta=\beta
kM/4\pi$ fixed, the mean-field approximation is exact
and the evolution of the one
particle distribution function is governed by the non-local
Smoluchowski equation
\begin{eqnarray}
{\partial\rho\over\partial
t}={\partial\over\partial\theta}\biggl\lbrack {1\over\xi}\biggl
(T{\partial \rho\over\partial
\theta}+\rho{\partial\Phi\over\partial\theta}\biggr
)\biggr\rbrack, \label{nlocal6}
\end{eqnarray}
where $\Phi$ is given by Eq. (\ref{mf7}). Alternatively, the
Smoluchowski equation (\ref{nlocal6}) can be obtained from the
non-local Kramers equation (\ref{nlocal3}) by using a Chapman-Enskog
expansion in power of $1/\xi$ (Chavanis et al. 2004a). In that approximation, the distribution function is
close to the Maxwellian
\begin{equation}
\label{nlocal5}
f(\theta,v,t)={1\over \sqrt{2\pi T}}\rho(\theta,t) e^{-{v^{2}\over 2T}}+O(\xi^{-1}),
\end{equation}
and the evolution of the density is governed by the non-local
Smoluchowski equation. The equations (\ref{nlocal3}) and
(\ref{nlocal6}) conserve mass and decrease the Boltzmann free energy,
i.e. $\dot F_{B}\le 0$. This is the canonical version of the H-theorem
(Chavanis 2003c,2004a).

\subsection{Linear stability}
\label{sec_bl}

The stationary solutions of Eq. (\ref{nlocal6}) are given by
Eq. (\ref{mb3}). They extremize the Boltzmann free energy
$F_{B}=E-TS_{B}$ with (\ref{tstab2}) and (\ref{tstab3}) at fixed mass
and temperature.  Furthermore, only stationary solutions that {\it
minimize} the free energy are linearly dynamically stable with respect
to the non-local Smoluchowski equation (Chavanis 2003c). Therefore,
thermodynamical and dynamical stability are clearly connected: the
stable stationary solutions of Eq. (\ref{nlocal6}) correspond to the
canonical statistical equilibrium states in the mean-field
approximation.

Considering a small perturbation $\delta\rho$ around a stationary
solution of Eq. (\ref{nlocal6}), we get
\begin{eqnarray}
{\partial\delta \rho\over\partial t}={\partial\over\partial\theta}\biggl\lbrack {1\over\xi}\biggl (T{\partial\delta\rho \over\partial \theta}+\delta\rho{\partial\Phi\over\partial\theta}+\rho{\partial\delta\Phi\over\partial\theta}\biggr )\biggr\rbrack.
\label{bl1}
\end{eqnarray}
Writing $\delta\rho\sim e^{\lambda t}$ and introducing the notation (\ref{tstab7}), we obtain the eigenvalue equation
\begin{eqnarray}
{d\over d\theta}\biggl ({1\over\rho}{dq\over d\theta}\biggr )+{k\over 2\pi T }\int_{0}^{2\pi}q(\theta')\cos(\theta-\theta')d\theta'={\lambda\xi\over T\rho} q,\nonumber\\
\label{bl2}
\end{eqnarray}
which is similar to the eigenvalue equations obtained in the preceding
sections. These eigenvalue equations all coincide at the point of
marginal stability $\lambda=0$ implying that the onset of the
instability is the same in all the models considered. Equation
(\ref{bl2}) is the counterpart of Eq. (27) of Chavanis et al. (2002)
for self-gravitating Brownian particles.  We note that the eigenvalue
equations (\ref{ejl7new}) and (\ref{bl2}) for the Euler model and the
Brownian model only differ by the substitution $\lambda^{2}\rightarrow
\lambda\xi$. Therefore, the results of Sec. \ref{sec_isog} can be directly
extended to the present context.

For the uniform phase, the most
destabilizing mode ($n=1$) is
\begin{eqnarray}
\delta\rho=a_{1}\cos\theta \ e^{\lambda t},
\label{bl3}
\end{eqnarray}
with a growth rate
\begin{eqnarray}
\lambda={1\over\xi} (T_{c}-T), \qquad T_{c}={kM\over 4\pi}.
\label{bl4}
\end{eqnarray}
For $T<T_{c}$, the perturbation grows expontentially while it is damped exponentially for $T>T_{c}$.

Considering now the clustered phase and using a perturbative approach
similar to that of Appendix \ref{sec_eigen} for $T\rightarrow
T_{c}^{-}$ (not detailed), we find that the perturbation is damped
exponentially with a rate
\begin{eqnarray}
\lambda=-{2\over\xi} (T_{c}-T). 
\label{bl5}
\end{eqnarray}
From Eqs. (\ref{bl4}) and (\ref{bl5}), we find that, starting from a homogeneous solution at $T=T_{c}^{-}$, there is first
an exponential growth on a timescale $\xi (T_{c}-T)^{-1}$. Then,
non-linear terms come into play and the system relaxes towards a
clustered state on a timescale $(1/2)\xi (T_{c}-T)^{-1}$.

\subsection{Local Fokker-Planck equation}
\label{sec_lfp}

\begin{figure}
\centering
\includegraphics[width=8cm]{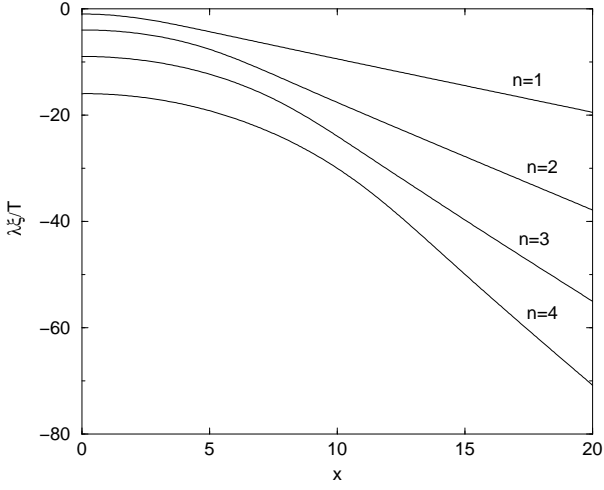}
\caption{Eigenvalues of the  Sturm-Liouville equation (\ref{lfp1a}). }
\label{lambdasturm}
\end{figure}

Before studying the non-local Smoluchowski equation
(\ref{nlocal6})-(\ref{mf7}) numerically, it may be of interest to
study a simplified problem where $\Phi$ is replaced by its equilibrium
value $\Phi=B\cos\theta$, where $B$ is fixed. Considering again the
strong friction limit, we obtain the classical (local) Smoluchowski
equation
\begin{eqnarray}
{\partial\rho\over\partial t}={\partial\over\partial\theta}\biggl\lbrack {1\over\xi}\biggl (T{\partial \rho\over\partial \theta}-B \rho\sin\theta\biggr )\biggr\rbrack.
\label{lfp1}
\end{eqnarray}
This Fokker-Planck equation also appears in the model of rotation of
dipoles in a constant electric field developed by Debye. The stationary
solutions of this equation are given by Eq. (\ref{mb3}). Considering a
perturbation $\delta\rho\sim e^{\lambda t}$ around a stationary solution
and setting $\delta\rho=dq/d\theta$, we obtain the eigenvalue equation
\begin{eqnarray}
{d\over d\theta}\biggl ({1\over\rho}{dq\over d\theta}\biggr )={\lambda\xi\over
T\rho}q.
\label{lfp1a}
\end{eqnarray}
This equation has the form of a Sturm-Liouville problem. For $B\ll 1$, so that $\rho$ is approximately uniform,
we find to leading order that the eigenvalues are
$\lambda_{n}=-n^{2}T/\xi$ where $n=1,2,...$ Using a procedure similar
to that of Appendix
\ref{sec_eigen} (not detailed), the next order correction is
$-\lambda_{n}\xi/T=n^{2}+n^{2}/\lbrack 2(4n^{2}-1)\rbrack x^{2}+...$
where $x=\beta B\ll 1$. The eigenvalues of the Sturm-Liouville
equation (\ref{lfp1a}) are evaluated numerically in
Fig. \ref{lambdasturm}.

\begin{figure}
\centering
\includegraphics[width=8cm]{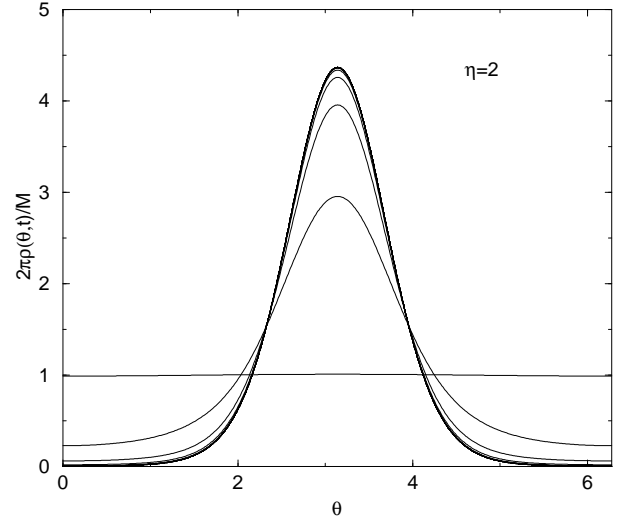}
\caption{Evolution of the density profile according to the local Fokker-Planck equation  for $\eta=2$ (corresponding to  $x=3.33$) starting from a homogeneous solution (numerical simulation). }
\label{NEWrhoteta2FPlocal}
\end{figure}

\begin{figure}
\centering
\includegraphics[width=8cm]{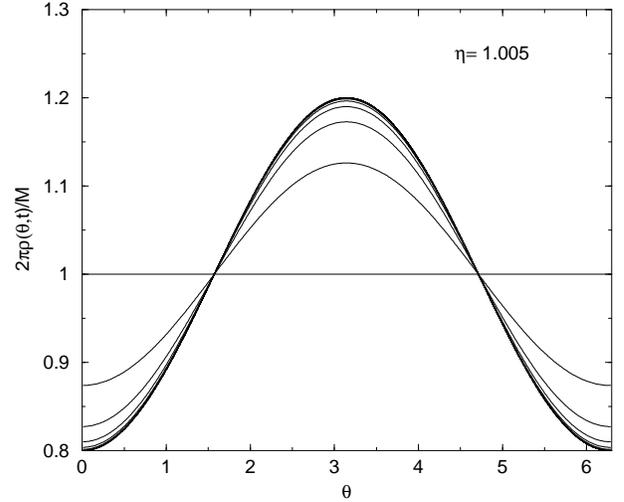}
\caption{Evolution of the density profile according to the local Fokker-Planck equation for $\eta=1.005$ (corresponding to  $x=0.2$) starting from a homogeneous solution. In this case, we have used the analytical solution (\ref{lfp2}) which is valid for $B\ll 1$.}
\label{fpanalytik}
\end{figure}

For $B\ll 1$, the Fokker-Planck equation (\ref{lfp1}) can be solved
analytically. Assuming that, initially, the density is uniform, we find
\begin{eqnarray}
\rho={M\over 2\pi}\biggl\lbrack 1+(e^{-t/\xi\beta}-1)\beta B\cos\theta\biggr\rbrack.
\label{lfp2}
\end{eqnarray}
For $t\gg t_{relax}=\xi\beta$ (relaxation time), the density reaches its equilibrium value (\ref{mb3}). For short times $t\ll 1$, we have 
\begin{eqnarray}
\rho={M\over 2\pi}\biggl (1-{1\over\xi}t B\cos\theta\biggr ).
\label{lfp2bis}
\end{eqnarray}
Starting from a homogeneous solution, there is first a linear growth
followed by an exponential relaxation towards the clustered state on
a timescale $\sim \xi/T$. This is illustrated numerically in
Fig. \ref{ant} and compared with the time evolution of the non-local
Fokker-Planck equation (\ref{nlocal6}).  The evolution is of course
quite different since, as we have seen, the non-local Fokker-Planck
equation displays an exponential growth on a timescale $\xi
(T_{c}-T)^{-1}$ followed by an exponential relaxation on a timescale
$(1/2)\xi (T_{c}-T)^{-1}$. These timescales diverge as we approach the
critical temperature $T_{c}$ while there is no critical temperature
for the local Fokker-Planck equation where $B$ is fixed. The numerical
solution of Eq. (\ref{lfp1}) is shown in Fig \ref{NEWrhoteta2FPlocal}
and the analytical solution (\ref{lfp2}) in Fig. \ref{fpanalytik} for
two different values of $\eta$.

We also note that the change of variables
\begin{eqnarray}
\rho=e^{-{1\over 2}\beta\Phi}\psi(\theta,t),
\label{lfp3}
\end{eqnarray}
transforms the Fokker-Planck equation (\ref{lfp1}) into a Schr\"odinger-like equation (with imaginary time)
\begin{eqnarray}
\xi {\partial\psi\over\partial t}=T{\partial^{2}\psi\over\partial\theta^{2}}+V(\theta)\psi,
\label{lfp4}
\end{eqnarray}
with
\begin{eqnarray}
V(\theta)={1\over 2}\biggl\lbrack \Delta\Phi-{\beta\over 2}(\nabla\Phi)^{2}\biggr \rbrack \psi.
\label{st4}
\end{eqnarray}
In our case
\begin{eqnarray}
V(\theta)=-{1\over 2}\biggl\lbrack  B\cos\theta+{\beta\over 2}B^{2}\sin^{2}\theta\biggr\rbrack.
\label{lfp5}
\end{eqnarray}
We can use this formalism to study the low temperature regime $T\rightarrow 0$. In that case, the equilibrium state is close to a Dirac peak centered on $\theta=\pi$ so that we can expand the potential to leading order in $x=\theta-\pi$. Eq. (\ref{lfp1}) becomes a Kramers-like equation 
\begin{eqnarray}
{\partial\rho\over\partial t}={\partial\over\partial x}\biggl\lbrack {1\over\xi}\biggl (T{\partial \rho\over\partial x}+ \rho B x\biggr )\biggr\rbrack.
\label{lfp1me}
\end{eqnarray}
The corresponding Schr\"odinger-like equation (\ref{lfp4}) is that of a harmonic oscillator 
\begin{eqnarray}
\xi{\partial\psi\over\partial t}=T{\partial^{2}\psi\over\partial x^{2}}+{1\over 2}(B-{1\over 2}\beta B^{2}x^{2})\psi.
\label{lfp1me2}
\end{eqnarray}
Therefore, for $T\rightarrow 0$ the eigenvalues are given by
$\lambda_{n}=-nB/\xi$, i.e. $\lambda_{n}\xi/T=-nx$ (with $n=1,2,...$)
and the corresponding eigenfunctions are the Hermite polynomials.

\begin{figure}
\centering
\includegraphics[width=8cm]{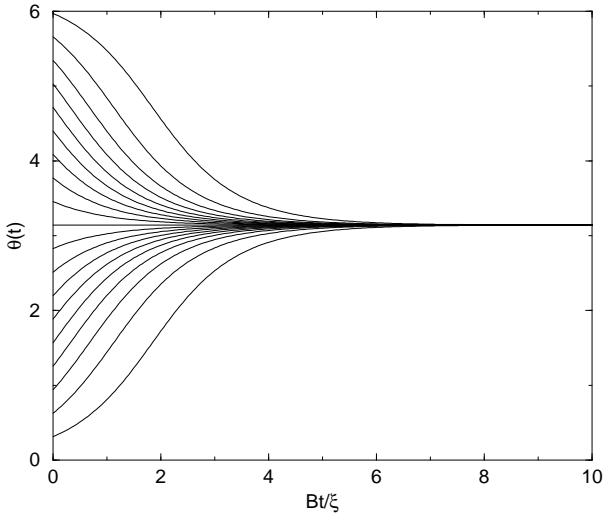}
\caption{Characteristics of Eq. (\ref{cara1}). For $t\rightarrow +\infty$, all
the particles converge at $\theta=\pi$ and a Dirac peak is formed.}
\label{carabrownian}
\end{figure}

\begin{figure}
\centering
\includegraphics[width=8cm]{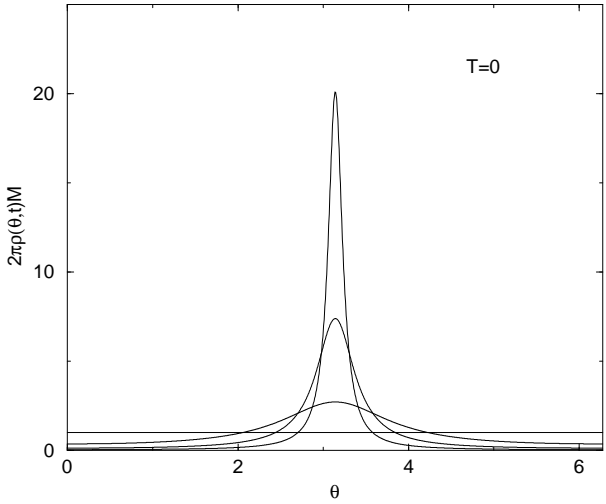}
\caption{Evolution of the density profile at $T=0$ starting from a homogeneous
distribution.}
\label{profilbrownT0}
\end{figure}

It is also possible to solve the problem exactly at $T=0$ by using the method of characteristics. Indeed, the Fokker-Planck equation becomes
\begin{eqnarray}
\xi{\partial\rho\over\partial t}+{\partial\over\partial\theta}(\rho B\sin\theta)=0,
\label{cara1}
\end{eqnarray} 
which is equivalent to an advection equation by a velocity field $v=(B/\xi)\sin\theta$. Therefore, the evolution is deterministic and the equation of motion of a particle is
\begin{eqnarray}
{d\theta\over dt}={1\over\xi}B\sin\theta.
\label{cara2}
\end{eqnarray} 
This equation of motion is readily solved and we get
\begin{eqnarray}
\theta(t)=2{\rm Arctan}\biggl \lbrack {\rm tan}\biggl ({\theta_{0}\over 2}\biggr )e^{Bt/\xi}\biggr\rbrack,
\label{cara3}
\end{eqnarray} 
where $\theta_{0}$ is the initial position of the particle. The
caracteristics are shown in Fig. \ref{carabrownian}. For $t\rightarrow
+\infty$, all the particles converge at $\theta=\pi$ and a Dirac peak
is formed. The density profile is determined by the condition
$\rho_{0}d\theta_{0}=\rho(\theta,t)d\theta$ yielding
$\rho(\theta,t)=\rho_{0}/(d\theta/d\theta_{0})$. From
Eq. (\ref{cara3}), we get
\begin{eqnarray}
\rho(\theta,t)={\rho_{0}\lbrack 1+{\rm tan}^{2}(\theta/2)\rbrack e^{-Bt/\xi}\over 1+{\rm tan}^{2}(\theta/2) e^{-2Bt/\xi}}.
\label{cara4}
\end{eqnarray} 
This profile is shown in Fig. \ref{profilbrownT0} at different times.

\begin{figure}
\centering
\includegraphics[width=8cm]{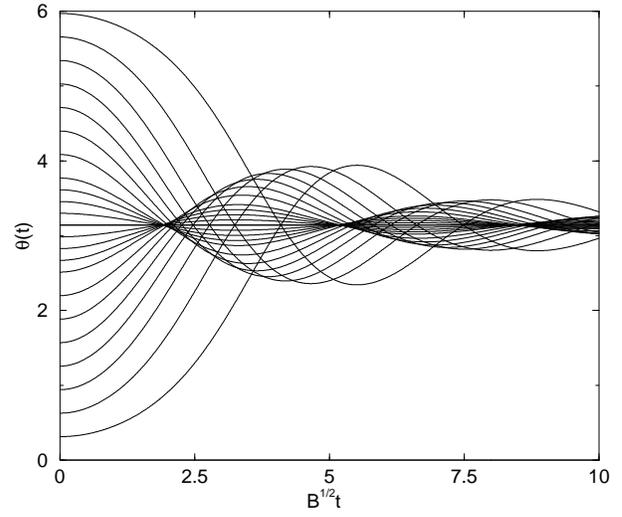}
\caption{Characteristics of the damped Euler equation (\ref{dameu}) for $p=0$ and $\Phi=B\cos\theta$. The ratio $\xi/B=1/2$. }
\label{caradamped}
\end{figure}

To make the link between Figs. \ref{carabrownian} and \ref{chevrons}, we can consider an intermediate hydrodynamical equation
\begin{eqnarray}
{\partial u\over\partial t}+u{\partial u\over\partial \theta}=-{1\over\rho}{\partial p\over\partial \theta}-{\partial \Phi\over\partial \theta}-\xi u.
\label{dameu}
\end{eqnarray}  
This could be called the damped Euler equation (see Chavanis 2003c). For
$\xi=0$ we recover the Euler equation (\ref{ej2}) and for
$\xi\rightarrow +\infty$, using the continuity equation (\ref{ej1}), we
recover the Smoluchowski equation (\ref{nlocal6}). For $p=0$ and
$\Phi=B\cos\theta$, Eq. (\ref{dameu}) can be solved by the method of
characteristics and the results are reported in
Fig. \ref{caradamped}. This clearly shows the passage from
Fig. \ref{chevrons} to Fig. \ref{carabrownian} as the friction parameter $\xi$ increases.

\subsection{Dynamics of Brownian particles in interaction}
\label{sec_dyn}

We now turn to the evolution of a system of Brownian particles in
interaction described by the $N$-coupled stochastic equations
(\ref{nlocal1}). Despite all the simplifications introduced, this
model remains a non-trivial and interesting model exhibiting a process
of self-organization. In particular, it shows the passage from a
homogeneous phase (disk-like) to a clustered phase (bar-like) under
the influence of long-range interactions.  Interestingly, the large
$N$ limit of this system is exactly described by an explicit kinetic
equation (\ref{nlocal3}) reducing to Eq.  (\ref{nlocal6}) for large
times. By contrast, the kinetic equation describing the evolution of
the HMF model (\ref{mf1}) towards statistical equilibrium is not known
as the collision term cancels out at the order $1/N$ within the
approximations usually considered (see Sec. \ref{sec_collisional}).

\begin{figure}
\centering
\includegraphics[width=8cm]{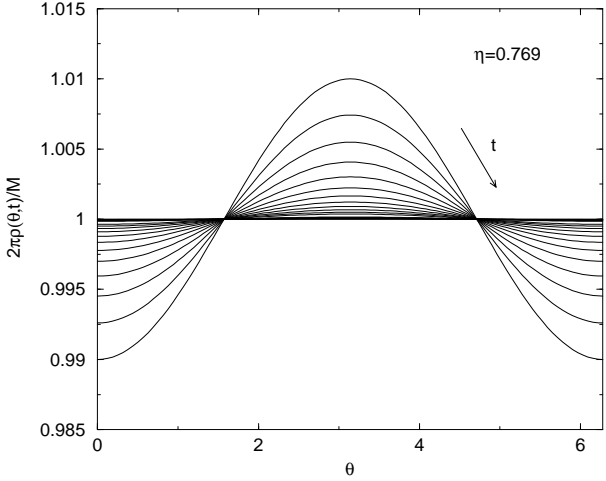}
\caption{Evolution of the density profile according to the non-local Fokker-Planck equation. For $\eta=0.769<1$, the homogeneous solution is stable. }
\label{NEWrhoteta0p769}
\end{figure}

\begin{figure}
\centering
\includegraphics[width=8cm]{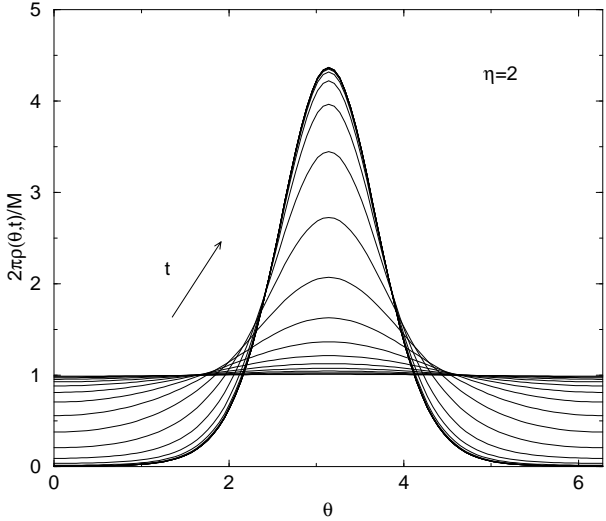}
\caption{For $\eta=2>1$, the homogeneous solution (disk-like) is unstable and the system  forms a cluster (bar-like). The evolution is longer than with the
local Fokker-Planck equation represented in
Fig. \ref{NEWrhoteta2FPlocal} as also shown in the next figure and
explained in the text.}
\label{NEWrhoteta2}
\end{figure}

\begin{figure}
\centering
\includegraphics[width=8cm]{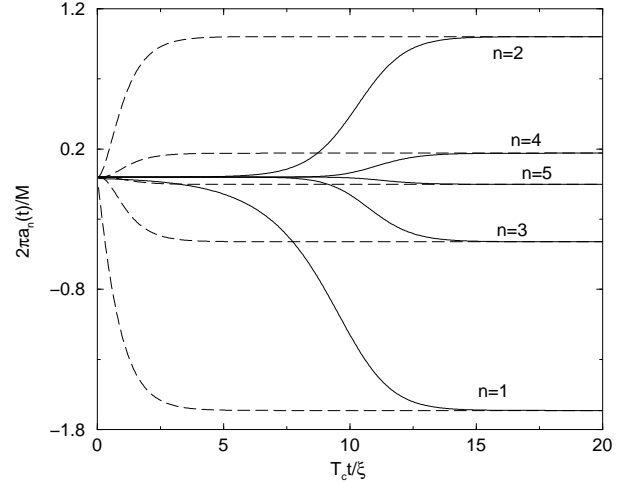}
\caption{Evolution of the different modes $a_{n}(t)$ for the non-local (full lines) and local (dashed lines) Fokker-Planck equations. The control parameters are  $\eta=2$, $x=3.33$.  }
\label{ant}
\end{figure}

For the Brownian gas, we have to solve the integro-differential
equation
\begin{eqnarray}
\xi{\partial\rho\over\partial t}=T{\partial^{2}\rho\over\partial\theta^{2}}+{k\over 2\pi}{\partial\over\partial\theta}\biggl\lbrace \rho\int_{0}^{2\pi}\sin(\theta-\theta')\rho(\theta',t)d\theta'\biggr\rbrace.\nonumber\\
\label{dyn1}
\end{eqnarray}
Substituting the decomposition 
\begin{eqnarray}
\rho=a_{0}(t)+\sum_{n=1}^{+\infty}a_{n}(t)\cos(n\theta)
\label{dyn2}
\end{eqnarray}
in Eq. (\ref{dyn1}), we find that
\begin{eqnarray}
a_{0}={M\over 2\pi},
\label{dyn3}
\end{eqnarray}
\begin{eqnarray}
\xi{da_{1}\over dt}+Ta_{1}={k\over 2}a_{1}\biggl ({M\over 2\pi}-{a_{2}\over 2}\biggr ),
\label{dyn4}
\end{eqnarray}
\begin{eqnarray}
\xi{da_{n}\over dt}+Tn^{2}a_{n}={k\over 4}n a_{1}(a_{n-1}-a_{n+1}), \qquad (n\ge 2).\qquad 
\label{dyn5}
\end{eqnarray}
We note that the coefficient $a_{1}(t)$ is related to the magnetization $B(t)$, defined in Eq. (\ref{mf9}), by the formula $B(t)=-{k\over 2}a_{1}(t)$. According to Eqs. (\ref{mb4}), (\ref{mb5}) and (\ref{mb6}), the coefficients $a_{n}$ are given at equilibrium by 
\begin{eqnarray}
a_{n}={M\over\pi}(-1)^{n}{I_{n}(\beta B)\over I_{0}(\beta B)}.
\label{anstatic}
\end{eqnarray}
Therefore, the static equations (\ref{dyn4}) and (\ref{dyn5}) with
$d/dt=0$ coincide with the recursive relations satisfied by the
Bessel functions $I_{n}(x)$.

For $T\rightarrow T_{c}^{-}$, the coefficients scale as $a_{n}\sim B^{n}$ with
$B\ll 1$. Therefore, after a transient regime of order $1/Tn^{2}$, Eq. (\ref{dyn5}) can be simplified in
\begin{eqnarray}
a_{n}={k\over 4T}{1\over n}a_{1}a_{n-1}.
\label{slave}
\end{eqnarray}
In particular, for $n=2$, we get $a_{2}={k\over 8T}a_{1}^{2}$. This
shows that the second mode is slaved to the first. Substituting this
in Eq. (\ref{dyn4}), we get
\begin{eqnarray}
\xi{da_{1}\over dt}+(T-T_{c})a_{1}=-{k^{2}\over 32T} a_{1}^{3}.
\label{dyn5a}
\end{eqnarray}  
This equation is readily solved (it may be convenient to use $p=a_{1}^{2}$ as a variable) and we obtain
\begin{eqnarray}
a_{1}^{2}(t)={A (T_{c}-T) e^{2(T_{c}-T)t/\xi}\over 1+{A k^{2}\over 32T}e^{2(T_{c}-T)t/\xi}},
\label{dyn5b}
\end{eqnarray}
where 
\begin{eqnarray}
A={a_{1}(0)^{2}\over (T_{c}-T)-{k^{2}a_{1}(0)^{2}\over 32 T}}.
\label{dyn5c}
\end{eqnarray}
For $t\rightarrow +\infty$, we get $a_{1}(\infty)^{2}={32T\over
k^{2}}(T_{c}-T)$, or equivalently $B^{2}=8T(T_{c}-T)$. We thus recover
the equilibrium result (\ref{tp4}) valid close to the critical
point. The approach to equilibrium for $t\rightarrow +\infty$ is
governed by $\delta a_{1}(t)/a_{1}(\infty)={16T\over
A^{2}k^{2}}e^{-2(T_{c}-T)t/\xi}$ yielding the damping rate
(\ref{bl5}).  On the other hand, for $t\ll t_{relax}$, one has
$a_{1}(t)=a_{1}(0)e^{(T_{c}-T)t/\xi}$ yielding the growth rate
(\ref{bl4}).  Finally, at $T=T_{c}$, Eq. (\ref{dyn5a}) leads to
\begin{eqnarray}
a_{1}(t)=\pm{1\over \sqrt{{1\over a_{1}(0)^{2}}+{k^{2}t\over 16\xi T_{c}}}}.
\label{dyn5bb}
\end{eqnarray}
so that the magnetization $B(t)=-{k\over 2}a_{1}(t)$ tends to zero
{\it algebraically} as $t^{-1/2}$ for $t\rightarrow +\infty$. Using
Eq. (\ref{slave}), the coefficients $a_{n}(t)$ are expressed in terms
of $a_{1}(t)$ by
\begin{eqnarray}
a_{n}=\biggl ({k\over 4T}\biggr )^{n-1}{1\over n!}a_{1}^{n}.
\label{dyn5q}
\end{eqnarray}
In the previous calculations, we have assumed for simplicity that the
density profile is symmetrical with respect to the $x$-axis. The
general case is treated in Appendix \ref{sec_gen}. Away from the
critical point, the non-local Smoluchowski equation has to be solved
numerically. Some numerical simulations are shown in
Figs. \ref{NEWrhoteta0p769}-\ref{ant}. Note also that for
$T\rightarrow 0$ and for sufficiently large times, the density is
peaked around $\theta=\pi$ and Eq. (\ref{dyn1}) becomes equivalent to
the local Kramers equation (\ref{lfp1me}) with $B=kM/2\pi$. Therefore,
for $T\rightarrow 0$, the eigenvalues of Eq. (\ref{bl2}) tend to
$\lambda_{n}=-nkM/2\pi\xi$ (harmonic oscillator). Substituting
$\xi\lambda\rightarrow \lambda^{2}$, we deduce that, for $T\rightarrow
0$, the eigenvalues of Eq. (\ref{ejl7new}) tend to $\lambda^{2}_{n}=-nkM/2\pi$
(they are represented in Fig. \ref{lambdaEuler}).

\section{The multi-species HMF model}
\label{sec_multi}

We finally briefly comment on the generalization of the preceding
results to the case of the multi-species HMF model. We thus return to
the Hamiltonian equations (\ref{mf1}) and account for the possibility
of having particles with different masses. Stellar systems also possess
a mass spectrum, so this generalization has a counterpart in
astrophysics.

Considering first the statistical equilibrium state, a straightforward
generalization of the counting analysis of Sec. \ref{sec_boltzmann} yields
\begin{eqnarray}
W(\lbrace n_{ia}\rbrace)=\prod_{a}N_{a}!\prod_{i}{\nu^{n_{ia}}\over n_{ia}!},
\label{multi1}
\end{eqnarray}
for the probability of the state $\lbrace n_{ia}\rbrace$, where
$n_{ia}$ gives the number of particles with mass $m_{a}$ in the $i$-th
macrocell. Therefore, the entropy $S=\ln W$ of the multi-species gas is
\begin{eqnarray}
S=-\sum_{a}\int {f_{a}\over m_{a}}\ln {f_{a}\over m_{a}} d\theta dv,
\label{multi2}
\end{eqnarray}
where $f_{a}(\theta,v)d\theta dv$ gives the total mass of particles of species $a$ in $(\theta,v)$. The distribution function of the whole assembly is
\begin{eqnarray}
f(\theta,v)=\sum_{a}f_{a}(\theta,v).
\label{multi3}
\end{eqnarray}
The statistical equilibrium state is obtained by maximizing the entropy (\ref{multi2}) while conserving the total energy $E$ and the mass $M_{a}$ of each species of particles. This yields 
\begin{eqnarray}
f_{a}=A_{a}'e^{-\beta m_{a}({v^{2}\over 2}+\Phi)},
\label{multi4}
\end{eqnarray}
which generalizes Eq. (\ref{mb2}). Note that the inverse temperature $\beta$
is the same for all species in accordance with the theorem of
equipartition of energy. This clearly leads to a mass {\it
seggregation} since the r.m.s. velocity of species $a$ decreases
with the mass: $\langle v^{2}\rangle_{a}=T/m_{a}$.  More precisely,
Eq. (\ref{multi4}) implies
\begin{eqnarray}
f_{a}(\epsilon)=C_{ab}\lbrack f_{b}(\epsilon)\rbrack^{m_{a}/m_{b}},
\label{multi5}
\end{eqnarray}
where $C_{ab}$ is a constant. Therefore, heavy particles will have the
tendency to occupy regions of low energy. Recall, by contrast, that
there is no mass seggregation in Lynden-Bell's statistical theory of
violent relaxation for collisionless systems (see
Sec. \ref{sec_vlasov}) since the mass of the particles does not appear
in the Vlasov equation (Lynden-Bell 1967). It would be of interest to
study these problems of mass seggregation with the HMF model for which
numerical simulations are simpler than with gravitational systems.

Considering now the collisional relaxation, a straightforward
generalization of the Lenard-Balescu equation to a multi-species
system yields
\begin{eqnarray}
{\partial f_{a}\over\partial t}={k^{2}\over 4}{\partial\over\partial v}\sum_{b}\int dv' {\delta(v-v')\over |\epsilon(1,v)|^{2}}\biggl (m_{b}f'_{b}{\partial 
f_{a}\over\partial v}-m_{a}f_{a}{\partial f'_{b}\over\partial v'}\biggr )\nonumber\\
\label{multi6}
\end{eqnarray}
with
\begin{eqnarray}
\epsilon(1,\omega)=1+{k\over 2}\sum_{b}\int {f'_{b}(v)\over v-\omega}dv. 
\label{epsmulti}
\end{eqnarray}
We now see that the collision term does not vanish anymore when there
are at least two different species. The diffusion and the friction
experienced by a particle of one species are caused by collisions with
particles of another species. Equation (\ref{multi6}) can be rewritten
in the suggestive form
\begin{eqnarray}
{\partial f_{a}\over\partial t}={\partial\over\partial v}\sum_{b}\biggl\lbrack D_{ab}{\partial f_{a}\over\partial v}-\overline{D}_{ab}{\partial f_{b}\over \partial v}\biggr\rbrack,
\label{multi7}
\end{eqnarray}
where 
\begin{eqnarray}
D_{ab}={k^{2}\over 4}{m_{b}f_{b}\over |\epsilon(1,v)|^{2}}, 
\label{multi8}
\end{eqnarray} 
is the diffusion coefficient for species $a$ due to collisions with species $b$ and 
\begin{eqnarray}
\overline{D}_{ab}={k^{2}\over 4}{m_{a}f_{a}\over |\epsilon(1,v)|^{2}},
\label{multi9}
\end{eqnarray} 
is an ``off-diagonal'' diffusion coefficient. It corresponds to a friction force
\begin{eqnarray}
\eta_{a}=-{\overline{D}_{ab}\over f_{a}}{\partial f_{b}\over\partial v}.
\label{multi10}
\end{eqnarray} 
These diffusion coefficients satisfy the relation
\begin{eqnarray}
{\overline{D}_{ab}\over m_{a}f_{a}}={{D}_{ab}\over m_{b}f_{b}}.
\label{multi11}
\end{eqnarray} 
These results are similar to those obtained by Dubin (2003) for the
multi-components point vortex gas in two dimensions. When the profile
of angular velocity is non-monotonic, the vorticity profile evolves
under the effect of long-range collisions caused by a process of
resonance (Dubin \& O'Neil 1988, Chavanis 2002). When the profile of
angular velocity is monotonic, there is no evolution for the
single-species system. An evolution is, however, possible for the
distribution function of each species in the multi-components system.
In a sense, at the order $1/N$, the kinetic theory of the point vortex
gas (evolution of the single-species system when the angular velocity
profile is non-monotonic) is intermediate between the kinetic theory
of stellar systems (evolution of the single-species system in any
case) and the kinetic theory of the HMF model (no evolution of the
single-species system).

Therefore, the results of Dubin (2003) can be directly transposed to the present context. In particular, we note that the total distribution function $\sum_{a}f_{a}(v,t)=f(v)$ is stationary so that the conservation of energy is trivially satisfied. On the other hand, a $H$-theorem can be proved for the entropy (\ref{multi2}), i.e. $\dot S\ge 0$. The equality $\dot S=0$ corresponds to  vanishing currents 
\begin{eqnarray}
J_{a}=-\sum_{b}\biggl\lbrack D_{ab}{\partial f_{a}\over\partial v}-\overline{D}_{ab}{\partial f_{b}\over \partial v}\biggr\rbrack=0,
\label{multi12}
\end{eqnarray} 
implying the following equilibrium relation between the densities
\begin{eqnarray}
f_{a}(v)=K_{ab}\lbrack f_{b}(v)\rbrack^{m_{a}/m_{b}},
\label{multi13}
\end{eqnarray}
where $K_{ab}$ is a constant independent on $v$. This equation is similar to 
Eq. (\ref{multi5}) but, here,  $f_{a}(v)$ is not necessarily the Maxwellian. Indeed, Eq. (\ref{multi13}) is satisfied by any distribution of the form $f_{a}(v)=A_{a}{\rm exp}\lbrack -\beta m_{a}\chi(v)\rbrack$, where $\chi(v)$ is determined by the initial conditions. 

We can use these results to study the relaxation of a test particle of mass $m$ in a bath of field particles with mass $m_{f}$. Neglecting collective effects for simplicity, we find that the equivalent of the Fokker-Planck equation (\ref{tb10}) is now
\begin{eqnarray}
{\partial P\over\partial t}={\partial\over\partial v}\biggl\lbrack 
D(v)\biggl ({\partial P\over\partial v}+\beta m  P v\biggr )\biggr\rbrack. \label{multi14}
\end{eqnarray}
with 
\begin{eqnarray}
D(v)={n k^{2}\over 4}m_{f}^{2}\biggl ({\beta m_{f}\over 2\pi}\biggr )^{1/2}e^{-\beta m_{f}{v^{2}\over 2}},\label{multi15}
\end{eqnarray}
where $n$ is the number density of field particles. The equilibrium distribution of the test particle is 
\begin{eqnarray}
P_{eq}(v)=\biggl ({\beta m\over 2\pi}\biggr )^{1/2}e^{-\beta m {v^{2}\over 2}}.
\label{multi15f}
\end{eqnarray}
The timescale of
collisional relaxation is
\begin{eqnarray}
t_{r}={v_{mf}^{3}\over 0.121 n m_{f}^{2}k^{2}},\label{multi16}
\end{eqnarray}
and $t_{r}'=2(m_{f}/m)t_{r}$. In dimensionless form, Eq. (\ref{multi14}) can be rewritten
\begin{eqnarray}
{\partial P\over\partial t}={1\over t_{R}}{\partial\over\partial w}\biggl\lbrack G(w)\biggl ({\partial P\over\partial w}+2{m\over m_{f}}Pw\biggr )\biggr\rbrack,
\label{multi17}
\end{eqnarray} 
with
\begin{eqnarray}
t_{R}={v_{m}^{3}\over 0.05 n m_{f}^{2}k^{2}}
\label{multi18}
\end{eqnarray}
and $G(w)=e^{-w^{2}}$. If we properly account for collective effects, the diffusion coefficient is given by 
\begin{eqnarray}
D(v)={{n k^{2}\over 4}m_{f}^{2}({\beta m_{f}\over 2\pi})^{1/2}e^{-\beta m_{f}{v^{2}\over 2}}\over \lbrack 1-\eta A(\sqrt{\beta m_{f}}v)\rbrack^{2}+{\pi\over 2}\eta^{2}\beta m_{f}v^{2}e^{-\beta m_{f}v^{2}}},\qquad \label{vde}
\end{eqnarray} 
where $\eta={kMm_{f}\over 4\pi T}$. On the other hand, for a distribution of the bath of the form $f_{0}(v)$, the Fokker-Planck equation is 
\begin{eqnarray}
\label{end1}{\partial P\over\partial t}={\partial\over\partial v}\biggl\lbrack 
D(v)\biggl ({\partial P\over\partial v}-{m\over m_{f}}  P {d\ln f_{0}\over dv}\biggr )\biggr\rbrack,
\end{eqnarray}
with $D(v)={k^{2}\over 4}m_{f}f_{0}(v)/|\epsilon(1,v)|^{2}$. We note
that the equilibrium distribution of the test particle is
$P_{eq}(v)=Af_{0}(v)^{m/m_{f}}$.

\section{Conclusion}
\label{sec_conclusion}

In this paper, we have given an exhaustive description of the HMF
model that recently appeared in statistical mechanics as a simple
model with long-range interactions similar to self-gravitating
systems. The originality of our approach is to offer an overview of
the subject and to see how different models (Hamiltonian, Brownian,
fluids,...) are related to each other. Other studies concentrate in
general on a specific aspect of the problem. We think that putting all
the models in parallel is illuminating because they are closely
connected to each other so that a unified (and aesthetic) description
can be given. These connections were previously noted by one of us
(P.H.C) in the case of 3D self-gravitating systems and it was natural
to extend these results to the HMF model. A more general
approach is given in Chavanis (2004b) for an arbitrary potential of
interaction in $D$ dimensions. The present paper can be seen as a
particular application of this general formalism for a one dimensional
potential truncated to one mode. The main interest of the HMF model in
this context is to yield simple explicit results.

Another originality of our approach is to emphasize the connection
between the HMF model and self-gravitating systems (and 2D vortices)
although this link is only sketched in other papers, except in the
early work of Inagaki. Many concepts and technics that are well-known
in astrophysics have been rediscovered for the HMF model, sometimes
with a different point of view. This is true in particular for the
notion of violent relaxation and metaequilibrium states. In
statistical mechanics, this has been approached via a notion of
``generalized thermodynamics'' (Tsallis 1988) although it was
understood early in astrophysics (Lynden-Bell 1967, Tremaine et
al. 1986) that these metaequilibrium states correspond to particular
stationary solutions of the Vlasov equations on a coarse-grained scale
(Chavanis et al. 1996, Chavanis 2003a). Thus, our dynamical
interpretation of Tsallis functional as a particular H-function
differs from the thermodynamical interpretation given by Boghosian
(1996), Latora et al. (2002) and Taruya \& Sakagami (2003).

Concerning the interest of the HMF model for astrophysicists, we have
shown that it exhibits the same types of behaviors as 3D
self-gravitating systems while being much simpler to study because it
is one dimensional and avoids complicated problems posed by the
divergence of the gravitational potential at short distances and the
absence of a large-scale confinement. Thus, it distinguishes what is
common to long-range interactions and what is specific to
gravity. This comparative study should bring new light in the
statistical mechanics of self-gravitating systems which has long been
a controversial subject. Other simplified models of gravity have been
introduced such as the parallel planar sheets of Camm (1950), the
concentric spherical shells of H\'enon (1971) or the toy models of
Lynden-Bell \& Lynden-Bell (1977) and Padmanabhan (1990). These toy
models have often allowed advances in the description of more
realistic self-gravitating systems that are difficult to study in full
detail. We think that, similarly, the HMF model should find its place
in the astrophysical literature. 

Another interest of the HMF model is to allow to study in great detail
what happens close to the critical point. In the HMF model, the
potential is truncated to one mode $n=1$ and there exists a critical
temperature $T_{c}=kM/4\pi$ below which the uniform phase is
unstable. For infinite homogeneous self-gravitating systems, there is
a continuous spectrum of modes but there exists a critical wavelength
$\lambda_{J}=(\pi T/Gm\rho)^{1/2}$ (depending on the temperature)
above which the system is unstable. Alternatively, if we fix the size
of the system (for example by placing it in a box of radius $R$), the
maximum wavelength is $R$ and the Jeans instability criterion now
determines a critical temperature $T_{c}\sim GmM/(\pi R)$ below which
the system becomes unstable (we have used $\lambda_{J}=R$ and
$\rho\sim M/R^{3}$). In fact, in the case of box-confined
gravitational systems, we must consider that the gaseous phase is
inhomogeneous and use a more precise stability analysis (Chavanis 2002b)
yielding $T_{c}=GmM/(2.52R)$. These remarks show that the critical
temperature in the HMF model plays exactly the same role as the
critical temperature in finite isothermal spheres. Now, in the
framework of the HMF model, it is possible to study how the mean-field
results are altered close to the critical point due to collective
effects. This is more complicated for self-gravitating systems because
they are inhomogeneous. However, on a qualitative point of view,
we expect similar results: divergence of the two-point correlation
function like in Eq. (\ref{corr2}), divergence of the force
auto-correlation function like in Eq. (\ref{corr4}), alteration of the
diffusion coefficient and increase of the relaxation time like in
Eq. (\ref{lb3new}), increase of the decorrelation time like in
Eq. (\ref{lq3})... These problems are difficult to study for
self-gravitating systems but they are of considerable importance. They
have never been discussed in detail
because it is usually implicitly assumed that the system size is much
smaller than the Jeans scale (or the temperature much larger than
$T_{c}$) so that collective effects are neglected and the dielectric
function is approximated as $\epsilon\simeq 1$. The present simplified
study is a first step to understand the failure of the mean-field
approximation close to the critical point and it can thus find
important applications in theoretical astrophysics. Note that fluctuations 
in isothermal spheres close to the critical point have been studied by 
Katz \& Okamoto (2000) and Chavanis (2005).

Finally, we have given an astrophysical application of the HMF model
in relation with the formation of bars in spiral
galaxies, following the original idea of Pichon and Lynden-Bell. This
simple model is consistent with the phenomenology of structure
formation which results from a competition between long-range
interactions (gravity) and thermal motion (velocity dispersion). It is
sometimes argued that the moment of inertia $\alpha^{-1}$ of stellar
orbits can be negative (Pichon 1994). In that case, we must consider
Eq. (\ref{mf1}) with $k<0$.  This corresponds to the repulsive
(anti-ferromagnetic) HMF model. Now, Barr\'e et al. (2002) have observed that
this model leads to the formation of a bicluster. In the stellar disk analogy, the equivalent of the  ``bicluster'' would be two bars perpendicular to each other. We do not know whether this type of structure is observed in astrophysics.

{\it Acknowledgements} One of us (P.H.C) is grateful to Donald
Lynden-Bell for inspiring discussions some years ago. He is also
grateful to Pr. D. Dubin for drawing his attention to his work on
non-neutral plasmas. We would like to thank T. Dauxois for interesting
discussions.

\appendix

\section{Estimate of the eigenvalue close to the critical point}
\label{sec_eigen}

For simplicity, we restrict ourselves to symmetrical perturbations
with respect to the $x$-axis. The general case is treated in Appendix
\ref{sec_gen} by another method.  In the canonical ensemble, we have
to study the eigenvalue equation $(V=0)$:
\begin{eqnarray}
\label{eigen1}
{d\over d\theta}\biggl ({1\over\rho(\theta)}{dq\over d\theta}\biggr )+{k\over 2\pi T}\int_{0}^{2\pi}q(\theta')\cos(\theta-\theta')d\theta'=2\lambda q,\nonumber\\
\end{eqnarray}
where $\rho(\theta)$ is the equilibrium density profile and
$q(\theta)$ is the perturbation. We consider the clustered phase close
to the critical point $T_{c}$. Thus, we can perform a systematic
expansion of the parameters in powers of $B\ll 1$ or, equivalently, in powers of $x=\beta B\ll 1$.

Using Eqs. (\ref{mb4}) and (\ref{mb5}), the density profile can be expanded as
\begin{equation}\label{eigen2}
{1\over\rho}={2\pi\over M}\biggl\lbrack 1+x\cos\theta+{x^{2}\over 4}(1+2\cos^{2}\theta)+...\biggr\rbrack.
\end{equation}
Substituting this result in Eq. (\ref{eigen1}), and using the expansion (\ref{tp2}) of the temperature, we obtain
\begin{eqnarray}\label{eigen3}
{d\over d\theta}\biggl \lbrace \pi\biggl\lbrack  1+x\cos\theta+{x^{2}\over 4}(1+2\cos^{2}\theta)\biggr\rbrack{dq\over d\theta}\biggr \rbrace\nonumber\\
+\biggl (1+{x^{2}\over 8}\biggr )\int_{0}^{2\pi}q(\theta')\cos(\theta-\theta')d\theta'=\mu x^{2} q.
\end{eqnarray}
where we have set $\lambda M=\mu x^{2}$ with $\mu=O(1)$. Here,
$\lambda$ refers to the largest eigenvalue of Eq. (\ref{eigen1}) which
is equal to zero for $x=0$ (the other eigenvalues are $\lambda_{n}
M=-\pi n^{2}+O(x)$ for $n\ge 2$). Furthermore, the following expansion
shows that the term of order $x$ vanishes so we have directly written
$\lambda\sim x^{2}$. According to Eq. (\ref{tp2}), we have
\begin{equation}\label{eigen4}
\lambda={8\mu\over M}\biggl ({\beta\over\beta_{c}}-1\biggr ),
\end{equation}
where $\mu$ has to be determined self-consistently. We thus expand the perturbation as
\begin{equation}\label{eigen5}
q(\theta)=q_{0}(\theta)+x \ q_{1}(\theta)+x^2 \ q_{2}(\theta)+...
\end{equation}
and we introduce the differential operator
\begin{equation}\label{eigen6}
{\cal L}q=\pi{d^{2}q\over d\theta^{2}}+\int_{0}^{2\pi}q(\theta')\cos(\theta-\theta')d\theta'.
\end{equation}
To order $0$, we have
\begin{equation}\label{eigen7}
{\cal L}q_{0}=0,
\end{equation}
yielding $q_{0}=\sin\theta$. To order $1$, we get
\begin{equation}\label{eigen8}
{\cal L}q_{1}=\pi\sin (2\theta),
\end{equation}
yielding
\begin{equation}\label{eigen9}
q_{1}=-{1\over 4}\sin (2\theta)+C\sin\theta,
\end{equation}
where $C$ is an arbitrary constant. Finally, to order $2$, we have after simplification
\begin{equation}\label{eigen10}
{\cal L}q_{2}=\biggl (\mu+{\pi\over 4}\biggr )\sin\theta+\pi C \sin(2\theta)-{3\pi\over 8}\sin (3\theta),
\end{equation}
yielding
\begin{equation}\label{eigen11}
\mu=-{\pi\over 4},
\end{equation}
and
\begin{equation}\label{eigen12}
q_{2}=D\sin\theta-{C\over 4}\sin (2\theta)+{1\over 24}\sin (3\theta),
\end{equation}
where $D$ is an arbitrary constant. Therefore, close to the critical point, the largest eigenvalue of Eq. (\ref{eigen1}) is 
\begin{equation}\label{eigen13}
\lambda M=-{\pi\over 4}x^{2}, \qquad {\rm or}\qquad \lambda M=-2\pi\biggl ({\beta\over\beta_{c}}-1\biggr ).
\end{equation}

We can obtain the expression of the eigenvalue by a slightly different
method. We consider the Hilbert space of $2\pi$-periodic continuous
real functions with scalar product
\begin{equation}\label{eigen14}
\langle f,g \rangle = {1\over 2 \pi}\int_{0}^{2\pi} f(\theta)g(\theta)d\theta.
\end{equation}
We note that the operator (\ref{eigen6}) is Hermitian in the sense that
\begin{equation}\label{eigen15}
 \langle {\cal L}f,g \rangle=\langle f,{\cal L}g \rangle.
\end{equation}
The equation obtained to second order can be written
\begin{equation}\label{eigen16}
{\cal L}q_{2}=g(\theta),
\end{equation}
with
\begin{equation}\label{eigen17}
g(\theta)=\biggl (\mu+{\pi\over 4}\biggr )\sin\theta+\pi C \sin(2\theta)-{3\pi\over 8}\sin (3\theta).
\end{equation}
We note $q_{0}$ the Kernel of ${\cal L}$, i.e. ${\cal L}q_{0}=0$. Then, we have the condition of solvability
\begin{equation}\label{eigen18}
\langle q_{0},g \rangle=0.
\end{equation}
Indeed,
\begin{equation}\label{eigen19}
\langle q_{0},g \rangle=\langle q_{0},{\cal L}q_{2} \rangle=\langle {\cal L}q_{0},q_{2}\rangle=0.
\end{equation}
In our case, $q_{0}=\sin\theta$, so that the condition  $\langle q_{0},g \rangle=0$ with (\ref{eigen17}) yields Eq. (\ref{eigen11}).

In the microcanonical ensemble, we have to study the eigenvalue
equation $(V\neq 0)$:
\begin{eqnarray}\label{eigen20}
{d\over d\theta}\biggl ({1\over\rho(\theta)}{dq\over d\theta}\biggr )+{k\over 2\pi T}\int_{0}^{2\pi}q(\theta')\cos(\theta-\theta')d\theta'\nonumber\\
={2V\over MT^{2}}{d\Phi\over d\theta}+2\lambda' q.
\end{eqnarray}
To leading order in $B\ll 1$, the term ${2V\over MT^{2}}{d\Phi\over d\theta}$ can be written ${2\pi\over M}x^{2}\sin\theta$. Therefore, Eq. (\ref{eigen20}) becomes
\begin{eqnarray}\label{eigen21}
{d\over d\theta}\biggl \lbrace \pi\biggl\lbrack  1+x\cos\theta+{x^{2}\over 4}(1+2\cos^{2}\theta)\biggr\rbrack{dq\over d\theta}\biggr \rbrace\qquad\qquad\qquad\nonumber\\
+\biggl (1+{x^{2}\over 8}\biggr )\int_{0}^{2\pi}q(\theta')\cos(\theta-\theta')d\theta'=\pi x^{2}\sin\theta+\mu' x^{2} q.\nonumber\\
\end{eqnarray}
The eigenvalue is now
\begin{equation}\label{eigen22}
\mu'=-{\pi\over 4}-\pi=-{5\pi\over 4},
\end{equation}
yielding
\begin{equation}\label{eigen23}
\lambda M=-{5\pi\over 4}x^{2}, \qquad {\rm or}\qquad \lambda M=-10\pi\biggl ({\beta\over\beta_{c}}-1\biggr ).
\end{equation}
Although the onset of instability is the same in the two ensembles, the eigenvalues differ in the condensed phase.

\section{Some useful identities}
\label{sec_idro}

For any system described by a distribution function $f(\theta,v)$, we define
 the density and the pressure by
\begin{eqnarray}
\rho=\int f dv, \qquad p=\int f v^{2}\ dv.
\label{idro1}
\end{eqnarray}
The kinetic temperature $T=p/\rho$ is equal to the velocity dispersion of the
particles. If the distribution function is of the form $f=f(\epsilon)$
with $\epsilon={v^{2}\over 2}+\Phi(\theta)$, we have
\begin{eqnarray}
{dp\over d\theta}=\int f'(\epsilon){d\Phi\over d\theta}v^{2}dv={d\Phi\over d\theta}\int {\partial f\over\partial v}vdv\nonumber\\
=-{d\Phi\over d\theta}\int fdv=-\rho {d\Phi\over d\theta},
\label{idro2}
\end{eqnarray}
which returns the condition of hydrostatic balance (\ref{hydro1}). This relation is equivalent to
\begin{eqnarray}
{p'(\rho)\over \rho}=-{1\over \rho'(\Phi)}.
\label{idro3}
\end{eqnarray}
Now, we note that
\begin{eqnarray}
{d\rho\over d\Phi}=\int f'(\epsilon)\ dv =\int {\partial f\over\partial v}{1\over v}\ dv.
\label{idro4}
\end{eqnarray}
This yields the important identity
\begin{eqnarray}
{p'(\rho)\over \rho}={-1\over \int {\partial f\over\partial v}{1\over v}\ dv}.
\label{idro5}
\end{eqnarray}
This identity is valid for both homogeneous and inhomogeneous systems. It may be useful to rederive it in the case of homogeneous systems since $\Phi=0$ in that case so that the above precedure is ill-determined.

We consider a homogeneous system described by the distribution
function $f=F(\beta {v^{2}\over 2}+\alpha)$ where $\alpha$ is a
Lagrange multiplier taking into account the conservation of mass
(normalization). In that case, $\rho=\rho(\alpha)$ and
$p=p(\alpha)$. Now,
\begin{eqnarray}
{dp\over d\alpha}=\int F'(\beta {v^{2}\over 2}+\alpha)v^{2}\ dv\qquad\qquad\qquad\nonumber\\
=\int {\partial\over\partial v}F(\beta {v^{2}\over 2}+\alpha) {v\over\beta}dv=-{\rho \over\beta},
\label{idro6}
\end{eqnarray}
and
\begin{eqnarray}
{d\rho\over d\alpha}=\int F'(\beta {v^{2}\over 2}+\alpha)\ dv\qquad\qquad\qquad\nonumber\\=\int {\partial\over\partial v}F(\beta {v^{2}\over 2}+\alpha) {1\over\beta v}dv=\int {f'(v)\over \beta v}dv .
\label{idro7}
\end{eqnarray}
Eliminating $\alpha$ from the foregoing relations, we obtain
\begin{eqnarray}
{p'(\rho)\over\rho}={-1\over \int {f'(v)\over v}dv},
\label{idro8}
\end{eqnarray}
which is consistent with Eq. (\ref{idro5}).  Now, introducing the velocity of
sound $c_{s}^{2}=p'(\rho)$ and using $\rho=M/2\pi$, we get
\begin{eqnarray}
c_{s}^{2}=-{M\over 2\pi}{1\over  \int_{-\infty}^{+\infty} {f'(v)\over v}dv}.
\label{idro9}
\end{eqnarray}

\section{General dispersion relation for a gaseous system}
\label{sec_disJ}

We consider the Euler equations
\begin{eqnarray}
{\partial \rho\over\partial t}+\nabla\cdot (\rho {\bf u})=0,
\label{disJ1}
\end{eqnarray}
\begin{eqnarray}
\rho\biggl \lbrack {\partial {\bf u}\over\partial t}+({\bf u}\cdot \nabla){\bf u}\biggr \rbrack=-\nabla p-\rho\nabla\Phi,
\label{disJ2}
\end{eqnarray}
in $D$ dimensions and for an arbitrary binary potential of interaction  $u(|{\bf r}-{\bf r}'|)$ such that
\begin{eqnarray}
\Phi({\bf r},t)=\int u({\bf r}-{\bf r}')\rho({\bf r}',t)d^{D}{\bf r}'.
\label{disJ3}
\end{eqnarray}
We also consider an arbitrary barotropic equation of state
$p=p(\rho)$.  Clearly, $\rho={\rm Cst.}$, ${\bf u}={\bf 0}$ and
$\Phi=0$ is a stationary solution of Eq. (\ref{disJ1})-(\ref{disJ3})
provided that $\int u({\bf x})d^{D}{\bf x}=0$ (for the gravitational
potential, this is not the case and we must advocate the Jeans
swindle). We shall restrict ourselves to such homogeneous
solutions. The linearized Euler equations are
\begin{eqnarray}
{\partial \delta \rho\over\partial t}+\rho\nabla\cdot \delta{\bf u}=0,
\label{disJ4}
\end{eqnarray}
\begin{eqnarray}
\rho {\partial \delta {\bf u}\over\partial t}=-p'(\rho)\nabla \delta\rho-\rho\nabla\delta \Phi,
\label{disJ5}
\end{eqnarray}
\begin{eqnarray}
\delta \Phi({\bf r},t)=\int u({\bf r}-{\bf r}')\delta \rho({\bf r}',t)d^{D}{\bf r}'.
\label{disJ6}
\end{eqnarray}
These equations can be combined to give
\begin{eqnarray}
{\partial^{2}\delta\rho\over\partial t^{2}}-c_{s}^{2}\Delta\delta\rho-\rho\Delta\delta\Phi=0,
\label{disJ7}
\end{eqnarray}
where we have introduced the velocity of sound
$c_{s}^{2}=p'(\rho)$. We introduce the Fourier transform of the
interaction potential such that
\begin{eqnarray}
u({\bf x})=\int e^{i{\bf k}\cdot {\bf x}}\hat{u}({\bf k})d^{D}{\bf k}.
\label{disJ8}
\end{eqnarray}
Taking the Fourier transform of Eqs. (\ref{disJ6}) and (\ref{disJ7}) with the 
convention $\delta \rho\sim e^{i({\bf k}\cdot {\bf r}-\omega t)}$, and combining the resulting expressions, we find that
\begin{eqnarray}
\omega^{2}=c_{s}^{2}k^{2}+(2\pi)^{D}\hat{u}({\bf k}) k^{2} \rho,
\label{disJ9}
\end{eqnarray}
which is the required dispersion relation. The system will be unstable if
\begin{eqnarray}
c_{s}^{2}+(2\pi)^{D}\hat{u}({\bf k})\rho<0.
\label{disJ9b}
\end{eqnarray}
In particular, it is necessary that $\hat{u}({\bf k})<0$ corresponding to attractive potentials. In that case, the condition of instability reads
\begin{eqnarray}
c_{s}^{2}<(c_{s}^{2})_{crit}\equiv (2\pi)^{D}\rho |\hat{u}({\bf k})|_{max}.
\label{disJ9c}
\end{eqnarray} 
Then, the unstable lengthscales are determined by
Eq. (\ref{disJ9b}). Various situations can happen (see Chavanis 2004b)
depending on the form of the potential $\hat{u}({\bf k})$. For the
gravitational interaction in $D=3$, using $\Delta u=4\pi G\delta({\bf
x})$, we have $(2\pi)^{3}\hat{u}({\bf k})=-4\pi G/k^{2}$. We recover
the usual formula (Binney \& Tremaine 1987)
\begin{eqnarray}
\omega^{2}=c_{s}^{2}k^{2}-4\pi G\rho.
\label{disJ10}
\end{eqnarray}
The system is always unstable ($(c_{s}^{2})_{crit}=\infty$) for sufficiently large wavelengths or equivalently for 
\begin{eqnarray}
k<k_{J}\equiv \biggl ({4\pi G\rho \over c_{s}^{2}}\biggr )^{1/2},
\label{disJ11}
\end{eqnarray}
where $k_{J}$ is the Jeans wave number.  For the HMF
model, $\hat{u}_{n}=0$ if $n\neq \pm 1$ and $\hat{u}_{\pm 1}=-{k\over
4\pi}$. In that case, Eq. (\ref{disJ9}) returns Eq. (\ref{ejl9}). The system is unstable for the mode $n=1$ if $c_{s}^{2}<kM/4\pi$.

\section{General dispersion relation for a stellar system}
\label{sec_disV}

We consider the Vlasov equation 
\begin{eqnarray}
{\partial f\over\partial t}+{\bf v}\cdot {\partial f\over\partial {\bf r}}-\nabla\Phi\cdot {\partial f\over\partial {\bf v}}=0,
\label{disV1}
\end{eqnarray}
in $D$ dimensions and for an arbitrary binary potential of interaction
$u(|{\bf r}-{\bf r}'|)$ as before. Clearly, a distribution function
$f=f_{0}({\bf v})$ which depends only on the velocity is a stationary
solution of Eq. (\ref{disV1}) under the same assumptions as before. We
shall restrict ourselves to such homogeneous solutions. The linearized
Vlasov equation is
\begin{eqnarray}
{\partial \delta f\over\partial t}+{\bf v}\cdot {\partial \delta f\over\partial {\bf r}}-\nabla\delta \Phi\cdot {\partial f_{0}\over\partial {\bf v}}=0,
\label{disV2}
\end{eqnarray}
which must be supplemented by
\begin{eqnarray}
\delta \Phi({\bf r},t)=\int u({\bf r}-{\bf r}')\delta \rho({\bf r}',t)d^{D}{\bf r}'.
\label{disV3}
\end{eqnarray}
Taking the Fourier transform of Eqs. (\ref{disV2}) and (\ref{disV3}) with the convention $\delta f\sim e^{i({\bf k}\cdot {\bf r}-\omega t)}$ and combining the resulting expressions, we find that
\begin{eqnarray}
(\omega-{\bf k}\cdot {\bf v})\delta\hat{f}+(2\pi)^{D}\hat{u}({\bf k}){\bf k}\cdot{\partial f_{0}\over\partial {\bf v}}\int\delta\hat{f}({\bf k},{\bf v},\omega)d^{D}{\bf v}=0.\nonumber\\
\label{disV4}
\end{eqnarray}
This can be rewritten
\begin{eqnarray}
\epsilon({\bf k},\omega)\equiv 1+(2\pi)^{D}\hat{u}({\bf k})\int {{\bf k}\cdot{\partial f_{0}\over\partial {\bf v}}\over \omega-{\bf k}\cdot {\bf v}}d^{D}{\bf v}=0,
\label{disV5}
\end{eqnarray}
which is the dispersion relation of the system. For the gravitational interaction in $D=3$, using $(2\pi)^{3}\hat{u}({\bf
k})=-4\pi G/k^{2}$, we recover the usual formula (Binney \& Tremaine 1987)
\begin{eqnarray}
1-{4\pi G\over k^{2}}\int {{\bf k}\cdot{\partial f_{0}\over\partial {\bf v}}\over \omega-{\bf k}\cdot {\bf v}}d^{D}{\bf v}=0.
\label{disV6}
\end{eqnarray} 
For the HMF model, we recover Eq. (\ref{ik1}).

\section{General dispersion relation for a Brownian system}
\label{sec_disB}

We finally consider a gas of Brownian particles in interaction described in the strong friction limit by the Smoluchowski equation 
\begin{eqnarray}
{\partial\rho\over\partial t}=\nabla\cdot \biggl\lbrack {1\over\xi}(T\nabla\rho+\rho\nabla\Phi)\biggr \rbrack,
\label{disB1}
\end{eqnarray} 
where $\Phi$ is given by Eq. (\ref{disJ3}). Under the same conditions as in Appendix \ref{sec_disJ}, the linearized Smoluchowski equation is
\begin{eqnarray}
{\partial\delta\rho\over\partial t}=\nabla\cdot \biggl\lbrack {1\over\xi}(T\nabla\delta\rho+\rho\nabla\delta\Phi)\biggr \rbrack.
\label{disB2}
\end{eqnarray} 
Taking the Fourier transform of Eq. (\ref{disB2}), we obtain the dispersion relation
\begin{eqnarray}
i\xi\omega=Tk^{2}+(2\pi)^{D}\hat{u}(k)k^{2}\rho,
\label{disB3}
\end{eqnarray} 
which can be compared to Eq. (\ref{disJ9}). In the gravitational case, we get
\begin{eqnarray}
i\xi\omega=Tk^{2}-4\pi G\rho.
\label{disB4}
\end{eqnarray} 
The stability condition coincides with the Jeans criterion  for an isothermal gas
\begin{eqnarray}
k<k_{J}\equiv \biggl ({4\pi G\rho \over T}\biggr )^{1/2}.
\label{disB5}
\end{eqnarray}
However, the stable modes are exponentially damped in the case of Brownian particles while they have an oscillatory nature in the case of gaseous systems. For a sinusoidal potential of interaction, we recover Eq. (\ref{bl4}).

\section{Generalization of the analytical solution (\ref{dyn5b}) 
to arbitrary perturbations}
\label{sec_gen}

In Sec. \ref{sec_dyn}, we have restricted our analysis to density
profiles that are always even,
i.e. $\rho(-\theta,t)=\rho(\theta,t)$. It is not difficult to relax
this hypothesis. Let us write the density profile in the general form
\begin{eqnarray}
\rho=\sum_{n=-\infty}^{+\infty}a_{n}(t)e^{in\theta}.
\label{gen1}
\end{eqnarray}
Substituting this decomposition in Eq. (\ref{dyn1}), we find that
\begin{eqnarray}
a_{0}={M\over 2\pi},
\label{gen2}
\end{eqnarray}
\begin{eqnarray}
\xi{da_{n}\over dt}+Tn^{2}a_{n}={k\over 2}n (a_{1}a_{n-1}-a_{-1}a_{n+1}).
\label{gen3}
\end{eqnarray}
To first order in  $T_{c}-T\ll T_{c}$, we can neglect $a_{n}$ with $|n|\ge 3$. We thus get 
\begin{eqnarray}
a_{\pm 2}={k\over 4T}a_{\pm 1}^{2}.
\label{gen4}
\end{eqnarray}
The equations for the modes $a_{\pm 1}$ are therefore
\begin{eqnarray}
\xi{da_{1}\over dt}+(T-T_{c})a_{1}=-{k^{2}\over 8T} a_{1}^{2}a_{-1},
\label{gen5}
\end{eqnarray}
\begin{eqnarray}
\xi{da_{-1}\over dt}+(T-T_{c})a_{-1}=-{k^{2}\over 8T} a_{-1}^{2}a_{1}.
\label{gen6}
\end{eqnarray}
At that point, it is convenient to introduce the variables $p=a_{1}a_{-1}$ and $X=a_{1}+a_{-1}$. They satisfy the differential equations
\begin{eqnarray}
\xi{dX\over dt}+(T-T_{c})X=-{k^{2}\over 8T}pX,
\label{gen7}
\end{eqnarray}
\begin{eqnarray}
\xi{dp\over dt}+2(T-T_{c})p=-{k^{2}\over 4T}p^{2}.
\label{gen8}
\end{eqnarray}
The equation for $p$ is readily solved and we obtain
\begin{eqnarray}
p(t)={2A (T_{c}-T) e^{2(T_{c}-T)t/\xi}\over 1+{A k^{2}\over 4T}e^{2(T_{c}-T)t/\xi}}.
\label{gen9}
\end{eqnarray}
Substituting this result in Eq. (\ref{gen7}) and solving the resulting equation, we get
\begin{eqnarray}
X(t)={B\over \sqrt{|{A k^{2}\over 4T}+e^{-2(T_{c}-T)t/\xi}|}}.
\label{gen10}
\end{eqnarray}
The modes $a_{1}$ and $a_{-1}$ are deduced from $X$ and $p$ by solving the second order equation $a^{2}-Xa+p=0$ yielding
\begin{eqnarray}
a_{\pm 1}(t)={X(t)\pm \sqrt{\Delta(t)}\over 2 }.
\label{gen11}
\end{eqnarray} 
The discriminant can be written
\begin{eqnarray}
\Delta={B^{2}-8|A|(T_{c}-T)\over |{k^{2}A\over 4T}+e^{-2(T_{c}-T)t/\xi}|}.
\label{gen12}
\end{eqnarray}
The constants of integration $A$ and $B$, which can be positive
or negative, are fixed by the initial condition. They must satisfy
$\Delta(0)\ge 0$, i.e. $B^{2}\ge 8|A|(T_{c}-T)$. Then, 
$\Delta(t)\ge 0$ at all times. If initially $a_{-1}(0)=a_{1}(0)$,
then $\Delta(0)=0$. By Eq. (\ref{gen12}), this implies that
$\Delta(t)=0$ for all times. Therefore, if the initial perturbation is
even, it remains even during all the evolution. We thus
recover the results of section \ref{sec_dyn} with $a_{1}$ equals to
$2a_{\pm 1}$ with the present notations. Finally, for $T=T_{c}$,
Eq. (\ref{gen7}) and (\ref{gen8}) lead to
\begin{eqnarray}
p(t)={1\over {{1\over p(0)}+{k^{2}t\over 4\xi T_{c}}}},\quad X(t)={X(0)\over \sqrt{{1}+{k^{2}p(0)t\over 4\xi T_{c}}}}.
\label{gen13}
\end{eqnarray}

We can also use this approach to study the dynamical stability of stationary solutions of Eq. (\ref{dyn1}). By truncation to order $2$ ($a_3=a_{-3}=0$), we obtain the equilibrium relations : $B=-k a_1^{e}$ (we suppose $a_1^e=a_{-1}^e$ without lack of generality), $a_2^e=a_{-2}^e={2\over k}(T_c-T)$  and $B^2 = 8T_c(T_c-T)$. These expressions are valid up to order 1 in $T_c-T$. Let us compute the four smallest eigenvalues corresponding to the relaxation towards equilibrium. Defining $a_n = a_n^e + \exp (\lambda t) \delta a_n$, we obtain after linearization $MY=0$ where $Y$ is a row vector of components $^T(\delta a_1,\delta a_{-1},\delta a_{2},\delta a_{-2})$ and 
\begin{eqnarray}
M=\left( \begin{array}{cccc}
-\xi\lambda +\Delta T & -\Delta T & -\sqrt{2T_{c}\Delta T} & 0\\
-\Delta T & -\xi\lambda +\Delta T & 0 & -\sqrt{2T_{c}\Delta T}\\
4\sqrt{2T_{c}\Delta T} & 0 & -\xi\lambda -4T & 0\\
0 & -4\sqrt{2T_{c}\Delta T} & 0 & -\xi\lambda -4T
\end{array}\right)\nonumber
\end{eqnarray}
where we have set $\Delta T=T_{c}-T$. The eigenvalues are the zeros
of the determinant of $M$. The two lowest eigenvalues scale as $(T_{c}-T)$ so we set  $\lambda={1\over\xi}({T}_{c}-{T})\lambda_{1}$. First dividing the two
first rows and the two first columns by $\sqrt{(T_{c}-T)}$, and
dividing the two last rows by $2\sqrt{2T_c}$ and the two last columns
by $\sqrt{2T_c}$ we obtain at leading order
\begin{eqnarray}
\det \left( \begin{array}{cccc}
-\lambda _{1}+1 & -1 & -1 & 0\\
-1 & -\lambda _{1}+1 & 0 & -1\\
2 & 0 & -1 & 0\\
0 & -2 & 0 & -1
\end{array}\right) =0\nonumber
\end{eqnarray}
which gives $\lambda ^{2}_{1}+2\lambda _{1}=0$. The smallest
eigenvalue \( \lambda_{1}=0 \) is the neutral mode associated to the
angle rotation invariance, whereas the smallest non zero eigenvalue is
$\lambda_{1}=-2$ or equivalently $\lambda =-{2\over\xi}(T_{c}-T)$.
The two consecutive eigenvalues are of order $0$. Considering the
determinant of $M$ at order $0$, we obtain two degenerate eigenvalues
$\lambda=-4T_{c}/\xi$. We note that this degeneracy will be removed at
next order.

\end{document}